\documentclass[aps,pra,reprint,amsmath,amssymb,amsfonts,superscriptaddress,longbibliography,nofootinbib]{revtex4-2}

\usepackage[T1]{fontenc}
\usepackage[colorlinks=true, allcolors=blue]{hyperref}
\usepackage{lipsum}
\usepackage{verbatim}
\usepackage{enumitem}
\usepackage{stmaryrd}
\usepackage{xspace}

\usepackage{bm}
\usepackage{amsthm}
\usepackage{array}
\usepackage{mathtools}

\usepackage{physics}
\usepackage{siunitx}

\usepackage{algorithmic}
\usepackage{algorithm}

\usepackage[table,svgnames,dvipsnames]{xcolor}
\usepackage{graphicx}
\usepackage{tikz}
\usepackage{quantikz}

\usepackage{tabularx}
\usepackage{booktabs}
\usepackage{multirow}
\usepackage{makecell}
\usepackage{dcolumn}

\DeclareMathOperator*{\maximize}{maximize}

\DeclareMathOperator{\poly}{poly}

\DeclareMathOperator{\wt}{wt}
\DeclareMathOperator{\symdiff}{\triangle}

\DeclareMathOperator{\vbeta}{\vb*{\beta}}

\DeclareMathOperator*{\prodr}{\overrightarrow{\prod}}
\DeclareMathOperator*{\prodl}{\overleftarrow{\prod}}

\usepackage{cleveref}
\crefname{equation}{Eq.}{Eqs.}
\crefname{section}{Section}{Sections}
\crefname{appendix}{Appendix}{Appendices}
\crefname{figure}{Figure}{Figures}
\crefname{table}{Table}{Tables}
\crefname{theorem}{Theorem}{Theorems}

\newcommand{\ie}{\textit{i}.\textit{e}.}
\newcommand{\eg}{\textit{e}.\textit{g}.}

\theoremstyle{definition}

\theoremstyle{plain}

\newtheorem{theorem}{Theorem}

\usepackage[many]{tcolorbox}
\tcolorboxenvironment{theorem}{
  colback=blue!5!white,
  colframe=blue,
  boxrule=0.5pt,
  boxsep=1pt,
  left=2pt,right=2pt,top=2pt,bottom=2pt,
  oversize=2pt,
  rounded corners,
  before skip=\topsep,
  after skip=\topsep,
}

\usepackage[caption=false]{subfig}
\newcolumntype{x}[1]{>{\centering\arraybackslash\hspace{0pt}}p{#1}}

\newcommand{\nocontentsline}[3]{}
\newcommand{\tocless}[2]{\vspace{4ex}\bgroup\let\addcontentsline=\nocontentsline#1{#2}\egroup}
\newcommand{\toclesslabel}[3]{\vspace{4ex}\bgroup\let\addcontentsline=\nocontentsline#1{#2\label{#3}}\egroup}

\newlength\replength
\newcommand\repfrac{.33}

\newcommand\rulewidth{.6pt}
\newcommand\tdashfill[1][\repfrac]{\cleaders\hbox to \replength{%
  \smash{\rule[\arraystretch\ht\strutbox]{\repfrac\replength}{\rulewidth}}}\hfill}
\newcommand\tabdashline{%
  \makebox[0pt][r]{\makebox[\tabcolsep]{\tdashfill\hfil}}\tdashfill\hfil%
  \makebox[0pt][l]{\makebox[\tabcolsep]{\tdashfill\hfil}}%
  \\[-\arraystretch\dimexpr\ht\strutbox+\dp\strutbox\relax]%
}
\newcommand\tdotfill[1][\repfrac]{\cleaders\hbox to \replength{%
  \smash{\raisebox{\arraystretch\dimexpr\ht\strutbox-.1ex\relax}{.}}}\hfill}

\usepackage[autostyle,english=american]{csquotes}
\MakeOuterQuote{"}

\appdef \turnpage {%
  \AddToHookNext{shipout/after}{%
    \global\pdfpageattr\expandafter{\the\pdfpageattr/Rotate 90}%
    \AddToHookNext{shipout/after}{%
      \global\pdfpageattr\expandafter{\the\pdfpageattr/Rotate 0}%
    }%
  }%
}

\usepackage{orcidlink}

\begin{document}

\title{Analytical Expressions for the Quantum Approximate Optimization Algorithm \\ and its Variants}

\author{Truman Yu \surname{Ng}\,\orcidlink{0009-0006-5036-2298}}
\thanks{These two authors contributed equally.}
\affiliation{Department of Physics, National University of Singapore, Singapore 117551, Republic of Singapore}
\affiliation{Quantum Innovation Centre (Q.InC), Agency for Science, Technology and Research (A*STAR), 2 Fusionopolis Way, Innovis \#08-03, Singapore 138634, Republic of Singapore\looseness=-1}

\author{Jin Ming \surname{Koh}\,\orcidlink{0000-0002-6130-5591}}
\thanks{These two authors contributed equally.}
\affiliation{Quantum Innovation Centre (Q.InC), Agency for Science, Technology and Research (A*STAR), 2 Fusionopolis Way, Innovis \#08-03, Singapore 138634, Republic of Singapore\looseness=-1}
\affiliation{Department of Physics, Harvard University, Cambridge, Massachusetts 02138, USA}

\author{Dax Enshan \surname{Koh}\,\orcidlink{0000-0002-8968-591X}}
\email{dax\_koh@ihpc.a-star.edu.sg}
\affiliation{Quantum Innovation Centre (Q.InC), Agency for Science, Technology and Research (A*STAR), 2 Fusionopolis Way, Innovis \#08-03, Singapore 138634, Republic of Singapore\looseness=-1}
\affiliation{Institute of High Performance Computing (IHPC), Agency for Science, Technology and Research (A*STAR), 1 Fusionopolis Way, \#16-16 Connexis, Singapore 138632, Republic of Singapore\looseness=-1}
\affiliation{Science, Mathematics and Technology Cluster, Singapore University of Technology and Design, 8 Somapah Road, Singapore 487372, Republic of Singapore\looseness=-1}

\begin{abstract}
The quantum approximate optimization algorithm (QAOA) is a near-term quantum algorithm aimed at solving combinatorial optimization problems. Since its introduction, various generalizations have emerged, spanning modifications to the initial state, phase unitaries, and mixer unitaries. In this work, we present an analytical study of broad families of QAOA variants. We begin by examining a family of QAOA with product mixers, which includes single-body mixers parametrized by multiple variational angles, and derive exact analytical expressions for the cost expectation on weighted problem graphs in the single-layer ansatz setting. We then analyze a family of QAOA that employs many-body Grover-type mixers, deriving analogous analytical expressions for weighted problem hypergraphs in the setting of arbitrarily many circuit ansatz layers. For both families, we allow individual phase angles for each node and edge (hyperedge) in the problem graph (hypergraph). Our results reveal that, in contrast to product mixers, the Grover mixer is sensitive to contributions from cycles of all lengths in the problem graph, exhibiting a form of non-locality. Our study advances the understanding of QAOA's behavior in general scenarios, providing a foundation for further theoretical exploration.
\end{abstract}

\maketitle

\toclesslabel\section{Introduction}{sec:intro}

\begin{figure*}
    \centering
    \begin{quantikz}
    \lstick{$\ket{\Omega(\vb*{\lambda}, \vb*{\omega})}$} & \gate[4, style={fill=red!10}]{U_{\mathrm{A}}(\vb*{\gamma}^{(1)})} & \gate[4, style={fill=blue!10}]{U_{\mathrm{B}}(\vb*{\beta}^{(1)})} & \ \ldots \ & \gate[4, style={fill=red!10}]{U_{\mathrm{A}}(\vb*{\gamma}^{(p)})} & \gate[4, style={fill=blue!10}]{U_{\mathrm{B}}(\vb*{\beta}^{(p)})} & \meter{} \\
    \lstick{$\ket{\Omega(\vb*{\lambda}, \vb*{\omega})}$} & & & \ \ldots \ & & & \meter{} \\
    \lstick{\hspace{-4.5em}$\vdots$} & & & \ \ldots \ & & & \hspace{0.5em}\vdots \\
    \lstick{$\ket{\Omega(\vb*{\lambda}, \vb*{\omega})}$} & & & \ \ldots \ & & & \meter{}
    \end{quantikz}
    \caption{\textbf{Circuit schematic for QAOA and its variants.} Following an initial product state preparation, alternating layers of phase unitaries $U_{\mathrm{A}}(\vb*{\gamma})$ and mixer unitaries $U_{\mathrm{B}}(\vb*{\beta})$ are applied, terminating with computational basis measurements. The unitaries are parametrized by vectors of angles $\vb*{\gamma}$ and $\vb*{\beta}$---see \cref{eq:overview/definition-UA-UB-general} for the general forms considered.}
    \label{fig:qaoa_circuit_diagram}
\end{figure*}
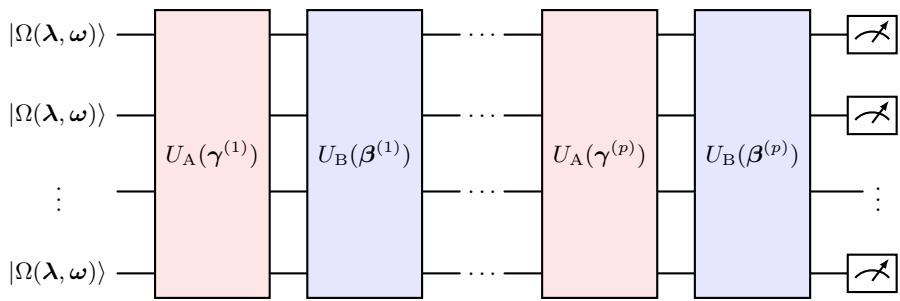

Quantum computation has garnered significant interest in recent years ~\cite{preskill2023quantum, dalzell2023quantum}. Among its numerous subfields~\cite{mcardle2020quantum, sun2024quantum, layden2023quantum, koh2023measurement, babbush2023exponential, koh2024realization, koh2022simulation, koh2022stabilizing, jafferis2022traversable, meglio2024quantum,leong2022variational,leong2023variational,conlon2024attainability}, quantum optimization has been particularly intensely studied due to its potential and applications~\cite{abbas2024challenges}. In the era of noisy intermediate-scale quantum (NISQ) devices~\cite{preskill2018quantum,cheng2023quantum,abughanem2024nisq}, 
a promising candidate for quantum optimization is the quantum approximate optimization algorithm (QAOA), first introduced by Farhi, Goldstone, and Gutmann in 2014~\cite{farhi2014quantum}. Since then, a broad generalization of QAOA's formalism, termed the quantum alternating operator ansatz~\cite{hadfield2019quantum}, has been proposed. To date, the application of QAOA to paradigmatic graph problems~\cite{sachdeva2024quantum, montanezbarrera2024universal, decross2022qubit, harrigan2021quantum,herrman2021impact,basso2022quantum,bae2024improved,weidinger2024performance,christiansen2024quantum,deller2023quantum}, Boolean satisfiability problems \cite{mandl2024amplitude,golden2023quantum,yu2023solution,zhang2024grover,boulebnane2024solving}, and more specialized domains~\cite{vikstal2020applying,brandhofer2022benchmarking, baker2022wasserstein, choi2020quantum} has been pursued, with varying degrees of success.

A key motivation for QAOA is its amenability to implementation via shallow circuits of simple structure---a necessity as NISQ devices are noisy \cite{bharti2022noisy,lau2022nisq}. Its structure consists of three main ingredients: an initial state, a diagonal phase unitary that encodes problem-specific information (\eg~the cost function), and a mixer unitary that generates superpositions. Starting from the initial state, QAOA alternates between applying phase and mixer unitaries over multiple cycles, concluding with computational basis measurements to yield solution samples \cite{choi2019tutorial,sturm2023theory}. 

Given current and near-term noise limitations \cite{xue2021effects, marshall2020characterizing,lotshaw2022scaling}, significant efforts have been directed at refining QAOA circuit designs to improve optimization performance while adhering to circuit depth constraints. These efforts include modifications to the phase unitaries \cite{wilkie2024quantum,langfitt2024phantom} and mixer unitaries, such as allowing arbitrary rotation axes for mixer unitaries, as in the free-angle mixer (FAM-) QAOA~\cite{PhysRevA.104.062428}, and assigning independent variational angles to individual cost terms to enhance the expressivity of the circuit ansatz~\cite{herrman2022multi,vijendran2024expressive,gaidai2024performance}, among other strategies~\cite{wang2020xy}. Other modifications at the algorithmic level include merging additional counterdiabatic driving into QAOA~\cite{PhysRevResearch.4.013141,PhysRevA.105.042415,PhysRevResearch.4.L042030,RevModPhys.91.045001,wurtz2022counterdiabaticity}, divide-and-conquer strategies \cite{zhou2023qaoa,junde2023large,bechtold2023investigating}, symmetry exploitation \cite{shaydulin2021classical,shaydulin2021exploiting,shi2022multiangle,nublein2024reducing}, and adaptive methods that adjust bias terms~\cite{PhysRevResearch.4.023249}, mixers~\cite{PhysRevResearch.4.033029,chalupnik2022augmenting}, and circuit operations~\cite{PhysRevLett.129.250502, PhysRevA.106.062414, mandl2024amplitude}.

The initial state of the algorithm can also be tuned to enhance performance \cite{tate2023bridging,tate2023warm}. A conventional choice for the input state is the uniform superposition over computational basis states. However, when equipped with additional information about the problem or potential solutions, tailored product states that are nonetheless easy to prepare can be advantageous \cite{tate2024theoretical,augustino2024strategies}. One such family of states are the warm-start states, which are utilized in the warm-start (WS-) QAOA~\cite{egger2021warm} variant. Further generalizations to precomputation-aided state preparation~\cite{wurtz2021classically} and leveraging entangled states~\cite{bartschi2020grover} have been reported. We refer readers to Ref.~\onlinecite{blekos2024review} for a review of the large multitude of QAOA variants proposed to date.

The original QAOA formulation, hereinafter referred to as vanilla QAOA, employs a non-entangling (\ie~product) mixer and is designed for solving unconstrained optimization problems, such as the \textsc{MaxCut} problem on graphs \cite{farhi2014quantum,guerreschi2019qaoa,leo2020quantum,wurtz2021maxcut,boulebnane2021predicting}. More sophisticated mixer designs have since broadened the applicability of QAOA to constrained problems, such as the \textsc{MaxIndependentSet} (MIS) problem. One such variant, termed GM-QAOA, employs entangling Grover-type mixers~\cite{bartschi2020grover, zhu2022multi}. In addition, recent studies have investigated pairing Grover-type mixers with thresholded phase Hamiltonians~\cite{golden2021threshold} and other types of constraint-enforcing mixers~\cite{fuchs2022constraint,fuchs2023optimal,larose2022mixer,PhysRevResearch.5.023071}.

While the landscape of QAOA variants has expanded significantly, only a few have been examined in detail from an analytical perspective. A systematic analytical study of QAOA is valuable as it provides an avenue for understanding and comparing the relative performance of different algorithm variants~\cite{herrman2022multi, vijendran2024expressive,bae2024recursive}, offers novel insights into QAOA behavior across various parameter regimes \cite{hadfield2022analytical,marwaha2021local,allcock2024dynamical,chernyavskiy2023entropic,montanaro2024quantum}, and helps determine globally optimal parameters or constrain the optimization space~\cite{sureshbabu2024parameter}. Given that simulating the output distribution of even a single-layer QAOA circuit can be difficult for classical computers, as suggested by complexity-theoretic arguments~\cite{farhi2016quantum, dalzell2020how, krovi2022average}, analytical approaches can be particularly useful. More fundamentally, an improved theoretical understanding of QAOA may provide broader insights into its behavior and computational power in general settings, beyond specific optimization problems, which are currently not well understood~\cite{blekos2024review}.

In prior literature, analytical results for the cost expectation for a single layer of the vanilla QAOA circuit ansatz have been reported for \textsc{MaxCut}~\cite{PhysRevA.97.022304,hadfield2018quantum}, MIS~\cite{brady2023iterative}, and the Ising Hamiltonian~\cite{ozaeta2022expectation,leontica2024exploring}. Similarly, single-layer cost expectations have been derived for multi-angle QAOA generalizations with non-entangling mixers~\cite{herrman2022multi,vijendran2024expressive}, where each vertex and edge in the problem graph has a unique variational angle, as well as for QAOA with modified phase operators~\cite{wilkie2024quantum,langfitt2024phantom}. Prior to this work, no exact analytical cost expectations were available for GM-QAOA, though approximate analytical expressions under ensemble averaging for an arbitrary number of layers had been reported~\cite{headley2023problem}. For simpler thresholded QAOA formulations, bounds on the cost expectation have also been derived~\cite{bridi2024analytical}. Generally, however, cost expectations for multiple QAOA layers are difficult to compute due to the exponentially growing number of terms (\ie~paths) involved. Beyond cost expectations, studies have also derived expressions for optimal variational angles~\cite{rabinovich2022progress, akshay2021parameter}.

In this work, we considerably generalize existing analytical results for cost expectations of QAOA on Ising-type optimization problems. First, we examine a family of QAOA employing product mixers parametrized by multiple variational angles---up to an angle per term in the cost Hamiltonian. This QAOA family, which we term PM-QAOA, subsumes the vanilla QAOA, WS-QAOA, FAM-QAOA, multi-angle (ma-) QAOA~\cite{herrman2022multi} and XQAOA~\cite{vijendran2024expressive} variants as special cases. We derive exact analytical solutions for edge-resolved and overall cost expectations of the PM-QAOA family on general problem graphs, accommodating vertex and edge weights, in the setting of a single circuit ansatz layer. 

We further study a second family of variants that use Grover-type mixers, termed GM-QAOA. Here, we consider general weighted problem hypergraphs, and derive analytical expressions for edge-resolved and overall cost expectations on arbitrary product initial states in the setting of multiple circuit ansatz layers. Our results distill connections between cost expectations and structural features of the problem graphs (or hypergraphs), such as their Euler subgraphs. In particular, we find that, in contrast to PM-QAOA, GM-QAOA is sensitive to contributions from cycles of all lengths in the problem graph, exhibiting a form of non-locality. Throughout our work, we generically allow the phase Hamiltonian of QAOA and the cost Hamiltonian of the optimization problem to be different, and accommodate independent variational angles for each term of the phase Hamiltonian for generality. 

Our approach contrasts with previous studies, such as Ref.~\onlinecite{sureshbabu2024parameter}, which examines the average-case behavior of QAOA under a distribution of variational angles, and Ref.~\onlinecite{headley2023problem}, which reports ensemble-averaged results for Grover-type QAOA. In contrast, we focus on deriving exact analytical results for any problem graph (or hypergraph) and any specified set of variational parameters.

The rest of our paper is organized as follows. In \cref{sec:overview}, we briefly review the formalism of QAOA. Thereafter, we present \textsc{MaxCut} as an introductory example in \cref{sec:intro-example-maxcut} before discussing PM-QAOA and GM-QAOA in \cref{sec:product,sec:grover} respectively. Finally, we conclude in \cref{sec:conclusion}, where we summarize our results and discuss open avenues for future research.

\tocless\section{Overview of QAOA
\label{sec:overview}}

\begin{table*}[!t]
    \centering
    \renewcommand*{\arraystretch}{1.2}
    \begin{tabular}{@{} p{4.4cm} p{4.3cm} p{4.3cm} p{4.3cm} @{}}
        \toprule 
        & \multicolumn{3}{c}{Variants} \\ 
        \cmidrule{2-4}
        Ingredients
            & Vanilla QAOA 
            & PM-QAOA (this work) 
            & GM-QAOA (this work) \\
        \midrule 
        Initial state $\ket{\psi_0}$ 
            & $\ket{s}$ 
            & $\ket{s}$ 
            & $\ket{\Omega(\vb*{\lambda},\vb*{\omega})}$ \\
        \midrule
        \makecell[cl]{Problem graph \\ $G = (V_G, E_G, w)$}
            & $\begin{aligned}
                & E_G = \{e\in 2^{V_G}: \abs{e} = 2\} \\
                & w : E_G \to 1
            \end{aligned}$
            & $\begin{aligned}
                & E_G = \{e\in 2^{V_G}: \abs{e} = 2\} \\
                & w = (a \in \mathbb{R}, h, J) \\
                & h : V_G \to \mathbb{R} \\
                & J : E_G \to \mathbb{R}
            \end{aligned}$
            & $\begin{aligned}
                & E_G \subseteq 2^{V_G} \\
                & w : E_G \to \mathbb{R}
            \end{aligned}$ \\
        \midrule
        Mixer Hamiltonian $B$ 
            & $\begin{aligned}
                \sum_{u = 1}^n X_u
            \end{aligned}$
            & $\begin{aligned}
                \sum_{u = 1}^n \vu{r}_u \cdot \vb{P}_u
            \end{aligned}$
            & $\begin{aligned}
                \ket{\Omega(\vb*{\lambda}, \vb*{\omega})} 
                \bra{\Omega(\vb*{\lambda}, \vb*{\omega})}
            \end{aligned}$ \\
        \midrule
        Number of layers $p$ 
            & $p = 1$ 
            & $p = 1$ 
            & $p \geq 1$ \\
        \midrule
        $\vb*{\gamma}$ angles 
            & $\begin{aligned}
                \gamma
            \end{aligned}$
            & $\begin{aligned}
                \begin{array}{lll}
                    \gamma_u & \text{for} & u \in V_G \\
                    \gamma_{uv} & \text{for} & \{u, v\} \in E_G
                \end{array}
            \end{aligned}$
            & $\begin{aligned}
                & (\vb*{\gamma}^{(1)}, \ldots, \vb*{\gamma}^{(p)}) \\[-4pt]
                & \begin{array}{lll}
                    \gamma^{(l)}_e & \text{for} & e \in E_G
                \end{array}
            \end{aligned}$ \\
        \midrule
        $\vb*{\beta}$ parameters 
            & $\begin{aligned}
                \beta
            \end{aligned}$
            & $\begin{aligned}
                \begin{array}{lll}
                    \beta_u & \text{for} & u \in V_G
                \end{array}
            \end{aligned}$
            & $\begin{aligned}
                (\beta^{(1)}, \ldots, \beta^{(p)})
            \end{aligned}$ \\
        \bottomrule
    \end{tabular}
    \caption{\textbf{QAOA variants for which analytical results for cost expectation $\expval{C}$ are available.} Prior literature have reported analytical expressions for vanilla QAOA in the contexts of \textsc{MaxCut}~\cite{PhysRevA.97.022304,hadfield2018quantum}, MIS~\cite{brady2023iterative}, and Ising-type problem Hamiltonians~\cite{ozaeta2022expectation}. Here, we consider two broad categories of QAOA variants using product mixers and Grover-type mixers~\cite{bartschi2020grover}, denoted as PM-QAOA and GM-QAOA respectively. The PM-QAOA category subsumes vanilla QAOA, WS-QAOA~\cite{egger2021warm}, FAM-QAOA~\cite{PhysRevA.104.062428}, ma-QAOA~\cite{herrman2022multi}, and XQAOA~\cite{vijendran2024expressive} as special cases. Analytical expressions for WS-QAOA and FAM-QAOA---and more broadly the general category of PM-QAOA---and for the GM-QAOA category are new results reported in this work. Prior work has reported approximate ensemble-averaged results for Grover-type QAOA~\cite{headley2023problem} whereas our work targets exact expressions.}
    \label{tab:QAOA_variants}
\end{table*}

To start, we provide an overview of the structure and formalism of QAOA, a quantum algorithm designed to solve combinatorial optimization problems. In such problems, the input consists of a collection of clauses $C_1,\ldots,C_\ell$ and an equinumerous collection of weights $c_1,\ldots,c_\ell$. Each clause $C_i$ is a Boolean expression over variables $x_,\ldots, x_n$, with each $x_i$ taking a value of either 0 or 1, and each weight $c_i$ is a real number. For a given assignment $\vb{x} =x_1 x_2 \ldots x_n \in \{0,1\}^n$, a clause is said to be \textit{satisfied} if substituting the values of $\vb{x}$ into the clause $C_i$ results in the clause evaluating to true. 

The goal of the optimization problem is to maximize the cost function 
\begin{equation}\begin{aligned}
    C(\vb{x}) = \sum_{j = 1}^\ell u_j(\vb{x}),
\end{aligned}\end{equation}
where $u_j(\vb{x}) = c_j$ if $\vb{x}$ satisfies $C_j$ and $u_j(\vb{x}) =0$ otherwise. In approximate optimization, one seeks an assignment $\vb{x}$ such that $C(\vb{x})$ is close to $\max_{\vb{x}} C(\vb{x})$, or equivalently, such that the approximation ratio $r(\vb{x}^*) = C(\vb{x}^*) / \max_{\vb{x}} C(\vb{x})$ is large.

Using $n$ qubits on a quantum computer, it is natural to associate each assignment $\vb{x}$ with the computational basis state $\ket{\vb{x}}$, and the objective function $C$ is accordingly promoted to a diagonal operator in this basis. Then, starting from an easy-to-prepare initial state $\ket{\psi_0}$, the QAOA circuit ansatz with $p \in \mathbb{Z}^+$ layers evolves the state through an alternating series of phase unitaries $U_{\mathrm{A}}$ and mixer unitaries $U_{\mathrm{B}}$,
\begin{align}
    \ket{\vb*{\gamma}, \vb*{\beta}} 
    &= U_{\mathrm{B}}(\vb*{\beta}^{(p)})
        U_{\mathrm{A}}(\vb*{\gamma}^{(p)})  \cdots 
        U_{\mathrm{B}}(\vb*{\beta}^{(1)}) 
        U_{\mathrm{A}}(\vb*{\gamma}^{(1)}) 
        \ket{\psi_0},
\end{align}
where $\vb*{\gamma} = (\vb*{\gamma}^{(1)}, \ldots, \vb*{\gamma}^{(p)})$ and $\vb*{\beta} = (\vb*{\beta}^{(1)}, \ldots, \vb*{\beta}^{(p)})$ are phase and mixer angles that parametrize the phase and mixer unitaries respectively. The QAOA protocol functions in two stages. First, in the training stage, by iteratively executing the circuit on a quantum computer and employing a classical optimizer, one variationally adjusts $(\vb*{\gamma}, \vb*{\beta})$ to maximize the measured objective expectation $\smash{\expval{C} = \mel{\vb*{\gamma}, \vb*{\beta}}{C}{\vb*{\gamma}, \vb*{\beta}}}$. Then, with $(\vb*{\gamma}, \vb*{\beta})$ fixed and with computational basis measurements, one samples assignments $\vb{x}$ over a number of shots, and the best-found $\vb{x}^*$ is chosen as the solution.

There is considerable freedom in the parametrized structure of the phase ($U_{\mathrm{A}}$) and mixer ($U_{\mathrm{B}}$) unitaries. Indeed, different choices define different variants of QAOA, as introduced in \cref{sec:intro}. For a unifying framework, we consider the general forms
\begin{equation}\begin{aligned}
    U_{\mathrm{A}}(\vb*{\gamma}^{(l)}) 
    &= e^{-i \vb*{\gamma}^{(l)} \cdot \vb{A}} 
    = \prod_{j = 1}^{\abs{\vb{A}}} 
        e^{-i \gamma_j^{(l)} A_j}, \\
    U_{\mathrm{B}}(\vb*{\beta}^{(l)}) 
    &= e^{-i \vb*{\beta}^{(l)} \cdot \vb{B}}
    = \prod_{j = 1}^{\abs{\vb{B}}} 
        e^{-i \beta_j^{(l)} B_j},
    \label{eq:overview/definition-UA-UB-general}
\end{aligned}\end{equation}
where $\vb{A}$ and $\vb{B}$ are sequences of phase and mixer Hamiltonian terms, each multiplied by a phase angle in $\vb*{\gamma}^{(l)}$ or a mixer angle in $\vb*{\beta}^{(l)}$ to form the unitaries, and we denote the overall Hamiltonians
\begin{equation}\begin{aligned}
    A = \sum_{j = 1}^{\abs{\vb{A}}} A_j,
    \qquad
    B = \sum_{j = 1}^{\abs{\vb{B}}} B_j.
\end{aligned}\end{equation}

These forms accommodate most QAOA variants, including the vanilla QAOA~\cite{farhi2014quantum}, WS-QAOA~\cite{egger2021warm}, FAM-QAOA~\cite{PhysRevA.104.062428}, and GM-QAOA~\cite{bartschi2020grover}; all of which we study in this work. In typical QAOA formulations, the phase and cost Hamiltonian terms, $\vb{A}$ and $\vb{C}$, are set to be identical, such that $A = C$. However, constrained optimization problems sometimes require $\vb{A} \neq \vb{C}$---an example is MIS, on which we elaborate in \cref{app:specialize/mis}. We therefore distinguish them for generality. 

We take $\ket{\psi_0}$ to be a general product state,
\begin{equation}
    \ket{\Omega(\vb*{\lambda}, \vb*{\omega})} 
    = \bigotimes_{i=1}^n R_Z(\lambda_i) R_Y(\omega_i) \ket{0},
    \label{eq:definition-Omega-state}
\end{equation}
where $\vb*{\lambda} \in [0, 2\pi)^n$ and $\vb*{\omega} \in [0, \pi]^n$ are fixed angles where $R_Z(\lambda) = \exp(-i Z \lambda / 2)$ and $R_Y(\omega) = \exp(-i Y \omega / 2)$ are rotations about the Pauli-$Z$ and $Y$ axes of the Bloch sphere respectively. As written, $\ket{\Omega(\vb*{\lambda}, \vb*{\omega})}$ spans all product states and is trivially preparable through single-qubit unitaries. Warm-start states used in variants of QAOA with classical precomputation~\cite{egger2021warm} are subsumed as special cases of $\ket{\Omega(\vb*{\lambda}, \vb*{\omega})}$. Additionally, we consider the equal superposition over all computational basis states, denoted as
\begin{equation}
    \ket{s} = \ket{+}^{\otimes n}
    = \frac{1}{\sqrt{2^n}} \sum_{\vb{x} \in \{0, 1\}^n} \ket{\vb{x}},
    \label{eq:definition-plus-state}
\end{equation}
which coincides with $\ket{\Omega(\vb*{\lambda}, \vb*{\omega})}$ with $\vb*{\lambda} = \vb{0}$ and $\vb*{\omega} = (\pi/2, \ldots, \pi/2)$ and is commonly used in QAOA without warm-start.

\tocless\section{Introductory Example---\\ \textsc{MaxCut} on Graphs
\label{sec:intro-example-maxcut}}

We first introduce simple versions of our results in the setting of the paradigmatic \textsc{MaxCut} problem. The general versions are presented later in \cref{sec:product,sec:grover}. Given an undirected graph $G$ with vertex set $V_G = [n]$ and edge set $E_G = \{\{u, v\}: u, v \in V_G\}$ comprising $m$ edges, the \textsc{MaxCut} problem seeks to partition the vertices into two disjoint sets $S_G \subset V_G$ and $V_G \setminus S_G$ such that the number of edges comprising a vertex in $S_G$ and the other in $V_G \setminus S_G$ is maximized. Explicitly, the classical formulation of the problem is
\begin{equation}\begin{aligned}
    & \maximize_{\vb{x} \in \{0, 1\}^n} \quad 
    C(\vb{x}) = \sum_{\{u, v\} \in E_G}
        x_u \oplus x_v,
    \label{eq:maxcut-classical-unweighted}
\end{aligned}\end{equation}
where $x_u \in \{0, 1\}$ is a binary variable associated with each vertex $u \in V_G$ labelling the partition that $u$ is assigned to, and $\oplus$ denotes the modulo-$2$ sum or XOR operation. A minor generalization of the problem is weighted \textsc{MaxCut}, for which a weight function $w: E_G \to \mathbb{R}$ on edges is associated with the graph $G$, and the objective is to maximize the total weight of edges crossing the cut,
\begin{equation}\begin{aligned}
    & \maximize_{\vb{x} \in \{0, 1\}^n} \quad 
    C(\vb{x}) = \sum_{\{u, v\} \in E_G}
        w_{uv} \left( x_u \oplus x_v \right),
    \label{eq:maxcut-classical-weighted}
\end{aligned}\end{equation}
where $w_{uv}$ is the weight of edge $\{u, v\}$. We denote the total weight of all edges by $W = \smash{\sum_{\{u, v\} \in E_G} w_{uv}}$. 

Associating each vertex with a qubit and promoting $C(\vb{x})$ to a diagonal operator in the computational basis of the $n$ qubits, the cost Hamiltonian $C$ straightforwardly comprises terms over the edges,
\begin{equation}\begin{aligned}
    C = \sum_{\{u, v\} \in E_G} C_{uv},
    \qquad 
    C_{uv} = \frac{w_{uv}}{2} \left( \mathbb{I} - Z_u Z_v \right),
    \label{eq:maxcut-hamiltonian}
\end{aligned}\end{equation}
where $\mathbb{I}$ is the identity and $Z_u$ denotes the Pauli-$Z$ operator acting on the $u^{\text{th}}$ qubit. The cost Hamiltonian terms form a vector $\vb{C} = [C_{uv}]_{\{u, v\} \in E_G}$. For simplicity in the present introductory context, we take the phase and cost Hamiltonians to be identical, $\vb{A} = \vb{C}$.

We consider two types of mixer unitaries, the product mixer which comprises independent single-qubit rotations and is non-entangling, and the Grover-type mixer which is a joint operation on all qubits and is entangling. We abbreviate QAOA with product and Grover-type mixers as PM-QAOA and GM-QAOA respectively. In the present \textsc{MaxCut} example, we restrict the QAOA circuit ansatz to comprise a single layer ($p = 1$), and we accordingly drop the layer labels on the mixer and phase angles.

\tocless\subsection{PM-QAOA for \textsc{MaxCut}
\label{sec:intro-example-maxcut/PM}}

The product mixer Hamiltonian we examine is of the form
\begin{equation}\begin{split}
    B^{\mathrm{PM}} = \sum_{u = 1}^n B^{\mathrm{PM}}_u,
    \qquad
    B^{\mathrm{PM}}_u = \vu{r}_u \cdot \vb{P}_u,
    \label{eq:product_mixer}
\end{split}\end{equation}
where each $\vu{r}_u = \smash{\mqty(r^X_u, r^Y_u, r^Z_u)} \in S^2$ is a unit vector of coefficients and $\vb{P}_u = \smash{\mqty(X_u, Y_u, Z_u)}$ are Pauli matrices that act on the $u^{\text{th}}$ qubit. To build the mixer unitary $U_{\mathrm{B}}(\vb*{\beta})$, a mixer angle $\beta_u \in \mathbb{R}$ is assigned to each mixer Hamiltonian term in
\begin{equation}\begin{split}
    \vb{B}^{\mathrm{PM}}
    = \mqty(
        B^{\mathrm{PM}}_1, 
        B^{\mathrm{PM}}_2, 
        \ldots, 
        B^{\mathrm{PM}}_n ),
    \label{eq:product_mixer_terms}
\end{split}\end{equation}
and accordingly $\vb*{\beta} = (\beta_1, \beta_2, \ldots, \beta_n)$ comprises $n$ angles. Likewise, each edge $\{u, v\}$ is associated with an angle $\gamma_{uv}$ and thus $\vb*{\gamma} \in \mathbb{R}^m$. The special case of single phase and mixer angles, as considered in the original formulations of vanilla QAOA~\cite{farhi2014quantum}, WS-QAOA~\cite{egger2021warm}, and FAM-QAOA~\cite{PhysRevA.104.062428}, is subsumed by setting $\vb*{\gamma} = (\gamma, \gamma, \ldots, \gamma)$ and $\vb*{\beta} = (\beta, \beta, \ldots, \beta)$. Our main result is presented in \cref{thm:intro-example/PM}. 

\vspace{8pt}
\begin{theorem}
    \label{thm:intro-example/PM}
    Single-layer PM-QAOA with the initial state $\ket{s}$ on the weighted \textsc{MaxCut} problem produces the edge-$\{u, v\}$ cost expectation value
    \begin{equation}\begin{split}
        \expval{C_{uv}}_{\ket{s}} 
        = \frac{w_{uv}}{2} \bigg[ 1 
            & - a^{XX}_{uv}(\vb*{\beta}) \xi^{XX}_{uv}(\vb*{\gamma})
            - a^{YY}_{uv}(\vb*{\beta}) \xi^{YY}_{uv}(\vb*{\gamma}) \\
            & - a^{YZ}_{uv}(\vb*{\beta}) \xi^{YZ}_{uv}(\vb*{\gamma})
            - a^{ZY}_{uv}(\vb*{\beta}) \xi^{ZY}_{uv}(\vb*{\gamma})
        \bigg],
        \label{eq:intro-example/PM/C-u-v-s-expectation}
    \end{split}\end{equation}
    where the coefficients $a^{PQ}_{uv}(\vb*{\beta})$ and $\xi^{PQ}_{uv}(\vb*{\gamma})$ for Pauli operators $P, Q \in \mathcal{P} = \{X, Y, Z\}$ are given in \cref{eq:intro-example/PM/a,eq:intro-example/PM/xi} respectively. The overall cost expectation is given by
    \begin{equation}\begin{aligned}
        \expval{C}_{\ket{s}} &= \sum_{\{u, v\} \in E_G} \expval{C_{uv}}_{\ket{s}}.
        \label{eq:intro-example/PM/C-s-expectation}
    \end{aligned}\end{equation}
\end{theorem}
\vspace{2pt}

In \cref{thm:intro-example/PM}, the coefficients $a^{PQ}_{uv}(\vb*{\beta})$ are
\begin{equation}\begin{split}
    a^{XX}_{uv}(\vb*{\beta}) 
        &= 4 \sin(\beta_u) \sin(\beta_v)        
            \Theta^{YX,-}_u(\beta_u)
            \Theta^{YX,-}_v(\beta_v), \\
    a^{YY}_{uv}(\vb*{\beta}) 
        &=  4 \sin(\beta_u) \sin(\beta_v)
            \Theta^{XY,+}_u(\beta_u)
            \Theta^{XY,+}_v(\beta_v), \\
    a^{YZ}_{uv}(\vb*{\beta}) 
        &= 2 \sin(\beta_u) \Theta^{XY,+}_u(\beta_u) \\
        & \quad \times \left[ 
            \cos^2(\beta_v) 
            - \sin^2(\beta_v) \left( 1 - 2(r^Z_v)^2 \right)
            \right],
    \label{eq:intro-example/PM/a}
\end{split}\end{equation}
where 
\begin{equation}\begin{split}
    \Theta^{PQ,\pm}_u(\beta_u) 
    = \cos(\beta_u)r_u^P \pm \sin(\beta_u) r_u^Q r_u^Z,
    \label{eq:intro-example/PM/Theta}
\end{split}\end{equation}
and the coefficients $\xi^{PQ}_{uv}(\vb*{\gamma})$ are
\begin{equation}\begin{split}
    \xi^{XX}_{uv}(\vb*{\gamma}) 
    &= \frac{1}{2} R^\bbslash_{uv}(\vb*{\gamma}) 
        R^\bbslash_{vu}(\vb*{\gamma}) \left[ 
            R^-_{uv}(\vb*{\gamma}) 
            + R^+_{uv}(\vb*{\gamma}) \right], \\
    \xi^{YY}_{uv}(\vb*{\gamma}) 
    &= \frac{1}{2} R^\bbslash_{uv}(\vb*{\gamma})
        R^\bbslash_{vu}(\vb*{\gamma}) \left[ 
            R^-_{uv}(\vb*{\gamma}) 
            - R^+_{uv}(\vb*{\gamma}) \right], \\
    \xi^{YZ}_{uv}(\vb*{\gamma}) 
    &= -R^\setminus_{uv}(\vb*{\gamma}) 
        \sin(w_{uv}\gamma_{uv}),
    \label{eq:intro-example/PM/xi}
\end{split}\end{equation}
where
\begin{equation}\begin{split}
    R^\setminus_{uv}(\vb*{\gamma}) 
    &= \prod_{g \in \mathcal{N}_{u \setminus v}} 
        \cos(w_{ug} \gamma_{ug}), \\
    R^\bbslash_{uv}(\vb*{\gamma}) 
    &= \prod_{g \in \mathcal{N}_{u \bbslash v}} 
        \cos(w_{ug} \gamma_{ug}), \\
    R^\pm_{uv}(\vb*{\gamma}) 
    &= \prod_{a \in \mathcal{N}_{uv}} 
        \cos(w_{ua}\gamma_{ua} \pm w_{va}\gamma_{va}).
    \label{eq:intro-example/PM/R}
\end{split}\end{equation}
Above, we note the decoupling of dependences of $\smash{\expval*{C_{uv}}}$ on the mixer and phase angles, $\vb*{\beta}$ and $\vb*{\gamma}$, separated in the $a$ and $\xi$ coefficients respectively. A symmetry property relates $a^{PQ}_{uv}(\vb*{\beta}) = a^{QP}_{vu}(\vb*{\beta})$ and $\xi^{PQ}_{uv}(\vb*{\gamma}) = \xi^{QP}_{vu}(\vb*{\gamma})$, which we show formally in \cref{app:product/symmetry}. 

\begin{figure}[!t]
    \centering
    \includegraphics[width=1.0\linewidth]{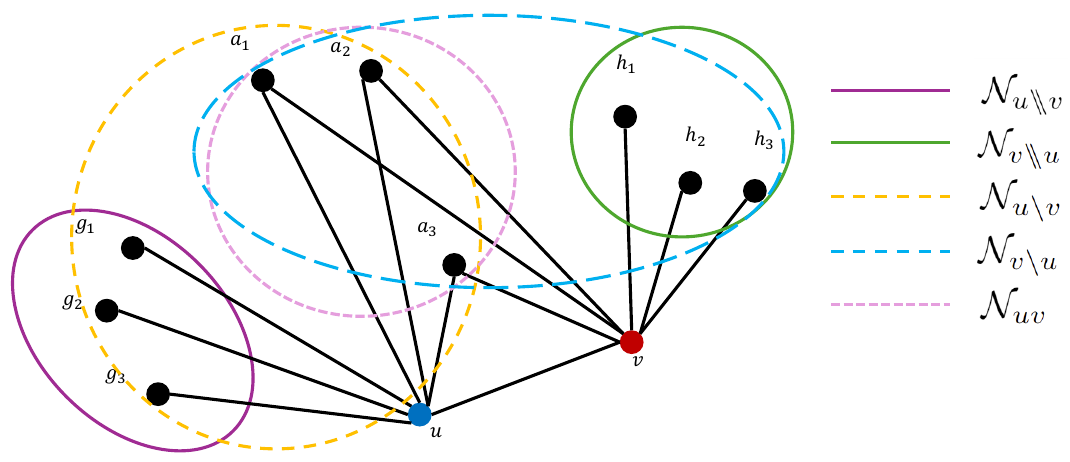}
    \caption{\textbf{Illustration of vertex neighborhoods used in the analysis of PM-QAOA.} An arbitrary edge $\{u, v\}$ is selected and key neighborhood sets of $u$ and $v$ are highlighted. $\mathcal{N}_{u}$ contains the neighbors of $u$, $\mathcal{N}_{u \setminus v}$ are the neighbors of $u$ excluding $v$, $\mathcal{N}_{u \bbslash v}$ are the neighbors of $u$ excluding neighbors of $v$, and vice versa for $v$. $\mathcal{N}_{uv}$ are the common neighbors of $u$ and $v$, which form triangles with edge $\{u, v\}$. See \cref{sec:intro-example-maxcut/PM} of the main text for more detailed discussion.}
    \label{fig:product_mixer_neighborhoods}
\end{figure}

In \cref{eq:intro-example/PM/R}, $\mathcal{N}_{u \setminus v} = \mathcal{N}_u \setminus \{v\}$ denotes the neighbors of $u$ excluding $v$, and $\mathcal{N}_{u \bbslash v} = \mathcal{N}_u \setminus \left( \mathcal{N}_v \cup \{v\} \right)$. That is, $\mathcal{N}_{u \bbslash v}$ comprises vertices incident to $u$ but which are not incident to $v$ to form a triangle containing $\{u, v\}$. Finally, $\mathcal{N}_{uv} = \mathcal{N}_u \cap \mathcal{N}_v$ is the set of neighbors of $u$ and $v$ that form triangles with $\{u, v\}$. A visualization of these neighborhood sets is provided in \cref{fig:product_mixer_neighborhoods}. 

We remark that, as defined, $\mathcal{N}_{u \setminus v} = \mathcal{N}_{u \bbslash v} \cup \mathcal{N}_{uv}$. That is, $\mathcal{N}_{u \setminus v}$ comprises a partitioning of neighbors of $u$ into a set that do not form triangles containing $\{u, v\}$ and those that do; such a partitioning enables the contributions of triangles to the cost expectation to be identified. In particular, $\xi^{XX}_{uv}(\vb*{\gamma})$ and $\xi^{YY}_{uv}(\vb*{\gamma})$ contain $R^\pm_{uv}(\vb*{\gamma})$ factors, which are products over $\mathcal{N}_{uv}$ and are precisely the cost contribution by triangles. Longer-length cycles in the problem graph $G$ beyond triangles do not contribute to cost expectation values.

Lastly, we observe that $\smash{\expval{C_{uv}}_{\ket{s}}}$ and $\smash{\expval{C}_{\ket{s}}}$ in \cref{thm:intro-example/PM} are computable in $\poly(n, m)$ time and $\order{1}$ space, that is, they are classically efficient to compute. Indeed, $a^{PQ}_{uv}(\vb*{\beta})$ can be evaluated in $\order{1}$ time and space, and $\xi^{PQ}_{uv}(\vb*{\gamma})$ likewise in $\poly(n, m)$ time and $\order{1}$ space.

We now give an overview of the main steps leading to \cref{thm:intro-example/PM}. The main quantity of interest is the cost expectation value of an edge $\{u, v\}$,
\begin{equation}\begin{split}
\expval{C_{u v}}_{\ket{s}} &= 
    \bra{s} 
    e^{+i \vb*{\gamma} \cdot \vb{A}} 
    e^{+i \vb*{\beta} \cdot \vb{B}^{\mathrm{PM}}} 
    \, C_{uv} \, \\
    & \qquad \qquad
    e^{-i \vb*{\beta} \cdot \vb{B}^{\mathrm{PM}}} 
    e^{-i \vb*{\gamma} \cdot \vb{A}}
    \ket{s}.
    \label{eq:intro-example/PM/explanation-C-u-v-s}
\end{split}\end{equation}

The identity term in $C_{uv}$ trivially commutes through, producing the $w_{uv} / 2$ scalar offset in \cref{eq:intro-example/PM/C-u-v-s-expectation}. It thus remains to consider $\smash{e^{+i \vb*{\beta} \cdot \vb{B}^{\mathrm{PM}}} Z_u Z_v e^{-i \vb*{\beta} \cdot \vb{B}^{\mathrm{PM}}}}$, which can be expressed as a linear combination $a^{PQ}_{uv}(\vb*{\beta}) P_u Q_v$ over $P, Q \in \mathcal{P} = \{X, Y, Z\}$ using the anti-commutation relations of the Pauli operators. Thus, ultimately we obtain
\begin{equation}
\expval{C_{u v}}_{\ket{s}} = 
    \sum_{P, Q \in \mathcal{P}}
    a^{PQ}_{uv}(\vb*{\beta})
    \underbrace{\bra{s} 
    e^{+i \vb*{\gamma} \cdot \vb{A}} 
    P_u Q_v e^{-i \vb*{\gamma} \cdot \vb{A}}
    \ket{s}}_{\xi^{PQ}_{uv}(\vb*{\gamma})}.
\end{equation}

To evaluate the $\xi^{PQ}_{uv}(\vb*{\gamma})$ coefficients, we note that the phase Hamiltonian terms $\vb{A} = \vb{C}$ commute and that $Z_u Z_v$ are involutory (see \cref{app:preliminaries/identities}), thus
\begin{equation}\begin{split}
    e^{\pm i \vb*{\gamma} \cdot \vb{A}}
    &= \prod_{\{u, v\} \in E_G} e^{\pm i \gamma_{uv} C_{uv}} \\
    &\sim \prod_{\{u, v\} \in E_G} \left[ 
            \cos(\frac{w_{uv} \gamma_{uv}}{2})\mathbb{I} 
                \right. \\
            & \qquad\qquad\qquad
            \pm \left. 
                i \sin(\frac{w_{uv} \gamma_{uv}}{2}) Z_u Z_v
            \right],
    \label{eq:intro-example/PM/exp-i-gamma-C}
\end{split}\end{equation}
where $\sim$ denotes equivalence up to a $\mathrm{U}(1)$ phase factor. Expanding the inner product yields terms of the forms
\begin{equation}\begin{split}
    & \mel{s}{e^{-i \vb*{\gamma} \cdot \vb{A}_u}
    e^{-i \vb*{\gamma} \cdot \vb{A}_v}
    P_u Q_v}{s}, \\
    & \mel{s}{e^{-i \gamma_{uv} w'_{uv} Z_u Z_v}
    e^{-i \vb*{\gamma} \cdot \vb{A}_u} P_u Q_v}{s}. 
\end{split}\end{equation}

The vector $\vb{A}_u$ ($\vb{A}_v$) contains phase Hamiltonian terms associated with edges comprising neighbors incident to vertex $u$ ($v$) on the problem graph. See \cref{app-eq:product/phase-unitary-expand-product} for expanded forms of $e^{-i \vb*{\gamma} \cdot \vb{A}_u}$ and $e^{-i \vb*{\gamma} \cdot \vb{A}_v}$. Evaluating these terms produce $\xi^{PQ}_{uv}(\vb*{\gamma})$, and equipped also with the corresponding $a^{PQ}_{uv}(\vb*{\beta})$ coefficients, the edge-resolved cost expectation $\expval{C_{u v}}$ and overall cost expectation $\expval{C}$ can be calculated through \cref{eq:intro-example/PM/C-u-v-s-expectation,eq:intro-example/PM/C-s-expectation}.

A complete derivation of our result in \cref{thm:intro-example/PM} and secondary results are detailed in \cref{app:specialize/maxcut}.

\tocless\subsection{GM-QAOA for \textsc{MaxCut}
\label{sec:intro-example-maxcut/GM}}

In contrast to the product mixer, the Grover-type mixer Hamiltonian we examine comprises a single term that acts jointly on all qubits,
\begin{equation}\begin{split}
    B^{\mathrm{GM}} = 
        \ket{\Omega(\vb*{\lambda},\vb*{\omega})}
        \bra{\Omega(\vb*{\lambda},\vb*{\omega})},
    \label{eq:intro-example/GM/grover-type-mixer}
\end{split}\end{equation}
where $\ket{\Omega(\vb*{\lambda},\vb*{\omega})}$ is the general product state defined in \cref{eq:definition-Omega-state}. Accordingly, the mixer unitary is parametrized by a single mixer angle $\beta \in \mathbb{R}$ due to the entangling nature of \cref{eq:intro-example/GM/grover-type-mixer}. Henceforth, we suppress the $(\vb*{\lambda},\vb*{\omega})$ labels for notational conciseness. In addition to $\ket{\Omega}$, we also examine the $\ket{s}$ product state as defined in \cref{eq:definition-plus-state}, which is a special case and enables simplifications. For generality, akin to PM-QAOA (see \cref{sec:intro-example-maxcut/PM}), we allow phase angles $\vb*{\gamma} \in \mathbb{R}^m$, such that each phase Hamiltonian term $A_{uv}$ for an edge $\{u, v\}$ is associated with an angle $\gamma_{uv}$. The more restricted scenario of having a single phase angle $\gamma$, as in the vanilla GM-QAOA formulations~\cite{bartschi2020grover,zhu2022multi}, is subsumed with $\vb*{\gamma} = (\gamma, \gamma, \ldots, \gamma)$. Our primary results are reported in \cref{thm:intro-example/GM/state-s,thm:intro-example/GM/state-Omega} below.

\vspace{8pt}
\begin{theorem}
    \label{thm:intro-example/GM/state-s}
    Single-layer GM-QAOA with the initial state $\ket{s}$ and mixer Hamiltonian $\ket{s} \bra{s}$ on the weighted \textsc{MaxCut} problem produces the edge-$\{u, v\}$ cost expectation value
    \begin{equation}\begin{split}
        \expval{C_{u v}}_{\ket{s}} = \frac{w_{uv}}{2}
            \left\{
            1
            - 2 \Re\left[ \left(e^{i \beta} - 1 \right)
                L_G(\vb*{\gamma}) L_G^{uv}(-\vb*{\gamma}) \right]
            \right\},
    \end{split}\end{equation}
    and the overall cost expectation value is given by
    \begin{equation}\begin{split}
        \expval{C}_{\ket{s}} = \frac{W}{2}
            - \Re\left[ \left(e^{i \beta} - 1 \right)
                L_G(\vb*{\gamma}) \overline{L}_G(-\vb*{\gamma}) \right],
    \end{split}\end{equation}
    where $L_G$, $L_G^{uv}$ for an edge $\{u, v\}$, and $\smash{\overline{L}_G}$ are structural factors dependent on cycles of the graph $G$ as defined in \cref{eq:intro-example/GM/definition-loop-factor-L,eq:intro-example/GM/definition-loop-factor-L-super}.
\end{theorem}
\vspace{2pt}
\begin{theorem}
    \label{thm:intro-example/GM/state-Omega}
    Single-layer GM-QAOA with the initial state $\ket{\Omega}$ and mixer Hamiltonian $\ket{\Omega} \bra{\Omega}$ on the weighted \textsc{MaxCut} problem produces the edge-$\{u, v\}$ cost expectation value
    \begin{equation}\begin{split}
        \expval{C_{u v}}_{\ket{\Omega}} = \frac{w_{uv}}{2} \bigg\{
            1
            & - \left[
                1 
                + 2 \left( 1 - \cos{\beta} \right) \abs{T_G(\vb*{\gamma})}^2
            \right] T_G^{uv}(\vb{0}) \\
            & - 2 \Re\left[ \left(e^{i \beta} - 1 \right)
                T_G(\vb*{\gamma}) T_G^{uv}(-\vb*{\gamma}) \right]
            \bigg\},
    \end{split}\end{equation}
    and the overall cost expectation value is given by
    \begin{equation}\begin{split}
        \expval{C}_{\ket{\Omega}} = \frac{W}{2}
            & - \frac{1}{2} \left[
                1 
                + 2 \left( 1 - \cos{\beta} \right) \abs{T_G(\vb*{\gamma})}^2
            \right] \overline{T}_G(\vb{0}) \\
            & - \Re\left[ \left(e^{i \beta} - 1 \right)
                T_G(\vb*{\gamma}) \overline{T}_G(-\vb*{\gamma}) \right],
    \end{split}\end{equation}
    where $T_G$, $T_G^{uv}$ for an edge $\{u, v\}$, and $\smash{\overline{T}_G}$ are structural factors dependent on subgraphs of the graph $G$ as defined in \cref{eq:intro-example/GM/definition-loop-factor-T,eq:intro-example/GM/definition-loop-factor-T-super}.
\end{theorem}
\vspace{2pt}

\begin{figure}
    \centering
    \includegraphics[width = \linewidth]{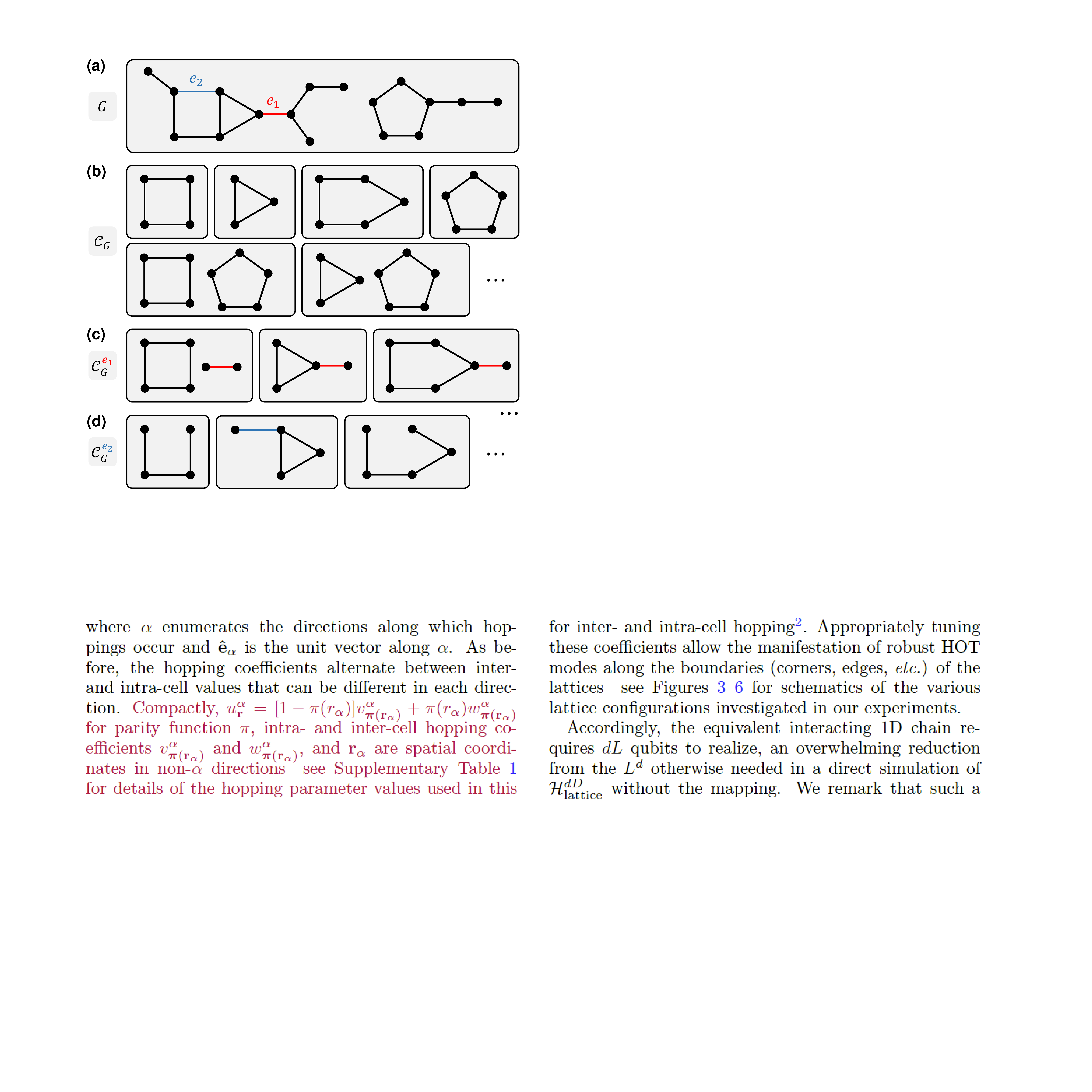}
    \caption{\textbf{Illustration of subgraph sets used in the analysis of GM-QAOA.} \textbf{(a)} An example graph $G$ with two arbitrary edges $e_1$ and $e_2$ highlighted in blue and red respectively. \textbf{(b)} The set $\mathcal{C}_G = \mathcal{C}_G^\varnothing$ comprises all even-regular subgraphs, equivalently Euler subgraphs, of $G$. This is equivalent to the cycle space of $G$, which is spanned by cycles in $G$ over the binary field. A subgraph $H \in \mathcal{C}_G$ can contain multiple disconnected cycles. \textbf{(c)} The set $\mathcal{C}_G^{e_1}$ comprises the symmetric difference between each subgraph in $\mathcal{C}_G$ and the edge $e_1$. The edge $e_1$ is added to each subgraph as it is not part of any cycle. \textbf{(d)} The set $\mathcal{C}_G^{e_2}$ likewise defined. The edge $e_2$ is added to each subgraph in $\mathcal{C}_G$ when absent and removed when already present. In the notation of \cref{sec:intro-example-maxcut/GM}, we write $\mathcal{C}_G^{uv}$ instead of $\mathcal{C}_G^e$ for an edge $e \in \{u, v\}$. Not drawn here is $\mathcal{H}_G$, the set of all subgraphs of $G$. In \cref{sec:grover} these definitions are generalized to refer to sets of subhypergraphs.}
    \label{fig:intro-example/illustrations}
\end{figure}

To express the structural factors in \cref{thm:intro-example/GM/state-s,thm:intro-example/GM/state-Omega}, it is most convenient to consider the subgraphs of $G$. We first discuss the simpler setting of the initial state $\ket{s}$ and mixer Hamiltonian $\ket{s} \bra{s}$. Let $\mathcal{C}_G$ be the set of even-regular subgraphs\footnote{We consider two subgraphs to be distinct when their edge sets are different.}, also termed Euler subgraphs, of $G$. Equivalently, $\mathcal{C}_G$ is the cycle space of $G$, which is the span of cycles in $G$ over the binary field. Moreover, we define $\mathcal{C}_G^{uv}$ to comprise the subgraphs in $\mathcal{C}_G$, but with edge $\{u, v\}$ added when absent or removed otherwise---see \cref{fig:intro-example/illustrations} for illustrations. Explicitly,
\begin{equation}\begin{split}
    \mathcal{C}_G^{uv} = \{H \symdiff \{\{u, v\}\}: H \in \mathcal{C}_G\},
\end{split}\end{equation}
where $\symdiff$ is the symmetric difference, herein understood to be acting on the edge sets of graphs. Thus, there is a bijection between $\mathcal{C}_G$ and $\mathcal{C}_G^{uv}$, and subgraphs in $\mathcal{C}_G^{uv}$ are even-regular except vertices $u$ and $v$ which are of odd degree. Then
\begin{equation}\begin{split}
    L_G(\vb*{\gamma})
    = & \left[ \smashoperator[r]{\prod_{\{u, v\} \in E_G}} \,\,\,
            \cos(\frac{w_{uv} \gamma_{uv}}{2}) \right] \\
        & \times
        \sum_{H \in \mathcal{C}_G}
        \left[ \smashoperator[r]{\prod_{\{u, v\} \in E_H}} \,\,\,
            i \tan(\frac{w_{uv} \gamma_{uv}}{2}) \right],
    \label{eq:intro-example/GM/definition-loop-factor-L}
\end{split}\end{equation}
and the definition of $L_G^{uv}(\vb*{\gamma})$ is identical to \cref{eq:intro-example/GM/definition-loop-factor-L} but with $\mathcal{C}_G^{uv}$ replacing $\mathcal{C}_G$. The super-structural factor $\overline{L}_G(\vb*{\gamma})$ is the weighted sum of $L_G^{uv}(\vb*{\gamma})$ over all edges, 
\begin{equation}\begin{split}
    \overline{L}_G(\vb*{\gamma})
    &= \sum_{\{u, v\} \in E_G} w_{uv} L_G^{uv}(\vb*{\gamma}).
    \label{eq:intro-example/GM/definition-loop-factor-L-super}
\end{split}\end{equation}

In the special case of an unweighted graph $G$, subsumed in our present formulation by setting $w_{uv} = 1$ uniformly for all edges, and with a single phase angle $\gamma$, the super-structural factor in \cref{eq:intro-example/GM/definition-loop-factor-L-super} can be expressed in a simpler form. By making use of the bijection between $\mathcal{C}_G$ and $\mathcal{C}_G^{uv}$, the factor $\overline{L}_G(\vb*{\gamma})$ can be transformed into a sum over $\mathcal{C}_G$, yielding after simplification
\begin{equation}\begin{split}
    \overline{L}_G(\vb*{\gamma})
    = & \cos[m](\frac{\gamma}{2}) 
        \sum_{k = 1}^m
        N_G(k) \cdot i^{k - 1} \cdot \bigg\{ \\
            & \qquad
            k \tan[k - 1](\frac{\gamma}{2})
            - (m - k)
            \tan[k + 1](\frac{\gamma}{2})
            \bigg\},
    \label{eq:intro-example/GM/unweighted-loop-factor-L-super}
\end{split}\end{equation}
where $N_G(k)$ is the the number of even-regular or Euler subgraphs in $G$ containing $k$ edges.

In the more general case of the initial state $\ket{\Omega}$ and mixer Hamiltonian $\ket{\Omega} \bra{\Omega}$, the structural factors depend not only on the cycle space of $G$ but generally on all subgraphs of $G$. Let $\mathcal{H}_G$ denote the set of subgraphs of $G$. Then
\begin{equation}\begin{split}
    T_G(\vb*{\gamma})
    = & \left[ \smashoperator[r]{\prod_{\{u, v\} \in E_G}} \,\,\,
            \cos(\frac{w_{uv} \gamma_{uv}}{2}) \right] \\
        & \times
        \sum_{H \in \mathcal{H}_G}
            \left[ \smashoperator[r]{\prod_{\{u, v\} \in E_H}} \,\,\,
                i \tan(\frac{w_{uv} \gamma_{uv}}{2}) \right] \\
        & \qquad \qquad \times
        \left[ \smashoperator[r]{\prod_{u \in V_H^{\mathrm{odd}}}} \,\,\,
            \cos{\omega_u} \right],
    \label{eq:intro-example/GM/definition-loop-factor-T}
\end{split}\end{equation}
where $V_H^{\mathrm{odd}}$ denotes the set of odd-degree vertices of the graph $H$. The definition of $T^{uv}_G(\vb*{\gamma})$ is identical to \cref{eq:intro-example/GM/definition-loop-factor-T} but with $V_H^{\mathrm{odd}} \symdiff \{\{u, v\}\}$ replacing $V_H^{\mathrm{odd}}$ in the product term. Likewise, the super-structural factor $\smash{\overline{L}_T(\vb*{\gamma})}$ is the weighted sum of $T_G^{uv}(\vb*{\gamma})$ over all edges, 
\begin{equation}\begin{split}
    \overline{T}_G(\vb*{\gamma})
    &= \sum_{\{u, v\} \in E_G} w_{uv} T_G^{uv}(\vb*{\gamma}).
    \label{eq:intro-example/GM/definition-loop-factor-T-super}
\end{split}\end{equation}

In the limit that $\vb*{\lambda} = \vb{0}$ and $\vb*{\omega} = (\pi / 2, \ldots, \pi / 2)$, which recovers $\ket{\Omega} = \ket{s}$, the expression for $T_G(\vb*{\gamma})$ in \cref{eq:intro-example/GM/definition-loop-factor-T} reduces to that for $L_G(\vb*{\gamma})$ in \cref{eq:intro-example/GM/definition-loop-factor-L}, as $\cos \omega_u = 0$ for all vertices $u \in V_G$ and the product over $V_H^{\mathrm{odd}}$ vanishes unless $V_H^{\mathrm{odd}}$ is empty---that is, unless the subgraph $H$ is even-regular. Thus, as observed above, $L_G(\vb*{\gamma})$ is dependent only on the even-regular or Euler subgraphs $\mathcal{C}_G$ of $G$. Likewise, the expression for $T^{uv}_G(\vb*{\gamma})$ reduces to that for $L^{uv}_G(\vb*{\gamma})$ discussed above. More explicitly, we observe that the product over $V_H^{\mathrm{odd}} \symdiff \{\{u, v\}\}$ vanishes unless $V_H^{\mathrm{odd}} \symdiff \{\{u, v\}\}$ is empty, which requires that $H$ is even-regular except vertices $u$ and $v$ that have odd degree. This coincides with the definition of the subgraphs $\mathcal{C}^{uv}_G$ of $G$, and indeed, $L^{uv}_G(\vb*{\gamma})$ is thus dependent only on $\mathcal{C}^{uv}_G$ as written.

We now give an overview of the main ideas that lead to the results in \cref{thm:intro-example/GM/state-s,thm:intro-example/GM/state-Omega}. To begin, the single-layer QAOA circuit produces the following edge-$\{u, v\}$ cost expectation value,
\begin{equation}\begin{split}
    \expval{C_{u v}}_{\ket{\Omega}} &= 
    \bra{\Omega} 
    e^{+i \vb*{\gamma} \cdot \vb{A}} 
    e^{+i \beta \ket{\Omega} \bra{\Omega}} 
    \, C_{uv} \, \\
    & \qquad \qquad
    e^{-i \beta \ket{\Omega} \bra{\Omega}} 
    e^{-i \vb*{\gamma} \cdot \vb{A}}
    \ket{\Omega},
    \label{eq:intro-example/GM/explanation-C-u-v-Omega}
\end{split}\end{equation}
but crucially, we observe that the Grover-type mixer Hamiltonian, of the form $\ket{\Omega} \bra{\Omega}$, is a projector and thus its exponential satisfies the decomposition $e^{\pm i \beta \ket{\Omega} \bra{\Omega}} = \mathbb{I} + (e^{\pm i \beta} - 1) \ket{\Omega} \bra{\Omega}$ (see \cref{app:preliminaries/identities}). This enables \cref{eq:intro-example/GM/explanation-C-u-v-Omega} to be re-written as a sum of $2 \times 2 = 4$ terms. Terms wherein $(e^{\pm i \beta} - 1) \ket{\Omega} \bra{\Omega}$ factors were inserted are broken into a product of two or more simpler expectation values with respect to $\ket{\Omega}$. Recall also that $C_{uv}$, as defined in \cref{eq:maxcut-hamiltonian}, comprises an identity $\mathbb{I}$ and a $Z_u Z_v$ diagonal operator; the former results only in an constant offset to the cost expectation. One thus finds that the only non-trivial factors to be evaluated are of the form
\begin{equation}\begin{split}
    \bra{\Omega} e^{\pm i \vb*{\gamma} \cdot \vb{A}} \ket{\Omega},
    \qquad 
    \bra{\Omega} e^{\pm i \vb*{\gamma} \cdot \vb{A}} Z_u Z_v \ket{\Omega}.
    \label{eq:intro-example/GM/non-trivial-factors}
\end{split}\end{equation}

The phase unitaries above are expanded in the same way as for PM-QAOA---see \cref{eq:intro-example/PM/exp-i-gamma-C}---by noting that all phase Hamiltonian terms $\vb{A} = \vb{C}$ commute and that $Z_u Z_v$ are involutory. Then one finds that the constituent terms of the non-trivial factors to be evaluated are all of the form $\mu(\vb{p}) = \smash{\bra{\Omega} \prod_{u \in V_G} Z_u^{p_u} \ket{\Omega}}$, where $p_u$ are integer powers for each vertex $u \in V_G$. 

In the simpler case of the $\ket{s}$ state instead of $\ket{\Omega}$, a phase-flip counting argument quickly reveals that the inner product $\mu(\vb{p})$ vanishes unless all powers $p_u$ are even. This directly motivates the definition of $\mathcal{C}_G$ as the set of even-regular subgraphs, and $\mathcal{C}_G^{uv}$ emerges in terms that contain $Z_u Z_v$ as in \cref{eq:intro-example/GM/non-trivial-factors}; all other subgraphs do not contribute. The expansion of the phase unitary leads to a sum over all contributing subgraphs, and trigonometric manipulation produces $L_G(\vb*{\gamma})$ and $L_G^{uv}(\vb*{\gamma})$ as presented in \cref{eq:intro-example/GM/definition-loop-factor-L}. In the more general case of $\ket{\Omega}$, more care is required in evaluating $\mu(\vb{p})$, and the inner product generically depends on the angles $\vb*{\omega}$ defining the state. In this setting, a clear separation of contributing and non-contributing subgraphs does not appear, and instead all subgraphs $\mathcal{H}_G$ must be accounted for, leading to $T_G(\vb*{\gamma})$ and $T_G^{uv}(\vb*{\gamma})$ as presented in \cref{eq:intro-example/GM/definition-loop-factor-T}.

A complete derivation of our results in \cref{thm:intro-example/GM/state-s,thm:intro-example/GM/state-Omega}, additional secondary results, and generalizations are provided in \cref{app:grover}.

\tocless\subsection{Non-locality in GM-QAOA vs.~PM-QAOA
\label{sec:intro-example/non-locality}}

Here we remark on a crucial difference between PM- and GM-QAOA. In PM-QAOA, the overall cost expectation $\expval{C}$ is sensitive only to triangles ($3$-cycles) in the problem graph and not to cycles of higher length. Indeed, at a finer-grained level, only triangles containing vertices $\{u, v\}$ can contribute to the $\{u, v\}$ edge-resolved cost expectation $\expval{C_{uv}}$, as observed in \cref{thm:intro-example/PM}, specifically \cref{eq:intro-example/PM/C-u-v-s-expectation} and the solutions for the coefficients in \cref{eq:intro-example/PM/xi}. 

In contrast, in GM-QAOA, given the same input state $\ket{s}$ for a fair comparison, the cost expectations are sensitive to $t$-cycles on the problem graph for arbitrarily large $t \geq 3$. This can be seen from \cref{thm:intro-example/GM/state-s} and the definition of the structural factors $L_G(\vb*{\gamma})$ and $L_G^{uv}(\vb*{\gamma})$ in \cref{eq:intro-example/GM/definition-loop-factor-L}. In particular, the edge-resolved cost expectation $\expval{C_{uv}}$ can be sensitive to long cycles that extend into regions of the graph far from the $\{u, v\}$ edge, such that structural changes to the graph in those distant regions would change $\expval{C_{uv}}$. This is a form of non-locality that is not present in PM-QAOA, which arises from the many-body entangling nature of the Grover-type mixer. Other mechanisms and types of non-local behavior have been examined in recent literature (see \eg~Ref.~\onlinecite{PhysRevLett.125.260505,patel2024reinforcement}), with the general finding that non-locality can aid in trickier optimization problems such as those on frustrated models or involving long-range many-body terms.

\tocless\section{QAOA with Product Mixer
\label{sec:product}}

In \cref{sec:intro-example-maxcut/PM} we have treated PM-QAOA on the \textsc{MaxCut} introductory problem. We now present the general formalism for PM-QAOA. Unlike \textsc{MaxCut} which only considers quadratic terms in the cost Hamiltonian, in general a quadratic unconstrained binary optimization (QUBO) problem may involve both linear and quadratic terms \cite{punnen2022quadratic}. Here, we examine this setting and allow a further generalization in which the phase and cost Hamiltonians need not be identical.

\tocless\subsection{Problem definitions
\label{sec:product/definitions}}

We consider a weighted undirected graph $G = \mqty(V_G, E_G, \{a, h, J\})$ with vertex set $V_G = [n]$, edge set $\smash{E_G = \{e\in 2^{V_G} : \abs{e} = 2\}}$ of size $m$, constant cost offset $a \in \mathbb{R}$, and vertex and edge weight functions $h: V_G \to \mathbb{R}$ and $J: E_G \to \mathbb{R}$. The associated cost Hamiltonian is
\begin{equation}\begin{split}
    C &= a \mathbb{I} 
        + \sum_{u \in V_G} C_u
        + \sum_{\{u, v\} \in E_G} C_{uv}, \\
    C_u &= h_u Z_u, \\
    C_{uv} &= J_{uv} Z_u Z_v,
    \label{eq:product/definitions/ising-ham-cost}
\end{split}\end{equation}
and we refer to the cost Hamiltonian terms $\vb{C} = \smash{[a \mathbb{I}]} \mathbin\Vert \smash{[C_u]_{u \in V_G}} \mathbin\Vert \smash{[C_{uv}]_{\{u, v\} \in E_G}}$ where $\mathbin\Vert$ denotes vector concatenation. We allow a phase Hamiltonian defined on the same graph structure $G$ but with generically different weights $\mqty(a', h', J')$,
\begin{equation}\begin{split}
    A &= a' \mathbb{I} 
        + \sum_{u \in V_G} A_u
        + \sum_{\{u, v\} \in E_G} A_{uv}, \\
    A_u &= h'_u Z_u, \\
    A_{uv} &= J'_{uv} Z_u Z_v,
    \label{eq:product/definitions/ising-ham-phase}
\end{split}\end{equation}
and likewise we refer to the phase Hamiltonian terms $\vb{A} = \smash{[a' \mathbb{I}]} \mathbin\Vert \smash{[A_u]_{u \in V_G}} \mathbin\Vert \smash{[A_{uv}]_{\{u, v\} \in E_G}}$. Accordingly, associating an angle with each term, we have phase angles $\vb*{\gamma} \in \mathbb{R}^{n + m + 1}$.

Different choices of cost weights $\mqty(a, h, J)$ yield cost Hamiltonians corresponding to various problem settings, inclusive of \textsc{MaxCut} in addition to the full range of QUBO problems, and different choices of phase weights $\mqty(a', h', J')$ enable adjustments to the QAOA circuit ansatz, for example to accommodate optimization constraints. Our primary result in this generalized setting is reported in \cref{thm:PM}.

\vspace{8pt}
\begin{theorem}
    Single-layer PM-QAOA with the initial state $\ket{s}$ on a weighted graph $G$ produces the vertex-$u$ and edge-$\{u, v\}$ cost expectation values, respectively $h_u F_u$ and $J_{uv} F_{uv}$, with
    \begin{equation}\begin{split}
        F_u &= \mel{s}{V^\dagger Z_u V}{s} 
        = \sum_{P \in \mathcal{P}} 
            a^P_u(\vb*{\beta}) 
            \xi^P_u(\vb*{\gamma}), \\
        F_{uv} &= \mel{s}{V^\dagger Z_u Z_v V}{s} 
        = \sum_{P,Q \in \mathcal{P}} 
            a^{PQ}_{uv}(\vb*{\beta}) 
            \xi^{PQ}_{uv}(\vb*{\gamma}),
        \label{eq:product/definition-F-u-F-uv}
    \end{split}\end{equation}
    where $\mathcal{P} = \{X, Y, Z\}$ are the non-identity Paulis and unitary $\smash{V = e^{-i \vb*{\beta} \cdot \vb{B}^{\mathrm{PM}}} e^{-i \vb*{\gamma} \cdot \vb{C}}}$. The coefficients $\xi^P_u(\vb*{\gamma})$ and $\xi^{PQ}_{uv}(\vb*{\gamma})$, $a^P_u(\vb*{\beta})$ and $a^{PQ}_{uv}(\vb*{\beta})$ are listed in \cref{app-tab:product/summary-xi-expressions,app-tab:product/summary-a-expressions}. The overall cost expectation is
    \begin{equation}\begin{split}
        \expval{C} &= 
        a
        + \sum_{u \in V_G} h_u F_u
        + \sum_{\{u, v\} \in E_G}
            J_{u v} F_{u v}.
        \label{eq:gen_costexp_C}
    \end{split}\end{equation}
\label{thm:PM}
\end{theorem}
\vspace{2pt}

As presented, \cref{thm:PM} is a natural generalization of \cref{thm:intro-example/PM} accommodating both one- and two-body terms in the phase Hamiltonian $A$ and cost Hamiltonian $C$, and distinguishing the weights for the $\vb{A}$ and $\vb{C}$ terms. The main steps leading to \cref{thm:PM} are similar to those discussed in \cref{sec:intro-example-maxcut/PM}, with the additional consideration of the one-body terms. A detailed derivation of our result can be found in \cref{app:product}.

We re-iterate that PM-QAOA subsumes vanilla QAOA, WS-QAOA, FAM-QAOA, and several other QAOA variants, in that these variants utilize mixers that are special cases of \cref{eq:product_mixer}. We provide combinations of the unit vectors $\vu{r}_u$ that reduce PM-QAOA to these special cases in \cref{tab:product_mixer_Ham_coeff}.

\begin{table}[!t]
    \centering
    \begin{tabular}{@{} p{3cm} p{1.7cm} p{1.7cm} p{1.7cm} @{}}
        \toprule 
        Variants & $r^X_u$ & $r^Y_u$ & $r^Z_u$ \\
        \midrule 
        Van.~QAOA~\cite{farhi2014quantum} 
            & 1 
            & 0 
            & 0 \\     
        WS-QAOA~\cite{egger2021warm} 
            & $-\sin \theta_u$ 
            & 0 
            & $-\cos \theta_u$ \\
        FAM-QAOA~\cite{PhysRevA.104.062428} 
            & $\cos \theta_u$ 
            & $-\sin \theta_u$ 
            & 0 \\
        \bottomrule
    \end{tabular}
    \caption{\textbf{Reduction of PM-QAOA to special cases.} Values for $\vu{r}_u = \left(r^X_u, r^Y_u, r^Z_u\right) \in S^2$ to be substituted into the PM-QAOA formalism [\cref{eq:product_mixer}] to obtain the special cases of vanilla QAOA, WS-QAOA, and FAM-QAOA, where angle $\theta_u \in \mathbb{R}$. The angles $\{\theta_u\}_{u \in V_G}$ parametrizing the initial state are fixed in WS-QAOA but are variational parameters in FAM-QAOA. Mappings for $\left(r^X_u, r^Y_u, r^Z_u\right)$ for ma-QAOA~\cite{herrman2022multi} and XQAOA~\cite{vijendran2024expressive} also exist but are bulkier.}
    \label{tab:product_mixer_Ham_coeff}
\end{table}

\tocless\section{QAOA with Grover-Type Mixer
\label{sec:grover}}

In \cref{sec:intro-example-maxcut}, we have discussed results for single-layer GM-QAOA on graphs---in particular, for the \textsc{MaxCut} Hamiltonian in \cref{eq:maxcut-hamiltonian} that accommodates arbitrary weights on edges of the graph. Here, we generalize the problem setting to weighted undirected hypergraphs, on which edges can connect an arbitrary number of vertices (not just two). Secondarily, we extend our analysis to multi-layer QAOA circuit ansatzes. In \cref{sec:grover/definitions} we formally introduce this generalized setting and in \cref{sec:grover/single-layer,sec:grover/multi-layer}, we report results for the single-layer and $p$-layer GM-QAOA respectively.

\tocless\subsection{Problem Definitions
\label{sec:grover/definitions}}

We consider a weighted undirected hypergraph $G = (V_G, E_G, w)$ with vertex set $V_G = [n]$, edge set $\smash{E_G \subseteq 2^{V_G}}$ of size $m$, and weight function $w: E_G \to \mathbb{R}$. Each edge $e \in E_G$ is a subset of $V_G$, and $w_e$ is the weight assigned to the edge. For ease of writing, we assume the empty edge $\varnothing$ is always present ($\varnothing \in E_G$) but with possibly zero weight. The cost Hamiltonian $C$ associated with $G$ is then
\begin{equation}\begin{split}
    C = \sum_{e \in E_G} C_e ,
    \qquad
    C_e = w_e Z_e,
\end{split}\end{equation}
where $C_e$ is the cost contribution by edge $e$. This definition of the cost Hamiltonian is fully general and encompasses every Ising-type problem definable on $G$. In particular, the empty edge $\varnothing$ generates a constant cost offset, and the $\abs{e} = 1$ and $\abs{e} = 2$ edges generate weighted cost terms acting on single vertices and pairs of vertices respectively, as present in ordinary Ising problems. The $\abs{e} > 3$ edges, if present, generate cost terms that are higher-weight Pauli-$Z$ operators---\ie~multi-body interactions when $C$ is interpreted as a physical spin Hamiltonian.

Consistent with the PM-QAOA setting in \cref{sec:product}, we take the phase Hamiltonian $A$ to be defined on the same hypergraph and to be of the same form as $C$, but we allow different weights to be assigned to the edges,
\begin{equation}\begin{split}
    A = \sum_{e \in E_G} A_e ,
    \qquad
    A_e = w'_e Z_e.
    \label{eq:grover/definitions/general-hypergraph-cost-phase}
\end{split}\end{equation}

As shorthand, we write the terms of the phase and cost Hamiltonians as $\vb{A} = [A_e]_{e \in E_G}$ and $\vb{C} = [C_e]_{e \in E_G}$. The Grover-type mixer we consider is as defined in \cref{eq:intro-example/GM/grover-type-mixer}. Correspondingly, in each layer $l  \in [p]$ of the QAOA circuit, there is a phase angle associated with each edge, $\smash{\vb*{\gamma}^{(l)}} \in \mathbb{R}^m$, and there is a single mixer angle $\smash{\beta^{(l)}} \in \mathbb{R}$.

\tocless\subsection{Single-Layer GM-QAOA
\label{sec:grover/single-layer}}

To start, we consider a single-layer QAOA circuit ($p = 1$), and we drop the layer superscripts on the phase and mixer angles, hereafter written $\vb*{\gamma}$ and $\beta$, for brevity. Our primary results are reported in \cref{thm:grover/single-layer/state-Omega,thm:grover/single-layer/state-s} below.

\vspace{8pt}
\begin{theorem}
    \label{thm:grover/single-layer/state-s}
    Single-layer GM-QAOA with the initial state $\ket{s}$ and mixer Hamiltonian $\ket{s} \bra{s}$ on a weighted hypergraph $G$ produces the edge-$e$ cost expectation value
    \begin{equation}\begin{split}
        \expval{C_e}_{\ket{s}}
        = 2 \Re\left[
                \left( e^{i \beta} - 1 \right) 
                L_G^\varnothing(\vb*{\gamma}) 
                L_G^e(-\vb*{\gamma}) \right],
    \end{split}\end{equation}
    which reduces to $w_\varnothing$ for $e = \varnothing$, and the overall cost expectation value is given by
    \begin{equation}\begin{split}
        \expval{C}_{\ket{s}}
        = & \left[
            1 
            + 2 \left( 1 - \cos{\beta} \right) \abs{L_G^\varnothing(\vb*{\gamma})}^2
        \right] w_\varnothing \\
        &
        + 2 \Re\left[
            \left( e^{i \beta} - 1 \right) 
            L_G^\varnothing(\vb*{\gamma}) 
            \overline{L}_G(-\vb*{\gamma}) \right],
    \end{split}\end{equation}
    where $L_G^e(\vb*{\gamma})$ for an edge $e$ and $\overline{L}_G$ are structural factors dependent on subhypergraphs of $G$, as defined in \cref{eq:grover/single-layer/definition-loop-factor-L,eq:grover/single-layer/definition-loop-factor-L-super}.
\end{theorem}
\vspace{2pt}
\begin{theorem}
    \label{thm:grover/single-layer/state-Omega}
    Single-layer GM-QAOA with the initial state $\ket{\Omega}$ and mixer Hamiltonian $\ket{\Omega} \bra{\Omega}$ on a weighted hypergraph $G$ produces the edge-$e$ cost expectation value
    \begin{equation}\begin{split}
        \expval{C_e}_{\ket{\Omega}}
        = w_e \bigg\{
        & \left[
            1 
            + 2 \left( 1 - \cos{\beta} \right) \abs{T_G^\varnothing(\vb*{\gamma})}^2
        \right] T_G^e(\vb{0}) \\
        & 
        + 2 \Re\left[
            \left( e^{i \beta} - 1 \right) 
            T_G^\varnothing(\vb*{\gamma}) 
            T_G^e(-\vb*{\gamma}) \right]
        \bigg\},
    \end{split}\end{equation}
    and the overall cost expectation value is given by
    \begin{equation}\begin{split}
        \expval{C}_{\ket{\Omega}}
        = & \left[
            1 
            + 2 \left( 1 - \cos{\beta} \right) \abs{T_G^\varnothing(\vb*{\gamma})}^2
        \right] \overline{T}_G(\vb{0}) \\
        &
        + 2 \Re\left[
            \left( e^{i \beta} - 1 \right) 
            T_G^\varnothing(\vb*{\gamma}) 
            \overline{T}_G(-\vb*{\gamma}) \right],
    \end{split}\end{equation}
    where $T_G^e(\vb*{\gamma})$ for an edge $e$ and $\overline{T}_G$ are structural factors dependent on subhypergraphs of $G$, as defined in \cref{eq:grover/single-layer/definition-loop-factor-T,eq:grover/single-layer/definition-loop-factor-T-super}.
\end{theorem}
\vspace{2pt}

To define the structural factors for the general case of the initial state $\ket{\Omega}$ and mixer Hamiltonian $\ket{\Omega} \bra{\Omega}$, as in \cref{thm:grover/single-layer/state-Omega}, we consider the set of all subhypergraphs of $G$, denoted $\mathcal{H}_G$. Each $H \in \mathcal{H}_G$ has an edge set $E_H \subseteq E_G$, and different $H, H' \in \mathcal{H}_G$ have distinct $\smash{E_H \neq E_{H'}}$. Then, analogous to the setting of graphs discussed in \cref{sec:intro-example-maxcut/GM} but now generalized for hypergraphs, we have the structural factors
\begin{equation}\begin{split}
    T_G^e(\vb*{\gamma})
    = & \left[ \smashoperator[r]{\prod_{f \in E_G}} \,\,\,
            \cos(\gamma_f w_f') \right] \\
        & \times \sum_{H \in \mathcal{H}_G}
        \left[ \smashoperator[r]{\prod_{f \in E_H}} \,\,\,
            i \tan(\gamma_f w_f') \right]
        \left[ \smashoperator[r]{\prod_{u \in V_{H \symdiff \{e\}}^{\mathrm{odd}}}} \,\,
            \cos{\omega_u} \right],
    \label{eq:grover/single-layer/definition-loop-factor-T}
\end{split}\end{equation}
where $\smash{V_{H \symdiff \{e\}}^{\mathrm{odd}}}$ is the set of odd-degree vertices in the symmetric difference between $H$ and $\{e\}$---that is, $H$ with edge $e$ added when absent and removed otherwise---and we remind that $\vb*{\omega} = [\omega_u]_{u \in V_G}$ are a set of fixed angles that parametrize the state $\ket{\Omega} = \ket{\Omega(\vb*{\lambda}, \vb*{\omega})}$. The other set of angles, $\vb*{\lambda}$, do not appear in \cref{eq:grover/single-layer/definition-loop-factor-T} and do not affect the cost expectations $\smash{\expval{C_e}_{\ket{\Omega}}}$ and $\smash{\expval{C}_{\ket{\Omega}}}$. In the case of an empty edge $e = \varnothing$, as in the structural factor $T_G^\varnothing(\vb*{\gamma})$, trivially $H \symdiff \{e\} = H$ and $\smash{V_{H \symdiff \{e\}}^{\mathrm{odd}}} = \smash{V_{H}^{\mathrm{odd}}}$ are the odd-degree vertices in $H$. Moreover, in the zero-angle case ($\vb*{\gamma} = \vb{0}$),
\begin{equation}\begin{split}
    T_G^e(\vb{0}) = \prod_{u \in e} \cos{\omega_u},
\end{split}\end{equation}
as the contributions from all non-empty $H \in \mathcal{H}_G$ vanish in \cref{eq:grover/single-layer/definition-loop-factor-T} due to the sine term in the products, and only the empty subhypergraph contributes.

The super-structural factor is similarly defined to be the cost-weighted sum over all structural factors,
\begin{equation}\begin{split}
    \overline{T}_G(\vb*{\gamma})
    = \sum_{e \in E_G} w_e T_G^e(\vb*{\gamma}).
    \label{eq:grover/single-layer/definition-loop-factor-T-super}
\end{split}\end{equation}

We next consider the set of even-regular subhypergraphs of $G$, denoted $\mathcal{C}_G \subset \mathcal{H}_G$. Each $H \in \mathcal{C}_G$ has an edge set $E_H \subseteq E_G$ and every vertex of $H$ is contained in an even number of edges; different $H, H' \in \mathcal{C}_G$ have distinct $E_H \neq E_{H'}$. We remark that unlike the more restricted case of graphs, an even-regular hypergraph is not guaranteed to admit an Euler tour (\ie~a tour that visits each edge exactly once) in each connected component---thus we do not term $H \in \mathcal{C}_G$ an Euler subhypergraph. In the same spirit as before, we define $\mathcal{C}_G^e$ to be the set of subhypergraphs $H$ of $G$ such that the symmetric difference $H \symdiff \{e\}$ is even-regular, or equivalently,
\begin{equation}\begin{split}
    \mathcal{C}_G^e = \{ H \symdiff \{e\}: H \in \mathcal{C}_G \},
    \label{eq:grover/single-layer/definition-C-G-e}
\end{split}\end{equation}
such that each $H \in \mathcal{C}_G^e$ is even-regular except vertices $u \in e$ which are odd-degree. Then  the structural factors
\begin{equation}\begin{split}
    L_G^e(\vb*{\gamma})
    &= \left[ \smashoperator[r]{\prod_{f \in E_G}} \,\,\,
            \cos(\gamma_f w_f') \right]
        \sum_{H \in \mathcal{C}_G^e}
        \left[ \smashoperator[r]{\prod_{f \in E_H}} \,\,
            i \tan(\gamma_f w_f') \right].
    \label{eq:grover/single-layer/definition-loop-factor-L}
\end{split}\end{equation}

In the case of an empty edge $e = \varnothing$, as in the structural factor $L_G^\varnothing(\vb*{\gamma})$, trivially $\mathcal{C}_G^\varnothing = \mathcal{C}_G$ in \cref{eq:grover/single-layer/definition-loop-factor-L} above, and moreover in the zero-angle case ($\vb*{\gamma} = \vb{0}$),
\begin{equation}\begin{split}
    L_G^e(\vb{0}) = \delta_{e \varnothing},
\end{split}\end{equation}
where $\delta$ denotes the Kronecker delta. Likewise, the super-structural factor is defined to be
\begin{equation}\begin{split}
    \overline{L}_G(\vb*{\gamma})
    = \sum_{e \in E_G} w_e L_G^e(\vb*{\gamma}).
    \label{eq:grover/single-layer/definition-loop-factor-L-super}
\end{split}\end{equation}

To revisit our prior results, substituting the setting of \textsc{MaxCut} as in \cref{sec:intro-example-maxcut} where $G$ is a graph with weights $w_{uv}$ on edges $\{u, v\} \in E_G$ and the cost and phase Hamiltonians are fixed to be identical ($\vb{A} = \vb{C}$), our more general \cref{thm:grover/single-layer/state-Omega,thm:grover/single-layer/state-s} reduce to \cref{thm:intro-example/GM/state-Omega,thm:intro-example/GM/state-s}, and the expressions for the structural factors in \cref{eq:grover/single-layer/definition-loop-factor-T,eq:grover/single-layer/definition-loop-factor-T-super,eq:grover/single-layer/definition-loop-factor-L,eq:grover/single-layer/definition-loop-factor-L-super} reduce to those previously reported in \cref{eq:intro-example/GM/definition-loop-factor-T,eq:intro-example/GM/definition-loop-factor-T-super,eq:intro-example/GM/definition-loop-factor-L-super,eq:intro-example/GM/definition-loop-factor-L}. To specialize to this problem setting, we straightforwardly fix the edge set of the hypergraph $G$ in our present setting to comprise the edges $\{u, v\}$ of the \textsc{MaxCut} graph alongside the empty edge, and the weights $\smash{w_{\{u, v\}}} = \smash{w'_{\{u, v\}}} = -w_{uv} / 2$ are inherited. The weight of the empty edge, which is a constant cost offset, is set at $w_\varnothing = w_\varnothing' = W / 2$ where $W$ is the total weight of the graph edges.

We remark that in a simpler setting of unweighted hypergraphs, wherein $w_e = w'_e = 1$ uniformly for all edges $e \in E_G$, and where only a single phase angle $\gamma$ is used, that is, $\vb*{\gamma} = (\gamma, \gamma, \ldots, \gamma)$, the bijection between $\mathcal{C}_G^e$ and $\mathcal{C}_G$ as expressed in \cref{eq:grover/single-layer/definition-C-G-e} enables a rewriting of the super-structural factor $\smash{\overline{L}_G(\vb*{\gamma})}$, along the same lines as the discussion in \cref{sec:intro-example-maxcut/GM}. This leads to the same expression for $\smash{\overline{L}_G(\vb*{\gamma})}$ as in \cref{eq:intro-example/GM/unweighted-loop-factor-L-super} but with $N_G(k)$ counting the number of even-regular subhypergraphs of $G$ containing $k$ edges. More details on this simplification and secondary results are given in \cref{app:grover/unweighted-hypergraph}.

The main ideas leading to our results in \cref{thm:grover/single-layer/state-Omega,thm:grover/single-layer/state-s} are similar to those discussed in \cref{sec:intro-example-maxcut/GM} for \cref{thm:intro-example/GM/state-Omega,thm:intro-example/GM/state-s}, here generalized for hypergraphs by relaxing the assumption that each edge connects exactly two vertices, and by distinguishing the weights for the phase $\vb{A}$ and cost $\vb{C}$ Hamiltonian terms. A complete derivation with additional secondary results are detailed in \cref{app:grover}.

\tocless\subsection{Multi-Layer GM-QAOA
\label{sec:grover/multi-layer}}

We now extend our analysis to QAOA circuit ansatzes containing multiple layers, and we accordingly re-introduce the layer superscripts on the phase $\smash{\vb*{\gamma}^{(l)}}$ and mixer $\smash{\beta^{(l)}}$ angles, for layers $l \in [p]$. Our primary result is reported in \cref{thm:grover/multi-layer}.

\vspace{8pt}
\begin{theorem}
    \label{thm:grover/multi-layer}
    $p$-layer GM-QAOA with the initial state $\ket{\Omega}$ and mixer Hamiltonian $\ket{\Omega} \bra{\Omega}$ on a weighted hypergraph $G$ produces the edge-$e$ cost expectation value
    \begin{equation}\begin{split}
        \expval{Z_e}_{\ket{\Omega}}
        = \sum_{\vb{f} \in \mathbb{F}_2^p}
            \sum_{\vb{g} \in \mathbb{F}_2^p}
            Q^{(e)}_{\vb{f} \vb{g}},
    \end{split}\end{equation}
    and the overall cost expectation value is given by
    \begin{equation}\begin{split}
        \expval{Z}_{\ket{\Omega}}
        = \sum_{\vb{f} \in \mathbb{F}_2^p}
            \sum_{\vb{g} \in \mathbb{F}_2^p}
            \overline{Q}_{\vb{f} \vb{g}},
    \end{split}\end{equation}
    where $\smash{Q^{(e)}_{\vb{f} \vb{g}}}$ and $\smash{\overline{Q}_{\vb{f} \vb{g}}}$ are expectation value contributions from quantum trajectories defined by bitstrings $\vb{f}, \vb{g}$, given in \cref{eq:grover/multi-layer/definition-Q-e-f-g,eq:grover/multi-layer/definition-Q-super-f-g}. Regarded as matrices, $\smash{Q^{(e)}}$ and $\smash{\overline{Q}}$ are Hermitian, that is, $\smash{Q^{(e)}_{\vb{f} \vb{g}} = [Q^{(e)}_{\vb{g} \vb{f}}]^*}$ and $\smash{\overline{Q}_{\vb{f} \vb{g}} = [\overline{Q}_{\vb{g} \vb{f}}]^*}$. \\[-0.4\baselineskip]

    With the initial state $\ket{s}$ and mixer Hamiltonian $\ket{s} \bra{s}$, the above statements hold but with structural factors $T^e_G, \overline{T}_G$ replaced by $L^e_G, \overline{L}_G$ in $\smash{Q^{(e)}_{\vb{f} \vb{g}}}$ and $\smash{\overline{Q}_{\vb{f} \vb{g}}}$, and diagonal entries $\smash{Q^{(e)}_{\vb{f} \vb{f}}}$ and $\smash{\overline{Q}_{\vb{f} \vb{f}}}$ vanish.
\end{theorem}
\vspace{2pt}

To express the expectation value contributions $\smash{Q^{(e)}_{\vb{f} \vb{g}}}$ in \cref{thm:grover/multi-layer}, it is convenient to first define $\vb{I}(\vb{f})$ to be the nonzero indices of a bitstring $\vb{f} \in \mathbb{F}_2^p$, that is,
\begin{equation}\begin{split}
    \vb{I}(\vb{f}) 
    = \left[ k = 1, 2, \ldots, p: f_k = 1 \right].
\end{split}\end{equation}

Therefore, as defined, $\vb{I}(\vb{0})$ is the empty vector containing no entries, and in general the length of $\vb{I}(\vb{f})$ always equals the Hamming weight of the bitstring $\vb{f}$, $\abs{\vb{I}(\vb{f})} = \wt(\vb{f})$. Conceptually, $\vb{I}(\vb{f})$ is merely an alternative representation of $\vb{f}$. Next, we define sums of the $\vb*{\gamma}^{(l)}$ angles over layer intervals corresponding to contiguous blocks of nonzero entries of $\vb{f}$, which can be written as
\begin{equation}\begin{split}
    \vb*{\Gamma}(\vb{f})
    = \left( 
        \sum_{l = 1}^{I(\vb{f})_1} \vb*{\gamma}^{(l)},
        \sum_{l = I(\vb{f})_1 + 1}^{I(\vb{f})_2} \vb*{\gamma}^{(l)},
        \ldots,
        \sum_{l = I(\vb{f})_{-1} + 1}^{p} \vb*{\gamma}^{(l)}
    \right),
\end{split}\end{equation}
where $I(\vb{f})_k$ refers to the $k^{\text{th}}$ entry of $\vb{I}(\vb{f})$, and $I(\vb{f})_{-1}$ denotes the last entry of $\vb{I}(\vb{f})$ when $\vb{I}(\vb{f})$ is non-empty and zero otherwise. Sums above that are invalid, because the referenced entries in their limits do not exist, are to be regarded as absent in the $\vb*{\Gamma}(\vb{f})$ vector. Thus, in general the length $\abs{\vb*{\Gamma}(\vb{f})} = \wt(\vb{f}) + 1$ and $\vb*{\Gamma}(\vb{f})$ is never empty. The last entry of $\vb*{\Gamma}(\vb{f})$ is a remainder term, summing over $\vb*{\gamma}^{(l)}$ angles in layers occurring after the aforementioned contiguous blocks in $\vb{f}$. 

With these notational preliminaries, for the initial state $\ket{\Omega}$ and mixer Hamiltonian $\ket{\Omega} \bra{\Omega}$, the contribution $\smash{Q^{(e)}_{\vb{f} \vb{g}}}$ can be succinctly written
\begin{equation}\begin{split}
    Q^{(e)}_{\vb{f} \vb{g}}
    &= R_{\vb{f}} \, R_{\vb{g}}^* \, T_G^e\left[\vb*{\Gamma}(\vb{f})_{-1} - \vb*{\Gamma}(\vb{g})_{-1}\right], \\
    R_{\vb{f}}
    &= \left( \prod_{l = 1}^p 
            \left( e^{i \beta_l} - 1 \right)^{f_l}
        \right)
        \left(
            \prod_{k = 1}^{\wt(\vb{f})}
            T_G^\varnothing\left[\vb*{\Gamma}(\vb{f})_k\right] 
        \right),
    \label{eq:grover/multi-layer/definition-Q-e-f-g}
\end{split}\end{equation}
where we remind that $T_G^e$ are structural factors dependent on the structure of the problem hypergraph $G$ as expressed in \cref{eq:grover/single-layer/definition-loop-factor-T}, here associated with entries of $\vb*{\Gamma}(\vb{f})$ as angle inputs. For a setting with initial state $\ket{s}$ and mixer Hamiltonian $\ket{s} \bra{s}$, one replaces $T_G^e \rightarrow L_G^e$ and $T_G^\varnothing \rightarrow L_G^\varnothing$ above. Finally, $\smash{\overline{Q}_{\vb{f} \vb{g}}}$ is a cost-weighted sum over these contributions,
\begin{equation}\begin{split}
    \overline{Q}_{\vb{f} \vb{g}}
    = \sum_{e \in E_G} w_e Q^{(e)}_{\vb{f} \vb{g}}.
    \label{eq:grover/multi-layer/definition-Q-super-f-g}
\end{split}\end{equation}

The main ideas underlying this extension to multi-layer QAOA circuit ansatzes build upon those discussed in \cref{sec:intro-example-maxcut/GM}. To re-iterate, most useful is the fact that the Grover-type mixer Hamiltonian, of the form $\ket{\Omega} \bra{\Omega}$, is a projector and satisfies the decomposition $e^{\pm i \beta \ket{\Omega} \bra{\Omega}} = \mathbb{I} + (e^{\pm i \beta} - 1) \ket{\Omega} \bra{\Omega}$. The multi-layer QAOA circuit produces the following edge-$e$ cost expectation value,
\begin{equation}\begin{split}
    \expval{C_e}_{\ket{\Omega}} = 
    \bra{\Omega} 
    & \left( \prodr_{l = 1}^p 
        e^{+i \vb*{\gamma}^{(l)} \cdot \vb{A}} 
        e^{+i \beta^{(l)} \ket{\Omega} \bra{\Omega}}
    \right)
    \, C_e \, \\
    & \left( \prodl_{l = 1}^p
        e^{-i \beta^{(l)} \ket{\Omega} \bra{\Omega}} 
        e^{-i \vb*{\gamma}^{(l)} \cdot \vb{A}}
    \right)
    \ket{\Omega},
    \label{eq:grover/multi-layer/explanation-C-u-v-Omega}
\end{split}\end{equation}
where $\prodr_{l = 1}^p M_l = M_1 M_2 \ldots M_p$ and $\prodl_{l = 1}^p M_l = M_p \ldots M_2 M_1$ denote products of generically non-commuting operators in the rightward and leftward directions respectively. The decomposition of each $\smash{e^{\pm i \beta^{(l)} \ket{\Omega} \bra{\Omega}}}$ factor inserts either an identity or a $\ket{\Omega} \bra{\Omega}$ projector up to a global $\mathrm{U}(1)$ phase; the latter breaks the $\smash{\expval{C_e}_{\ket{\Omega}}}$ inner product into smaller ones, and former enables the two adjacent phase layers to be combined with $\vb*{\gamma}$ angles added together. The choices of an identity or a $\ket{\Omega} \bra{\Omega}$ projector, respectively recorded by bitstrings $\vb{f} \in \mathbb{F}_2^p$ and $\vb{g} \in \mathbb{F}_2^p$ for the rightward and leftward products, in effect specifies a quantum trajectory, and the contribution of the trajectory to $\smash{\expval{C_e}_{\ket{\Omega}}}$ is $\smash{Q^{(e)}_{\vb{f} \vb{g}}}$. For each trajectory, all constituent expectation values are of the forms listed in \cref{eq:intro-example/GM/non-trivial-factors}, which we have previously analyzed; thus $\smash{Q^{(e)}_{\vb{f} \vb{g}}}$ reduces to a combination of $T_G^e$ structural factors. The remaining notational definitions of $\vb{I}(\vb{f})$ and $\vb{\Gamma}(\vb{f})$ are motivated by the book-keeping needs of this structure. 

A complete derivation of our results in \cref{thm:grover/multi-layer} and additional secondary results are provided in \cref{app:grover}.

\tocless\section{Conclusion
\label{sec:conclusion}}

In this work, we have examined the PM- and GM-QAOA families, the former using a non-entangling product mixer and the latter an entangling many-body mixer. For PM-QAOA, we derived analytical expressions for vertex-resolved, edge-resolved, and overall cost expectation values on Ising cost Hamiltonians, representable as weighted graphs, in the multi-angle regime for a single circuit ansatz layer and the initial product state $\ket{s}$. The PM-QAOA family subsumes vanilla QAOA~\cite{farhi2014quantum}, WS-QAOA~\cite{egger2021warm}, FAM-QAOA~\cite{PhysRevA.104.062428}, ma-QAOA~\cite{herrman2022multi}, and XQAOA~\cite{vijendran2024expressive} as special cases; therefore, PM-QAOA is a unifying framework spanning a broad space of non-entangling mixers. Naturally, our results apply to any of the subsumed variants upon specialization, and in their generic form describe interpolations between (and generalizations from) these variants. For example, several studies~\cite{herrman2022multi, vijendran2024expressive, sureshbabu2024parameter, wilkie2024quantum,langfitt2024phantom} have built upon analytical results to better understand the behavior of QAOA.

For GM-QAOA, we examined multiple phase angles and a single mixer angle on Ising-type cost Hamiltonians, representable as weighted hypergraphs, and derived hyperedge-resolved and overall cost expectation values for the general product initial state $\ket{\Omega(\vb*{\lambda}, \vb*{\omega})}$. These Ising-type hypergraph problems subsume every Ising Hamiltonian but also accommodate $b$-body terms for $b > 2$. We further generalized our results for multiple circuit ansatz layers. The entangling many-body nature of the GM-QAOA mixer results in cycles of arbitrarily large lengths contributing to cost expectations, which presents a form of non-locality (see \cref{sec:intro-example/non-locality}). This is in contrast to PM-QAOA where only triangles ($3$-cycles) contribute to cost expectations. This non-local feature of GM-QAOA is captured in structural factors of the problem hypergraphs in our formulation.

We remark that it is yet unclear whether QAOA can provide a decisive quantum advantage over classical optimization~\cite{blekos2024review}. Beyond numerical benchmarking \cite{shaydulin2019evaluating,willsch2020benchmarking,lotshaw2021empirical,schwagerl2024benchmarking,singhal2024performance,hao2024end}, analytical studies into the behavior and performance of QAOA in general problem settings can be helpful in unveiling potential avenues for quantum utility. In this vein, there are several potential directions for future work. Our treatment could be modified to compute the Gibbs objective function~\cite{PhysRevResearch.2.023074}, conditional value-at-risk (CVaR)~\cite{barkoutsos2020improving,kolotouros2022evolving,barron2024provable}, or other proxies \cite{larkin2022evaluation} for QAOA performance instead of the cost expectation. Extensions to determine optimal variational angles and parameter optimization techniques~\cite{akshay2021parameter, sureshbabu2024parameter} may also be of relevance. Lastly, theoretically examining the effects of noise and error mitigation strategies~\cite{shaydulin2021error,sud2022dual,pellow2024effects,abbas2024challenges, weidinger2023error, koh2024readout, marshall2020characterizing, wang2021noise, xue2021effects,he2024performance,sack2024large} on QAOA can be meaningful, especially in the non-fault-tolerant quantum era of the present and near term.

\vspace{0.5cm}
{\begin{center}
    \textbf{ACKNOWLEDGEMENTS}
\end{center}
\vspace{0.2cm}

This research is supported by the National Research Foundation, Singapore, and the Agency for Science, Technology and Research (A*STAR), Singapore, under its Quantum Engineering Programme (NRF2021-QEP2-02-P03); A*STAR C230917003; and A*STAR under the Central Research Fund (CRF) Award for Use-Inspired Basic Research (UIBR) and the Quantum Innovation Centre (Q.InC) Strategic Research and Translational Thrust. T.Y.N and J.M.K.~are grateful for support from the A*STAR Graduate Academy. 

\let\oldaddcontentsline\addcontentsline
\renewcommand{\addcontentsline}[3]{}
\bibliography{references}
\let\addcontentsline\oldaddcontentsline

\clearpage 
\pagebreak

\appendix
\onecolumngrid

\begin{center}
{\large\textbf{Supplemental Information} \par}
\end{center}

\tableofcontents

\clearpage
\pagebreak

\section{Preliminaries}
\label{app:preliminaries}

\subsection{Notation}
\label{app:preliminaries/notation}

We define the following notational conventions for the succeeding derivations. First, in regard to binary vectors:
\begin{itemize}[itemsep=0.15\baselineskip]
    \item $\mathbb{F}^n_2 = \{0, 1\}^n$ is the set of binary vectors of length $n$, for any $n \in \mathbb{N}$.
    \item The length of a binary vector $\vb{s}$, that is, the number of bits in $\vb{s}$, is denoted $\abs{\vb{s}}$.
    \item The Hamming weight of a binary vector $\vb{s}$, that is, the number of nonzero bits in $\vb{s}$, is denoted $\wt(\vb{s})$.
\end{itemize}

Next in regard to operators, matrices and general vectors:
\begin{itemize}[itemsep=0.15\baselineskip]
    \item Vectors are written with boldface, for example $\vb{v}$. Matrices are not bolded, for example $A$.
    \item Entries are indexed with subscripts. For example $v_j$ is the $\smash{j^\text{th}}$ entry of a vector $\vb{v}$, and $A_{ij}$ is the $\smash{(i, j)^\text{th}}$ entry of a matrix $A$.
    \item $\smash{\norm{\vb{v}}}$ is the Euclidean norm of a vector $\vb{v}$. 
    \item $A_u$ for any single-qubit operator $A$ and qubit $u$ refers to $A$ applied on that qubit. If there are other qubits present in the system then the identity operation is implicitly implied on those qubits.
    \item $A_e$ for any single-qubit operator $A$ and set of qubits $e$ refers to $A$ supported on those qubits, that is,
    \begin{equation}\begin{split}
        A_e = \prod_{u \in e} A_u.
    \end{split}\end{equation}
\end{itemize}

We use conventional notation for Pauli matrices,
\begin{equation}\begin{split}
    X = \mqty[
        0 & 1 \\
        1 & 0
    ], \qquad 
    Y = \mqty[
        0 & -i \\
        i & 0
    ], \qquad 
    Z = \mqty[
        1 & 0 \\
        0 & -1
    ],
\end{split}\end{equation}
and we define $\mathcal{P} = \{X, Y, Z\}$ to be the set of non-identity Pauli operators acting on a single qubit. We denote the identity as $\mathbb{I}$ with dimensions inferred from context. The $d$-fold tensor product of an object $A$, for instance quantum state, vector, or matrix, is denoted $A^{\otimes d}$. The Kronecker product $\delta_{p q}$ is of unit value when $p = q$ and zero otherwise. 

To avoid ambiguity in expressing products of non-commuting operators, we define
\begin{equation}\begin{split}
    \prodr_{k = 1}^n A_k = A_1 A_2 \cdots A_n,
    \qquad
    \prodl_{k = 1}^n A_k = A_n \cdots A_2 A_1.
\end{split}\end{equation}

\subsection{Useful identities}
\label{app:preliminaries/identities}

We recall that the Pauli operators satisfy an orthogonality property,
\begin{equation}\begin{split}
    \frac{1}{2^n} \tr(P Q) = \delta_{P Q},
    \label{eq:app/preliminaries/identities/pauli-orthogonality}
\end{split}\end{equation}
where $P$ and $Q$ are tensor products of Pauli operators over $n$ qubits, and $\dim(P) = \dim(Q) = 2^n$ is their dimensionality. 

We make use of the following identities regarding matrix exponentials \cite{nielsen2010quantum}:
\begin{itemize}
    
    \item For any projector $P^2 = P$ and any $a \in \mathbb{C}$, one has
    \begin{equation}\begin{split}
        e^{a P}
        = \sum_{k = 0}^\infty \frac{1}{k!} a^k P^k
        = \mathbb{I} + P \sum_{k = 1}^\infty \frac{1}{k!} a^k
        = \mathbb{I} + \left( e^a - 1 \right) P,
        \label{eq:app/preliminaries/identities/exp-projector}
    \end{split}\end{equation}
    noting that the matrix exponential power series expansion is convergent for any matrix.
    
    \item For any involutory operator $A$, \ie~$A^2 = \mathbb{I}$, and any $\theta \in \mathbb{R}$, one has
    \begin{equation}\begin{split}
        e^{i \theta A}
        = \sum_{k = 0}^\infty \frac{1}{k!} (i \theta)^k A^k
        = \sum_{k = 0}^\infty \frac{1}{(2k)!} (i \theta)^{2k} \mathbb{I}
            + \sum_{k = 0}^\infty \frac{1}{(2k + 1)!} (i \theta)^{2k + 1} A
        = \cos{\theta} \, \mathbb{I} + i \sin{\theta} \, A.
        \label{eq:app/preliminaries/identities/exp-involutory-cos-sin}
    \end{split}\end{equation}
    
\end{itemize}

Also, the following expansion of a product of binary sums into a sum of products is useful,
\begin{equation}\begin{split}
    \prod_{k = 1}^K \left( a_k + b_k \right)
    = \sum_{\vb{f} \in \mathbb{F}_2^2}
        \prod_{k = 1}^K a_k^{f_k} b_k^{1 - f_k},
    \label{eq:app/preliminaries/identities/prod-sum-rewriting-binary}
\end{split}\end{equation}
where the bitvector $\vb{f}$ specifies whether $a_k$ or $b_k$ is chosen for each term when the product is multiplied out. More generally, for a product of $R$-term sums,
\begin{equation}\begin{split}
    \prod_{k = 1}^K \sum_{r = 1}^R a^{(k)}_r
    = \sum_{\vb{f} \in \mathbb{F}_2^R}
        \prod_{k = 1}^K \prod_{r = 1}^R
        \left( a^{(k)}_r \right)^{\delta_{f_k r}}.
    \label{eq:app/preliminaries/identities/prod-sum-rewriting-general}
\end{split}\end{equation}
A more specialized identity~\cite{vijendran2024expressive} useful in derivations is
\begin{equation}\begin{aligned}
    \frac{1}{2} \left[ \prod_{i = 1}^n \cos(x_i - y_i) + \prod_{i = 1}^n \cos(x_i + y_i) \right] 
    &= \sum_{\substack{\vb*{\mu} \in \mathbb{F}^n_2 \\ \wt(\vb*{\mu}) \, \mathrm{even}}} 
        \prod_{i = 1}^n 
            \cos^{1 - \mu_i}(x_i) \cos^{1 - \mu_i}(y_i) 
            \sin^{\mu_i}(x_i) \sin^{\mu_i}(y_i), \\
    \frac{1}{2} \left[ \prod_{i=1}^n \cos(x_i - y_i) - \prod_{i = 1}^n \cos(x_i + y_i)\right] 
    &= \sum_{\substack{\vb*{\mu} \in \mathbb{F}^n_2 \\ \wt(\vb*{\mu}) \, \mathrm{odd}}}
        \prod_{i = 1}^n 
            \cos^{1 - \mu_i}(x_i) \cos^{1 - \mu_i}(y_i) 
            \sin^{\mu_i}(x_i) \sin^{\mu_i}(y_i).
    \label{eq:app/preliminaries/identities/trig-addsub-prod-expansion}
\end{aligned}\end{equation}
for any $n \in \mathbb{Z}^+$ and $\vb{x}, \vb{y} \in \mathbb{R}^n$. Lastly, we note that
\begin{equation}\begin{aligned}
    \ket{s} \bra{s} = \bigotimes_{u = 1}^n \frac{1}{2}(\mathbb{I} + X_u),
    \label{eq:app/preliminaries/identities/plus-density-matrix-expansion}
\end{aligned}\end{equation}
where $\ket{s} = \ket{+}^{\otimes n}$ is the $n$-qubit product state, as considered in \cref{eq:definition-plus-state} of the main text.

\clearpage
\pagebreak

\section{Derivation for QAOA for Product Mixers}
\label{app:product}

Here, we provide a detailed derivation of our results for single-layer PM-QAOA on arbitrary weighted undirected graphs. We adopt the same setting as in \cref{sec:product}. That is, we consider a weighted undirected graph $G = \mqty(V_G, E_G, \{a, h, J\})$ with vertex set $V_G = [n]$, edge set $\smash{E_G = \{e\in 2^{V_G} : \abs{e} = 2\}}$ of size $m$, constant cost offset $a \in \mathbb{R}$, and vertex and edge weight functions $h: V_G \to \mathbb{R}$ and $J: E_G \to \mathbb{R}$. The associated cost Hamiltonian is
\begin{equation}\begin{split}
    C = a \mathbb{I} 
        + \sum_{u \in V_G} C_u
        + \sum_{\{u, v\} \in E_G} C_{uv},
    \qquad
    C_u = h_u Z_u,
    \qquad
    C_{uv} = J_{uv} Z_u Z_v,
\end{split}\end{equation}
and the phase Hamiltonian is
\begin{equation}\begin{split}
    A = a' \mathbb{I} 
        + \sum_{u \in V_G} A_u
        + \sum_{\{u, v\} \in E_G} A_{uv},
    \qquad
    A_u = h'_u Z_u,
    \qquad
    A_{uv} = J'_{uv} Z_u Z_v,
\end{split}\end{equation}
as written in \cref{eq:product/definitions/ising-ham-cost,eq:product/definitions/ising-ham-phase} of the main text. We denote the cost and phase Hamiltonian terms
\begin{equation}\begin{split}
    \vb{C} = [a \mathbb{I}] 
        \mathbin\Vert [C_u]_{u \in V_G} 
        \mathbin\Vert [C_{uv}]_{\{u, v\} \in E_G},
    \qquad
    \vb{A} = [a' \mathbb{I}] 
        \mathbin\Vert [A_u]_{u \in V_G} 
        \mathbin\Vert [A_{uv}]_{\{u, v\} \in E_G},
\end{split}\end{equation}
where $\mathbin\Vert$ denotes vector concatenation. 

The mixer Hamiltonian, as written in \cref{eq:product_mixer,eq:product_mixer_terms} of the main text, are
\begin{equation}
    B^{\mathrm{PM}} = \sum_{u = 1}^n B^{\mathrm{PM}}_u,
    \qquad
    B^{\mathrm{PM}}_u = \vu{r}_u \cdot \vb{P}_u,
    \qquad
    \vb{B}^{\mathrm{PM}}
    = \mqty(
        B^{\mathrm{PM}}_1, 
        B^{\mathrm{PM}}_2, 
        \ldots, 
        B^{\mathrm{PM}}_n ),
\end{equation}
where each $\vu{r}_u = \smash{\mqty(r^X_u, r^Y_u, r^Z_u)} \in S^2$ is a unit vector of coefficients and $\vb{P}_u = \smash{\mqty(X_u, Y_u, Z_u)}$ are Pauli matrices that act on the $u^{\text{th}}$ qubit. Note that $\vu{r}_u$ being a unit vector is without loss of generality as norm rescalings can be absorbed into the mixer angles $\vb*{\beta}$.

The QAOA circuit ansatz is accordingly parametrized by mixer angles $\vb*{\beta} \in \mathbb{R}^n$ and phase angles $\vb*{\gamma} \in \mathbb{R}^{n + m + 1}$, with a mixer angle associated with each mixer Hamiltonian term, and a phase angle associated with each phase Hamiltonian term. We take the initial state of the circuit to be the product state $\ket{s}$ as defined in \cref{eq:definition-plus-state}.

\subsection{Symmetry property}
\label{app:product/symmetry}

To simplify succeeding calculations, we first establish the symmetry properties
\begin{equation}\begin{split}
    a^{PQ}_{uv}(\vb*{\beta}) = a^{QP}_{vu}(\vb*{\beta}),
    \qquad 
    \xi^{PQ}_{uv}(\vb*{\gamma}) = \xi^{QP}_{vu}(\vb*{\gamma}),
    \label{app-eq:/product/symmetry/property}
\end{split}\end{equation}
for any edge $\{u, v\} \in E_G$ and Pauli operators $P \in \mathcal{P}$ and $Q \in \mathcal{P}$. The second symmetry property is quickly seen from the definition of the $\xi^{PQ}_{uv}(\vb*{\gamma})$ coefficients, as in \cref{app-eq:xi_eq},
\begin{equation}\begin{split}
    \xi^{PQ}_{uv}(\vb*{\gamma})
    = \mel{s}{
            e^{i \vb*{\gamma} \cdot \vb{A}} 
            P_u Q_v 
            e^{-i \vb*{\gamma} \cdot \vb{A}}}
        {s}
    = \mel{s}{
            e^{i \vb*{\gamma} \cdot \vb{A}} 
            Q_v P_u  
            e^{-i \vb*{\gamma} \cdot \vb{A}}}
        {s}
    = \xi^{QP}_{vu}(\vb*{\gamma}),
\end{split}\end{equation}
since $P_u Q_v = P_v Q_u$ for any two operators $P$ and $Q$, and vertices $u \neq v$. Using the same fact, we note also that
\begin{equation}\begin{split}
    e^{i \vb*{\beta} \cdot \vb{B}^{\mathrm{PM}}}
        Z_u Z_v e^{-i \vb*{\beta} \cdot \vb{B}^{\mathrm{PM}}}
    = e^{i \vb*{\beta} \cdot \vb{B}^{\mathrm{PM}}}
        Z_v Z_u e^{-i \vb*{\beta} \cdot \vb{B}^{\mathrm{PM}}},
\end{split}\end{equation}
and invoking the definition of the $a^{PQ}_{uv}(\vb*{\beta})$ coefficients, as in \cref{app-eq:conj-Zu-Zu-Zv}, this corresponds to
\begin{equation}\begin{split}
    \sum_{P, Q \in \mathcal{P}} 
        a^{PQ}_{uv}(\vb*{\beta}) P_u Q_v
    = \sum_{P, Q \in \mathcal{P}} 
        a^{PQ}_{vu}(\vb*{\beta}) P_v Q_u.
\end{split}\end{equation}
Choosing any $A_u, B_v \in \mathcal{P}$ and applying the orthogonality property of Pauli operators in \cref{eq:app/preliminaries/identities/pauli-orthogonality}, we obtain
\begin{equation}\begin{split}
    \tr( A_u B_v 
        \sum_{P, Q \in \mathcal{P}} 
        a^{PQ}_{uv}(\vb*{\beta}) P_u Q_v )
    = \tr( A_u B_v 
        \sum_{P, Q \in \mathcal{P}} 
        a^{PQ}_{vu}(\vb*{\beta}) P_v Q_u ) 
    \Longrightarrow
    a^{AB}_{uv}(\vb*{\beta})
    = a^{BA}_{vu}(\vb*{\beta}),
\end{split}\end{equation}
which proves the first symmetry property.

\clearpage 
\pagebreak

\subsection{Calculating \texorpdfstring{$\expval{C}$}{<C>} on general weighted graphs (Proof of \texorpdfstring{\cref{thm:PM}}{})}
\label{app:product/expval-C}

\begin{table*}[!ht]
    \centering
    \setlength{\tabcolsep}{0pt}
    \renewcommand*{\arraystretch}{1.2}
    \newcommand\tstrut{\rule{0pt}{12pt}}
    \newcommand\bstrut{\rule[-8pt]{0pt}{0pt}}
    \begin{tabular}[t]{p{1.5cm} p{16cm}}
        \toprule
        $P$ & $\xi^P_u(\vb*{\gamma})$ \\
        \midrule
        $X$ &
            $\begin{aligned}
                \tstrut
                \cos(2 h'_u \gamma_u) 
                \cos(2 J'_{uv} \gamma_{uv})
                R^\setminus_{uv}(\vb*{\gamma})
                \bstrut
            \end{aligned}$ \\
        \tabdashline & \tabdashline
        $Y$ &
            $\begin{aligned}
                \tstrut
                \sin(2 h'_u \gamma_u)
                \cos(2 J'_{uv} \gamma_{uv})
                R^\setminus_{uv}(\vb*{\gamma})
                \bstrut
            \end{aligned}$ \\
        \bottomrule
    \end{tabular} \\[1.2\baselineskip]
    \begin{tabular}[t]{p{1.5cm} p{16cm}}
        \toprule
        $PQ$ & $\xi_{uv}^{PQ}(\vb*{\gamma})$ \\
        \midrule
        $XX$ &
            $\begin{aligned}
                \tstrut
                \frac{1}{2}
                R^\bbslash_{uv}(\vb*{\gamma})
                R^\bbslash_{vu}(\vb*{\gamma})
                \left\{
                    \cos(2 h'_u \gamma_u)
                    \cos(2 h'_v \gamma_v)
                    \left[ R^-_{uv}(\vb*{\gamma}) + R^+_{uv}(\vb*{\gamma}) \right]
                    + \sin(2 h'_u \gamma_u)
                    \sin(2 h'_v \gamma_v)
                    \left[ R^-_{uv}(\vb*{\gamma}) - R^+_{uv}(\vb*{\gamma}) \right]
                \right\}
                \bstrut
            \end{aligned}$ \\
        \tabdashline & \tabdashline
        $XY$ &
            $\begin{aligned}
                \tstrut
                \frac{1}{2}
                R^\bbslash_{uv}(\vb*{\gamma})
                R^\bbslash_{vu}(\vb*{\gamma})
                \left\{
                    \cos(2 h'_u \gamma_u)
                    \sin(2 h'_v \gamma_v)
                    \left[ R^-_{uv}(\vb*{\gamma}) + R^+_{uv}(\vb*{\gamma}) \right]
                    - \sin(2 h'_u \gamma_u)
                    \cos(2 h'_v \gamma_v)
                    \left[ R^-_{uv}(\vb*{\gamma}) - R^+_{uv}(\vb*{\gamma}) \right]
                \right\}
                \bstrut
            \end{aligned}$ \\
        \tabdashline & \tabdashline
        $YY$ &
            $\begin{aligned}
                \tstrut
                \frac{1}{2}
                R^\bbslash_{uv}(\vb*{\gamma})
                R^\bbslash_{vu}(\vb*{\gamma})
                \left\{
                    \cos(2 h'_u \gamma_u)
                    \cos(2 h'_v \gamma_v)
                    \left[ R^-_{uv}(\vb*{\gamma}) - R^+_{uv}(\vb*{\gamma}) \right]
                    + \sin(2 h'_u \gamma_u)
                    \sin(2 h'_v \gamma_v)
                    \left[ R^-_{uv}(\vb*{\gamma}) + R^+_{uv}(\vb*{\gamma}) \right]
                \right\}
                \bstrut
            \end{aligned}$ \\
        \tabdashline & \tabdashline
        $XZ$ &
            $\begin{aligned}
                \tstrut
                -\sin(2 h'_u \gamma_u)
                \sin(2 J'_{uv} \gamma_{uv})
                R^\setminus_{uv}(\vb*{\gamma})
                \bstrut
            \end{aligned}$ \\
        \tabdashline & \tabdashline
        $YZ$ &
            $\begin{aligned}
                \tstrut
                \cos(2 h'_u \gamma_u) 
                \sin(2 J'_{uv} \gamma_{uv})
                R^\setminus_{uv}(\vb*{\gamma})
                \bstrut
            \end{aligned}$ \\
        \bottomrule
    \end{tabular}
    \caption{\textbf{List of $\xi^P_u(\vb*{\gamma})$ and $\xi^{PQ}_{uv}(\vb*{\gamma})$ expressions for PM-QAOA in the setting of general weighted graphs.} Coefficients for the diagonal Paulis $\smash{\xi^{Z}_{u}(\vb*{\gamma})} = \smash{\xi^{ZZ}_{uv}(\vb*{\gamma})} = 0$. By the symmetry property in \cref{app-eq:/product/symmetry/property}, the coefficients $\smash{\xi^{YX}_{uv}(\vb*{\gamma})}$, $\smash{\xi^{ZX}_{uv}(\vb*{\gamma})}$, and $\smash{\xi^{ZY}_{uv}(\vb*{\gamma})}$ can be obtained from $\smash{\xi^{XY}_{uv}(\vb*{\gamma})}$, $\smash{\xi^{XZ}_{uv}(\vb*{\gamma})}$, and $\smash{\xi^{YZ}_{uv}(\vb*{\gamma})}$ respectively by swapping the $u$ and $v$ vertices. The functions $\smash{R^\setminus_{uv}(\vb*{\gamma})}$, $\smash{R^\bbslash_{uv}(\vb*{\gamma})}$, and $\smash{R^\pm_{uv}(\vb*{\gamma})}$ are defined in \cref{app-eq:product/PM/R}. We remind that these expressions are slightly dissimilar to those in \cref{eq:intro-example/PM/xi,eq:intro-example/PM/R}, which have been simplified for the particular problem setting of \textsc{MaxCut} discussed in \cref{sec:intro-example-maxcut}.}
    \label{app-tab:product/summary-xi-expressions}
\end{table*}

\begin{table*}[!ht]
    \centering
    \setlength{\tabcolsep}{0pt}
    \renewcommand*{\arraystretch}{1.2}
    \newcommand\tstrut{\rule{0pt}{12pt}}
    \newcommand\bstrut{\rule[-8pt]{0pt}{0pt}}
    \begin{tabular}[t]{p{1.5cm} p{16cm}}
        \toprule
        $P$ & $a^P_u(\vb*{\beta})$ \\
        \midrule
        $X$ &
            $\begin{aligned}
                \tstrut
                \Phi^{XY,-}_u(\beta_u) 
                \bstrut
            \end{aligned}$ \\
        \tabdashline & \tabdashline
        $Y$ &
            $\begin{aligned}
                \tstrut
                \Phi^{YX,+}_u(\beta_u)
                \bstrut
            \end{aligned}$ \\
        \tabdashline & \tabdashline
        $Z$ &
            $\begin{aligned}
                \tstrut
                \cos^2(\beta_u) - \sin^2(\beta_u) \left[ 1 - 2 \big(r^Z_u\big)^2 \right]
                \bstrut
            \end{aligned}$ \\
        \bottomrule
    \end{tabular} \\[1.2\baselineskip]
    \begin{tabular}[t]{p{1.5cm} p{16cm}}
        \toprule
        $PQ$ & $a^{PQ}_{uv}(\vb*{\beta})$ \\
        \midrule
        $XX$ &
            $\begin{aligned}
                \tstrut
                4 \sin(\beta_u) \sin(\beta_v) 
                    \Theta^{YX,-}_u(\beta_u)
                    \Theta^{YX,-}_v(\beta_v)
                \bstrut
            \end{aligned}$ \\
        \tabdashline & \tabdashline
        $XY$ &
            $\begin{aligned}
                \tstrut
                -4 \sin(\beta_u) \sin(\beta_v)
                    \Theta^{YX,-}_u(\beta_u)
                    \Theta^{XY,+}_v(\beta_v)
                \bstrut
            \end{aligned}$ \\
        \tabdashline & \tabdashline
        $YY$ &
            $\begin{aligned}
                \tstrut
                4 \sin(\beta_u) \sin(\beta_v) 
                    \Theta^{XY,+}_u(\beta_u)
                    \Theta^{XY,+}_v(\beta_v)
                \bstrut
            \end{aligned}$ \\
        \tabdashline & \tabdashline
        $XZ$ &
            $\begin{aligned}
                \tstrut
                2 \sin^2(\beta_u) r^X_u r^Z_u \left[
                    \cos(2\beta_v) 
                    + 2\sin^2(\beta_v) \big(r^Z_v\big)^2 
                \right]
                + \sin(2\beta_u) r^Y_u \left\{
                    \sin^2(\beta_v) \left[ 1 - 2 \big(r^Z_v\big)^2 \right]
                    -\cos^2(\beta_v) \right\}
                \bstrut
            \end{aligned}$ \\
        \tabdashline & \tabdashline
        $YZ$ &
            $\begin{aligned}
                \tstrut
                2 \sin(\beta_u) 
                \Theta^{XY,+}_u (\beta_u) 
                \left\{ 
                    \cos^2(\beta_v) - \sin^2(\beta_v) 
                    \left[1 - 2 \big(r^Z_v\big)^2 \right]
                \right\}
                \bstrut
            \end{aligned}$ \\
        \tabdashline & \tabdashline
        $ZZ$ &
            $\begin{aligned}
                \tstrut
                & \cos^2(\beta_u) \cos(2\beta_v)
                    -\cos^2(\beta_v) \sin^2(\beta_u) \\
                & \qquad + \sin^2(\beta_u) \left( 
                    2 \cos^2(\beta_v) \big(r^Z_u\big)^2
                    + \sin^2(\beta_v) \left\{
                        \big(r^X_v\big)^2 \big(r^Y_u\big)^2
                        + \big(r^X_u\big)^2 \left[ 1 + \big(r^Y_v\big)^2 \right]
                        - \left[ \big(r^X_v\big)^2 + \big(r^Y_v\big)^2 \right] \big(r^Z_u\big)^2 
                    \right\} \right) \\
                & \qquad + \sin^2(\beta_v) \big(r^Z_v\big)^2 \left\{ 
                    2 \cos^2(\beta_u)
                    + \sin^2(\beta_u) \left[
                        \big(r^Z_u\big)^2 - 2 \big(r^X_u\big)^2 - \big(r^Y_u\big)^2 \right] 
                    \right\}
                \bstrut
            \end{aligned}$ \\
        \bottomrule
    \end{tabular}
    \caption{\textbf{List of $a^P_u(\vb*{\beta})$ and $a^{PQ}_{uv}(\vb*{\beta})$ expressions for PM-QAOA.} By the symmetry property in \cref{app-eq:/product/symmetry/property}, the coefficients $\smash{a^{YX}_{uv}(\vb*{\beta})}$, $\smash{a^{ZX}_{uv}(\vb*{\beta})}$, and $\smash{a^{ZY}_{uv}(\vb*{\beta})}$ can be obtained from $\smash{a^{XY}_{uv}(\vb*{\beta})}$, $\smash{a^{XZ}_{uv}(\vb*{\beta})}$, and $\smash{a^{YZ}_{uv}(\vb*{\beta})}$ respectively by swapping the $u$ and $v$ vertices. The functions $\Phi^{PQ,\pm}_{uv}(\beta)$ and $\Theta^{PQ,\pm}_u(\beta)$ are defined in \cref{eq:Phi_Theta_eq}. Parts of these expressions relevant to the particular problem setting of \textsc{MaxCut} appear in \cref{eq:intro-example/PM/a} of the main text.}
    \label{app-tab:product/summary-a-expressions}
\end{table*}

\clearpage
\pagebreak

We seek to evaluate the cost expectation
\begin{equation}\begin{split}
    \expval{C} &= a \expval{\mathbb{I}}
        + \sum_{u \in V_G} \expval{C_u}
        + \sum_{\{u, v\} \in E_G} \expval{C_{uv}} \\
    &= a 
        + \sum_{u \in V_G} h_u 
            \underbrace{\mel{s}{V^\dagger Z_u V}{s}}_{F_u} 
        + \sum_{\{u, v\} \in E_G} J_{uv} 
            \underbrace{\mel{s}{V^\dagger Z_u Z_v V}{s}}_{F_{uv}},
    \label{app-eq:app/product/expval-C-F-u-F-uv}
\end{split}\end{equation}
where $V$ is the QAOA circuit unitary,
\begin{equation}\begin{split}
    V = e^{-i \vb*{\beta} \cdot \vb{B}^{\mathrm{PM}}} 
        e^{-i \vb*{\gamma} \cdot \vb{A}}.
\end{split}\end{equation}

Selecting an arbitrary edge $\{u, v\} \in E_G$, we define the following neighborhood sets:
\begin{itemize}[itemsep=0pt]
    \item $\mathcal{N}_u$ is the neighborhood of vertex $u$.
    \item $\mathcal{N}_v$ is the neighborhood of vertex $v$.
    \item $\mathcal{N}_{u \bbslash v} 
    = \mathcal{N}_u \setminus \left( \mathcal{N}_v \cup \{v\} \right)
    = \{ g_1, \ldots, g_b \}$ are neighbors of $u$ that are not incident to $v$ or $v$ itself, or equivalently, are of distance at least $2$ from $v$.
    \item $\mathcal{N}_{v \bbslash u} 
    = \mathcal{N}_v \setminus \left( \mathcal{N}_u \cup \{u\} \right)
    = \{ h_1, \ldots, h_c \}$ are neighbors of $v$ that are not incident to $u$ or $u$ itself, or equivalently, are of distance at least $2$ from $u$.
    \item $\mathcal{N}_{uv} 
    = \mathcal{N}_u\cap\mathcal{N}_v 
    = \{a_1, \ldots, a_f\} 
    = \{g_{b+1}, \ldots, g_d\} 
    = \{h_{c+1}, \ldots, h_e\}$ are vertices that are incident to both $u$ and $v$, or equivalently, that form triangles together with the $\{u, v\}$ edge.
    \item $\mathcal{N}_{u\setminus v} 
    = \mathcal{N}_u\setminus\{v\} 
    = \mathcal{N}_{u\bbslash v}\cup\mathcal{N}_{uv} 
    = \{g_1, \ldots, g_b, a_1, \ldots, a_f\} 
    = \{g_1, \ldots, g_d\}$ are neighbors of $u$ excluding $v$.
    \item $\mathcal{N}_{v\setminus u} 
    = \mathcal{N}_v\setminus\{u\} 
    = \mathcal{N}_{v\bbslash u}\cup\mathcal{N}_{uv} 
    = \{h_1, \ldots, h_c, a_1, \ldots, a_f\} 
    = \{h_1, \ldots, h_e\}$ are neighbors of $v$ excluding $u$.
\end{itemize}

Thus, by the definitions above, $d$ denotes $\deg(u) - 1$, $e$ denotes $\deg(v) - 1$, and $f$ is the number of triangles in $G$ that contain the edge $\{u, v\}$. This notation for neighborhood sizes is exactly consistent with Refs.~\cite{PhysRevA.97.022304,herrman2022multi,vijendran2024expressive}. We provide an illustration of the various neighborhoods in \cref{fig:product_mixer_neighborhoods}.

The phase Hamiltonian can then be rewritten as
\begin{equation}\begin{split}
    A = a' \mathbb{I}
        + A^{(1)}
        + A^{(2)}_{uv}
        + A^{(2)}_{u}
        + A^{(2)}_{v}
        + A^{(2)}_{0},
\end{split}\end{equation}
where $\smash{A^{(1)}}$ is the sum of phase Hamiltonian terms associated with vertices, $\smash{A^{(2)}_{uv}}$ is the term associated with the edge $\{u, v\}$ itself, $\smash{A^{(2)}_{u}}$ is the sum of terms associated with edges containing neighbors incident to $u$ but not $v$ and likewise for $\smash{A^{(2)}_{v}}$, and $\smash{A^{(2)}_{0}}$ is the sum of terms associated with edges containing neighbors incident to neither $u$ nor $v$. Their corresponding vectors of constituent terms, $\smash{\vb{A}^{(1)}}$, $\smash{\vb{A}^{(2)}_{uv}}$, $\smash{\vb{A}^{(2)}_{u}}$, $\smash{\vb{A}^{(2)}_{v}}$, $\smash{\vb{A}^{(2)}_{0}}$, can be regarded as $\vb{A}$ but with non-participating entries zeroed. This rewriting is in essence an edge $\{u, v\}$-centric reorganization, so that terms are partitioned according to whether their associated edge contains $u$ or $v$, neither, or both. The phase unitary is then
\begin{equation}\begin{split}
    e^{-i \vb*{\gamma} \cdot \vb{A}} 
    &\sim e^{-i \vb*{\gamma} \cdot \vb{A}^{(1)}}
        e^{-i \vb*{\gamma} \cdot \vb{A}^{(2)}_{uv}}
        e^{-i \vb*{\gamma} \cdot \vb{A}^{(2)}_{u}}
        e^{-i \vb*{\gamma} \cdot \vb{A}^{(2)}_{v}}
        e^{-i \vb*{\gamma} \cdot \vb{A}^{(2)}_{0}} \\
    &= \left( \prod_{u \in V_G} e^{-i \gamma_u h'_u Z_u} \right) 
        \left( e^{-i \gamma_{uv} J'_{uv} Z_u Z_v} \right) \\
        & \qquad \qquad \qquad \times
        \left( \prod_{j = 1}^d e^{-i \gamma_{u w_j} J'_{u w_j} Z_u Z_{w_j}} \right)
        \left( \prod_{j = 1}^e e^{-i \gamma_{v w_j} J'_{v w_j} Z_v Z_{w_j}} \right)
        \left( \prod_{\substack{
                \{a, b\} \in E_G \\ 
                \{a, b\} \cap \{u, v\} = \varnothing}} 
            e^{-i \gamma_{ab} J'_{ab} Z_a Z_b} \right),
    \label{app-eq:product/phase-unitary-expand-product}
\end{split}\end{equation}
where $\sim$ denotes equivalence up to a $\mathrm{U}(1)$ phase factor associated with the $a' \mathbb{I}$ constant term in $A$, which is irrelevant in the computation of $\expval{C}$ since it is always cancelled by its inverse in the constituent $F_u$ and $F_{uv}$ terms. 

The $\smash{A^{(2)}_u}$ and $\smash{A^{(2)}_v}$ phase Hamiltonian terms accounting for the neighbors of $u$ and $v$ can be further organized into terms that form a triangle with the edge $\{u, v\}$ and those that do not. That is,
\begin{equation}\begin{split}
    A^{(2)}_u 
    &= \sum_{j = 1}^d J'_{u g_j} \gamma_{u g_j} Z_u Z_{g_j} 
    = \sum_{j = 1}^b J'_{u g_j} \gamma_{u g_j} Z_u Z_{g_j}
        +\sum_{j = 1}^f J'_{u a_j} \gamma_{u a_j} Z_u Z_{a_j}, \\
    A^{(2)}_v  
    &= \sum_{j = 1}^e J'_{v h_j} \gamma_{v h_j} Z_v Z_{h_j} 
    = \sum_{j = 1}^c J'_{v h_j} \gamma_{v h_j} Z_v Z_{h_j}
        +\sum_{j = 1}^f J'_{v a_j} \gamma_{v a_j} Z_v Z_{a_j}.
    \label{app-eq:product/A-2-u-A-2-v-repartition}
\end{split}\end{equation}

We are now ready to begin calculations. First, we note it is possible to write
\begin{equation}\begin{split}
    e^{i \vb*{\beta} \cdot \vb{B}^{\mathrm{PM}}} 
        Z_u 
        e^{-i \vb*{\beta} \cdot \vb{B}^{\mathrm{PM}}} 
    = \sum_{P \in \mathcal{P}} a^P_u(\vb*{\beta}) P_u,
    \qquad
    e^{i \vb*{\beta} \cdot \vb{B}^{\mathrm{PM}}}
        Z_u Z_v 
        e^{-i \vb*{\beta} \cdot \vb{B}^{\mathrm{PM}}} 
    = \sum_{P, Q \in \mathcal{P}} a^{PQ}_{uv}(\vb*{\beta}) P_u Q_v,
    \label{app-eq:conj-Zu-Zu-Zv}
\end{split}\end{equation}
as the set of Pauli operators on vertices $u$ and $v$ form a basis for all operators on those qubits, and $\vb{B}^{\mathrm{PM}}$ is the product mixer and thus $\smash{e^{\pm i \vb*{\beta} \cdot \vb{B}^{\mathrm{PM}}}}$ does not grow the support of operators upon acting on them (via conjugation). The identity operators ($P$ and $Q$) do not occur in the expansion. Here, it is useful to expand
\begin{equation}\begin{split}
    e^{-i \vb*{\beta} \cdot \vb*{B}^{\mathrm{PM}}} 
    = e^{-i \sum_{u = 1}^n \beta_u(\vu{r}_u \cdot \vb{P}_u)}
    = \prod_{u = 1}^n e^{-i \beta_u (\vu{r}_u \cdot \vb{P}_u)}
    = \prod_{u = 1}^n \left[ 
        \cos(\beta_u) \mathbb{I} 
        - i \sin(\beta_u) \left( \vu{r}_u \cdot \vb{P}_u \right) \right],
\end{split}\end{equation}
where we have used the identity for involutory operators in \cref{eq:app/preliminaries/identities/exp-involutory-cos-sin}. 

To evaluate the coefficients $\smash{a^P_u(\vb*{\beta})}$ and $\smash{a^{PQ}_{uv}(\vb*{\beta})}$, we examine the left-hand sides of \cref{app-eq:conj-Zu-Zu-Zv}, and push $Z_u$ and $Z_u Z_v$ rightward through $\smash{e^{-i \vb*{\beta} \cdot \vb*{B}^{\mathrm{PM}}}}$, keeping note of possible anti-commutation relations between Pauli operators which contribute to the coefficients. As Pauli operators on different qubits commute, anti-commutation relations can arise only from mixer unitary terms acting on vertices $u$ and $v$. Thus, $\smash{a^P_u(\vb*{\beta})}$ can be dependent only on parameters associated with the vertex $u$, namely $\smash{\{\beta_u, \vu{r}_u\}}$, and $\smash{a^{PQ}_{uv}(\vb*{\beta})}$ can be dependent only on parameters associated with $u$ and $v$, namely $\smash{\{\beta_u, \beta_v, \vu{r}_u, \vu{r}_v\}}$. We provide a complete list of explicit expressions for $\smash{a^P_u(\vb*{\beta})}$ and $\smash{a^{PQ}_{uv}(\vb*{\beta})}$ computed in this manner in \cref{app-tab:product/summary-a-expressions}, where for brevity we have used the shorthand
\begin{equation}\begin{split}
    \Phi^{PQ,\pm}_u(\beta_u) 
    &= 2 \sin^2(\beta_u) r^P_u r^Z_u \pm \sin(2 \beta_u) r^Q_u, \\
    \Theta^{PQ,\pm}_u(\beta_u) 
    &= \cos(\beta_u) r_u^P \pm \sin(\beta_u) r_u^Q r_u^Z.
    \label{eq:Phi_Theta_eq}
\end{split}\end{equation}
Parts of \cref{app-tab:product/summary-a-expressions,eq:Phi_Theta_eq} appear in \cref{eq:intro-example/PM/a,eq:intro-example/PM/Theta} of the main text. Next, treating the cost expectation terms $F_u$ and $F_{uv}$ in \cref{app-eq:app/product/expval-C-F-u-F-uv} with the expansion in \cref{app-eq:conj-Zu-Zu-Zv}, we write
\begin{equation}\begin{split}
    F_u 
    &= \mel{s}{V^\dag Z_u V}{s}
    = \sum_{P \in \mathcal{P}} a^P_u(\vb*{\beta}) 
        \underbrace{\mel{s}{
            e^{i \vb*{\gamma} \cdot \vb{A}} 
            P_u
            e^{-i \vb*{\gamma} \cdot \vb{A}}
        }{s}}_{\xi^P_u(\vb*{\gamma})}, \\
    F_{uv}
    &= \mel{s}{V^\dag Z_u Z_v V}{s}
    = \sum_{P, Q \in \mathcal{P}} a^{PQ}_{uv}(\vb*{\beta}) 
        \underbrace{\mel{s}{
            e^{i \vb*{\gamma} \cdot \vb{A}} 
            P_u Q_v
            e^{-i \vb*{\gamma} \cdot \vb{A}}
        }{s}}_{\xi^{PQ}_{uv}(\vb*{\gamma})},
    \label{app-eq:Fu-Fuv}
\end{split}\end{equation}
and we further label
\begin{equation}\begin{split}
    \xi^P_u(\vb*{\gamma}) 
    = \mel{s}{\underbrace{
            e^{i \vb*{\gamma} \cdot \vb{A}} 
            P_u 
            e^{-i \vb*{\gamma} \cdot \vb{A}}
        }_{\eta^P_u(\vb*{\gamma})}}{s},
    \qquad
    \xi^{PQ}_{uv}(\vb*{\gamma}) 
    = \mel{s}{\underbrace{
            e^{i \vb*{\gamma} \cdot\vb{A}} 
            P_u Q_v 
            e^{-i \vb*{\gamma} \cdot \vb{A}}
        }_{\eta^{PQ}_{uv}(\vb*{\gamma})}}{s}.
    \label{app-eq:xi_eq}
\end{split}\end{equation}
Writing out the $e^{\pm i \vb*{\gamma} \cdot \vb{A}}$ operators as in \cref{app-eq:product/phase-unitary-expand-product}, we observe
\begin{equation}\begin{split}
    \eta^P_u(\vb*{\gamma}) 
    = V_u^\dag(\vb*{\gamma})  
        P_u
        V_u(\vb*{\gamma}),
    \qquad
    \eta^{PQ}_{uv}(\vb*{\gamma}) 
    = V_{uv}^\dag(\vb*{\gamma})
        P_u Q_v 
        V_{uv}(\vb*{\gamma}),
    \label{eq:eta_eqs}
\end{split}\end{equation}
where
\begin{equation}\begin{split}
    V_u(\vb*{\gamma})
    &= e^{-i \gamma_u h'_u Z_u}
    e^{-i \gamma_{uv} J'_{uv} Z_u Z_v}
        e^{-i \vb*{\gamma} \cdot \vb{A}^{(2)}_u}, \\
    V_{uv}(\vb*{\gamma})
    &=  e^{-i \gamma_u h'_u Z_u}
        e^{-i \gamma_v h'_v Z_v}
        e^{-i \gamma_{uv} J'_{uv} Z_u Z_v}
        e^{-i \vb*{\gamma} \cdot \vb{A}^{(2)}_u}
        e^{-i \vb*{\gamma} \cdot \vb{A}^{(2)}_v}.
\end{split}\end{equation}

In consolidating $V_u(\vb*{\gamma})$ and $V_{uv}(\vb*{\gamma})$, we have noted that phase unitary terms associated with edges that do not contain vertices $u$ and/or $v$ commute through $P_u$ and $P_u Q_v$ and cancel into the identity---hence $V_u(\vb*{\gamma})$ retains only terms associated with edges containing $u$, and $V_{uv}(\vb*{\gamma})$ retains only terms associated with edges containing $u$ or $v$. Examining \cref{eq:eta_eqs}, we now push $P_u$ rightward through $V_u(\vb*{\gamma})$ and $P_u Q_v$ rightward through $V_{uv}(\vb*{\gamma})$. That is, we seek to find $\Gamma_u^P(\vb*{\gamma})$ and $\Gamma_{uv}^{PQ}(\vb*{\gamma})$, where
\begin{equation}\begin{split}
    \eta^P_u(\vb*{\gamma}) = \Gamma_u^P(\vb*{\gamma}) P_u
    \Longleftrightarrow 
    \Gamma_u^P(\vb*{\gamma}) = \eta^P_u(\vb*{\gamma}) P_u,
    \qquad
    \eta^{PQ}_{uv}(\vb*{\gamma}) = \Gamma_{uv}^{PQ}(\vb*{\gamma}) P_u Q_v
    \Longleftrightarrow 
    \Gamma_{uv}^{PQ}(\vb*{\gamma}) = \eta^{PQ}_{uv}(\vb*{\gamma}) P_u Q_v.
    \label{app-eq:product/definition-Gamma-P-u-PQ-uv}
\end{split}\end{equation}
Accounting for possible anti-commutation relations of $P_u$ with the $Z_u$ and $Z_u Z_v$ phase unitary terms in $V_u(\vb*{\gamma})$, and likewise anti-commutation relations of $P_u Q_v$ with the $Z_u$ and $Z_u Z_v$ phase unitary terms in $V_{uv}(\vb*{\gamma})$, we obtain
\begin{equation}\begin{split}
    \Gamma^X_u(\vb*{\gamma}) 
    &= \Gamma^Y_u(\vb*{\gamma}) 
    = e^{2i \gamma_u h'_u Z_u} 
        e^{2i \gamma_{uv} J'_{uv} Z_u Z_v} 
        e^{2i \vb*{\gamma} \cdot \vb{A}^{(2)}_u}, \\
    \Gamma^{XX}_{uv}(\vb*{\gamma}) 
    &= \Gamma^{XY}_{uv}(\vb*{\gamma}) 
    = \Gamma^{YX}_{uv}(\vb*{\gamma}) 
    = \Gamma^{YY}_{uv}(\vb*{\gamma}) 
    = e^{2i \gamma_u h'_u Z_u} e^{2i \gamma_v h'_v Z_v}
        e^{2i \vb*{\gamma} \cdot \vb{A}^{(2)}_u}
        e^{2i \vb*{\gamma} \cdot \vb{A}^{(2)}_v}, \\
    \Gamma^{XZ}_{uv}(\vb*{\gamma}) 
    &= \Gamma^{YZ}_{uv}(\vb*{\gamma}) 
    = e^{2i \gamma_u h'_u Z_u}
        e^{2i \gamma_{uv} J'_{uv} Z_u Z_v}
        e^{2i \vb*{\gamma} \cdot \vb{A}^{(2)}_u}, \\
    \Gamma^{ZX}_{uv}(\vb*{\gamma}) 
    &= \Gamma^{ZY}_{uv}(\vb*{\gamma}) 
    = e^{2i \gamma_v h'_v Z_v} 
        e^{2i \gamma_{uv} J'_{uv} Z_u Z_v}
        e^{2i \vb*{\gamma} \cdot \vb{A}^{(2)}_v}, \\
    \Gamma^Z_u(\vb*{\gamma}) 
    &= \Gamma^{ZZ}_{uv}(\vb*{\gamma}) 
    = \mathbb{I}.
    \label{eq:Gamma_eqs}
\end{split}\end{equation}
Noting that Pauli-$Z$ operators are involutory, we can expand $\smash{e^{2i \vb*{\gamma} \cdot \vb{A}^{(2)}_u}}$ and $\smash{e^{2i \vb*{\gamma} \cdot \vb{A}^{(2)}_v}}$ using \cref{eq:app/preliminaries/identities/exp-involutory-cos-sin} followed by \cref{eq:app/preliminaries/identities/prod-sum-rewriting-binary}, to obtain
\begin{equation}\begin{split}
    e^{2i \vb*{\gamma} \cdot \vb{A}^{(2)}_u} 
    &= \prod_{j = 1}^d \left[
        \cos(2 J'_{u g_j} \gamma_{u g_j}) \mathbb{I}
        + i \sin(2 J'_{u g_j} \gamma_{u g_j}) Z_u Z_{g_j}
        \right] \\
    &= \sum_{\vb{x} \in \mathbb{F}_2^d}
        \left\{ \prod_{j = 1}^d 
            \cos[1 - x_j](2 J'_{u g_j} \gamma_{u g_j})
            \left[ i \sin(2 J'_{u g_j} \gamma_{u g_j}) \right]^{x_j}
        \right\} 
        Z_u^{\wt(\vb{x})}
        \left( \prod_{j = 1}^d Z_{g_j}^{x_j} \right), \\
    e^{2i \vb*{\gamma} \cdot \vb{A}^{(2)}_v} 
    &= \prod_{j = 1}^e \left[
        \cos(2 J'_{v h_j} \gamma_{v h_j}) \mathbb{I}
        + i \sin(2 J'_{v h_j} \gamma_{v h_j}) Z_v Z_{h_j}
        \right] \\
    &= \sum_{\vb{x} \in \mathbb{F}_2^e}
        \left\{ \prod_{j = 1}^d 
            \cos[1 - x_j](2 J'_{v h_j} \gamma_{v h_j})
            \left[ i \sin(2 J'_{v h_j} \gamma_{v h_j}) \right]^{x_j}
        \right\} 
        Z_v^{\wt(\vb{x})}
        \left( \prod_{j = 1}^d Z_{h_j}^{x_j} \right).
    \label{app-eq:product/expansion-A-u-A-v-1}
\end{split}\end{equation}
An alternative expansion can be written using the re-organization of phase Hamiltonian terms expressed in \cref{app-eq:product/A-2-u-A-2-v-repartition}, which gives
\begin{equation}\begin{split}
    e^{2i \vb*{\gamma} \cdot \vb{A}^{(2)}_u}
    &= \sum_{\vb*{\alpha} \in \mathbb{F}_2^b}
        \sum_{\vb*{\mu} \in \mathbb{F}_2^f}
        \left\{ \prod_{j = 1}^b 
            \cos[1 - \alpha_j](2 J'_{u g_j} \gamma_{u g_j})
            \left[ i \sin(2 J'_{u g_j} \gamma_{u g_j}) \right]^{\alpha_j}
        \right\}
        \left\{ \prod_{j = 1}^f 
            \cos[1 - \mu_j](2 J'_{u a_j} \gamma_{u a_j})
            \left[ i \sin(2 J'_{u a_j} \gamma_{u a_j}) \right]^{\mu_j}
        \right\} \\
        & \qquad \qquad \qquad \times
        Z_u^{\wt(\vb*{\alpha}) + \wt(\vb*{\mu})}
        \left( \prod_{j = 1}^b Z_{g_j}^{\alpha_j} \right)
        \left( \prod_{j = 1}^f Z_{a_j}^{\mu_j} \right), \\
    e^{2i \vb*{\gamma} \cdot \vb{A}^{(2)}_v}
    &= \sum_{\vb*{\beta} \in \mathbb{F}_2^c}
        \sum_{\vb*{\nu} \in \mathbb{F}_2^f}
        \left\{ \prod_{j = 1}^c 
            \cos[1 - \beta_j](2 J'_{v h_j} \gamma_{v h_j})
            \left[ i \sin(2 J'_{v h_j} \gamma_{v h_j}) \right]^{\beta_j}
        \right\}
        \left\{ \prod_{j = 1}^f 
            \cos[1 - \nu_j](2 J'_{v a_j} \gamma_{v a_j})
            \left[ i \sin(2 J'_{v a_j} \gamma_{v a_j}) \right]^{\nu_j}
        \right\} \\
        & \qquad \qquad \qquad \times
        Z_u^{\wt(\vb*{\beta}) + \wt(\vb*{\nu})}
        \left( \prod_{j = 1}^b Z_{h_j}^{\beta_j} \right)
        \left( \prod_{j = 1}^f Z_{a_j}^{\nu_j} \right).
    \label{app-eq:product/expansion-A-u-A-v-2}
\end{split}\end{equation}
Now we are ready to evaluate $\smash{\xi^P_u(\vb*{\gamma})}$ and $\smash{\xi^{PQ}_{uv}(\vb*{\gamma})}$. For brevity, we define the shorthand
\begin{equation}\begin{split}
    R^\setminus_{uv}(\vb*{\gamma})
    &= \prod_{g \in \mathcal{N}_{u \setminus v}} 
        \cos(2 J'_{ug} \gamma_{ug}),
    \\
    R^\bbslash_{uv}(\vb*{\gamma}) 
    &= \prod_{g \in \mathcal{N}_{u \bbslash v}} 
        \cos(2 J'_{ug} \gamma_{ug}),
    \\
    R^\pm_{uv}(\vb*{\gamma}) 
    &= \prod_{a \in \mathcal{N}_{uv}} 
        \cos[2 \left( J'_{ua} \gamma_{ua} \pm J'_{va} \gamma_{va} \right)].
    \label{app-eq:product/PM/R}
\end{split}\end{equation}
Generally, we substitute \cref{eq:Gamma_eqs} into \cref{app-eq:product/definition-Gamma-P-u-PQ-uv} to obtain $\smash{\eta^P_u(\vb*{\gamma})}$ and $\smash{\eta^{PQ}_{uv}(\vb*{\gamma})}$, and then evaluate the inner products in \cref{app-eq:xi_eq}. Following the equivalences in \cref{eq:Gamma_eqs}, we break the Pauli operators $P_u$ and $P_u Q_v$ by categories, up to the symmetry property as established in \cref{app-eq:/product/symmetry/property}.

\begin{itemize}

    \item $P \in \{X, Y\}$. Then invoking the expansion of the phase unitary in \cref{app-eq:product/expansion-A-u-A-v-1},
    \begin{equation}\begin{split}
        \xi^P_u(\vb*{\gamma}) 
        &= \mel{s}{
            e^{2 i \gamma_u h'_u Z_u} 
            e^{2 i \gamma_{uv} J'_{uv} Z_u Z_v}
            e^{2 i \vb*{\gamma} \cdot \vb{A}^{(2)}_u}
            P_u}
            {s} \\
        &= \sum_{k \in \mathbb{F}_2} 
            \sum_{s \in \mathbb{F}_2}
            \sum_{\vb{x} \in \mathbb{F}_2^d}
            \cos[1 - k](2 h'_u \gamma_u) \cos[1 - s](2 J'_{uv} \gamma_{uv})
            \left[ i \sin(2 h'_u \gamma_u) \right]^k \left[ i \sin(2 J'_{uv} \gamma_{uv}) \right]^s \\
            & \qquad \times
            \left\{ \prod_{j = 1}^d 
                \cos[1 - x_j](2 J'_{u g_j} \gamma_{u g_j})
                \left[ i \sin(2 J'_{u g_j} \gamma_{u g_j}) \right]^{x_j}
            \right\}
            \underbrace{
                \tr( \ket{s} \bra{s}
                    Z_u^{k + s + \wt(\vb{x})}
                    P_u Z_v^s 
                    \prod_{j = 1}^d Z_{g_j}^{x_j} )
            }_{\psi^P},
        \label{app-eq:product/xi-P-X-Y-initial}
    \end{split}\end{equation}
    where we have denoted the trace term as $\psi^P$. Expanding $\ket{s} \bra{s}$ using \cref{eq:app/preliminaries/identities/plus-density-matrix-expansion} and noting the orthogonality property of Pauli operators as in \cref{eq:app/preliminaries/identities/pauli-orthogonality},
    \begin{equation}\begin{split}
        \psi^P
        &= \tr(
                \frac{\mathbb{I} + X_u}{2} 
                Z_u^{k + s + \wt(\vb{x})} 
                P_u)
            \underbrace{\tr(
                \frac{\mathbb{I} + X_v}{2}
                Z_v^s)}_{\delta_{s0}}
            \underbrace{\prod_{j = 1}^d \tr(
                \frac{\mathbb{I} + X_{g_j}}{2}
                Z_{g_j}^{x_i})}_{\delta_{\vb{x} \vb{0}}} \\
        &= \tr(
                \frac{\mathbb{I} + X_u}{2}
                Z_u^k
                P_u)
            \delta_{s0} \delta_{\vb{x} \vb{0}}
        = \left(
                \delta_{PX} \delta_{k0}
                -i \delta_{PY} \delta_{k1}
            \right) \delta_{s0} \delta_{\vb{x} \vb{0}}.
        \label{app-eq:product/psi-P-X-Y}
    \end{split}\end{equation}

    Substituting \cref{app-eq:product/psi-P-X-Y} into \cref{app-eq:product/xi-P-X-Y-initial}, we obtain
    \begin{equation}\begin{split}
        \xi^X_u(\vb*{\gamma}) 
        &= \cos(2 h'_u \gamma_u)
            \cos(2 J'_{uv} \gamma_{uv}) 
            \prod_{j = 1}^d \cos(2 J'_{u g_j} \gamma_{u g_j})
        = \cos(2 h'_u \gamma_u) 
            \cos(2 J'_{uv} \gamma_{uv}) 
            R^\setminus_{uv}(\vb*{\gamma}), \\
        \xi^Y_u(\vb*{\gamma}) 
        &= \sin(2 h'_u\gamma_u) 
            \cos(2 J'_{uv} \gamma_{uv})
            \prod_{j = 1}^d \cos(2 J'_{ug_j} \gamma_{u g_j})
        = \sin(2 h'_u \gamma_u)
            \cos(2 J'_{uv} \gamma_{uv})
            R^\setminus_{uv}(\vb*{\gamma}).
        \label{app-eq:product/xi-P-X-Y-final}
    \end{split}\end{equation}

    \item $P \in \{Z\}$. Then trivially
    \begin{equation}\begin{split}
        \xi^Z_u(\vb*{\gamma}) 
        = \tr(\ket{s} \bra{s} Z_u)
        = \left[ \prod_{q \neq u}
                \tr(\frac{\mathbb{I} + X_q}{2}) \right] 
            \tr(\frac{\mathbb{I} + X_u}{2} Z_u)
        = 0.
        \label{app-eq/product/xi-Z}
    \end{split}\end{equation}

    \item $PQ \in \{XX, XY, YX, YY\}$. Then invoking the expansion of the phase unitary in \cref{app-eq:product/expansion-A-u-A-v-2},
    \begin{equation}\begin{split}
        \xi^{PQ}_{uv}(\vb*{\gamma}) 
        &= \mel{s}{
            e^{2i \gamma_u h'_u Z_u} e^{2i \gamma_v h'_v Z_v}
            e^{2i \vb*{\gamma} \cdot \vb{A}^{(2)}_u}
            e^{2i \vb*{\gamma} \cdot \vb{A}^{(2)}_v}
            P_u Q_v}
            {s} \\
        &= \sum_{k \in \mathbb{F}_2} 
            \sum_{l \in \mathbb{F}_2}
            \sum_{\vb*{\alpha} \in \mathbb{F}_2^b}
            \sum_{\vb*{\beta} \in \mathbb{F}_2^c}
            \sum_{\vb*{\mu} \in \mathbb{F}_2^f}
            \sum_{\vb*{\nu} \in \mathbb{F}_2^f}
            \cos[1 - k](2 h'_u \gamma_u) 
            \cos[1 - l](2 h'_v \gamma_v)
            \left[ i \sin(2 h'_u \gamma_u) \right]^k 
            \left[ i \sin(2 h'_v \gamma_v) \right]^l \\
            & \qquad \times
            \left\{ \prod_{j = 1}^b
                \cos[1 - \alpha_j](2 J'_{u g_j} \gamma_{u g_j})
                \left[ i \sin(2 J'_{u g_j} \gamma_{u g_j}) \right]^{\alpha_j}
            \right\}
            \left\{ \prod_{j = 1}^c 
                \cos[1 - \beta_j](2 J'_{v h_j} \gamma_{v h_j})
                \left[ i \sin(2 J'_{v h_j} \gamma_{v h_j}) \right]^{\beta_j}
            \right\} \\
            & \qquad \times
            \left\{ \prod_{j = 1}^f
                \cos[1 - \mu_j](2 J'_{u a_j} \gamma_{u a_j})
                \left[ i \sin(2 J'_{u a_j} \gamma_{u a_j}) \right]^{\mu_j}
            \right\}
            \left\{ \prod_{j = 1}^f
                \cos[1 - \nu_j](2 J'_{v a_j} \gamma_{v a_j})
                \left[ i \sin(2 J'_{v a_j} \gamma_{v a_j}) \right]^{\nu_j}
            \right\} \\
            & \qquad \times
            \underbrace{
                \tr( \ket{s} \bra{s}
                    Z_u^{k + \wt(\vb*{\alpha}) + \wt(\vb*{\mu})}
                    P_u
                    Z_v^{l + \wt(\vb*{\beta}) + \wt(\vb*{\nu})}
                    Q_v
                    \prod_{j = 1}^b Z_{g_j}^{\alpha_j}
                    \prod_{j = 1}^c Z_{h_j}^{\beta_j}
                    \prod_{j = 1}^f Z_{a_j}^{\mu_j + \nu_j} )
            }_{\psi^{PQ}},
        \label{app-eq:product/xi-PQ-XX-XY-YY-initial}
    \end{split}\end{equation}
    where we have denoted the trace term as $\psi^{PQ}$. In the same spirit as before, we expand
    \begin{equation}\begin{split}
        \psi^{PQ}
        &= \tr(
                \frac{\mathbb{I} + X_u}{2}
                Z_u^{k + \wt(\vb*{\alpha}) + \wt(\vb*{\mu})}
                P_u)
            \tr(
                \frac{\mathbb{I} + X_v}{2}
                Z_v^{l + \wt(\vb*{\beta}) + \wt(\vb*{\nu})}
                Q_v) \\
            & \qquad \times
            \underbrace{ \prod_{j = 1}^b \tr(
                \frac{\mathbb{I} + X_{g_j}}{2}
                Z_{g_j}^{\alpha_j})
            }_{\delta_{\vb*{\alpha} \vb{0}}}
            \underbrace{ \prod_{j = 1}^c \tr(
                \frac{\mathbb{I} + X_{h_j}}{2}
                Z_{h_j}^{\beta_j})
            }_{\delta_{\vb*{\beta} \vb{0}}}
            \underbrace{ \prod_{j = 1}^f \tr(
                \frac{\mathbb{I} + X_{a_j}}{2}
                Z_{a_j}^{\mu_i + \nu_j})}_{\delta_{\vb*{\mu} \vb*{\nu}}} \\
        &= \underbrace{ \tr(
                \frac{\mathbb{I} + X_u}{2}
                Z_u^{k + \wt(\vb*{\mu})} 
                P_u)
            }_{\phi^P}
            \underbrace{ \tr(
                \frac{\mathbb{I} + X_v}{2}
                Z_v^{l + \wt(\vb*{\mu})}
                Q_v)}_{\phi^Q}
            \delta_{\vb*{\alpha} \vb{0}} 
            \delta_{\vb*{\beta} \vb{0}} 
            \delta_{\vb*{\mu} \vb*{\nu}}.
        \label{app-eq:product/psi-PQ-XX-XY-YY-1}
    \end{split}\end{equation}

    Moreover
    \begin{equation}\begin{split}
        \phi^P
        &= \delta_{PX} \left\{ 
                \delta_{k0} \pi^{(\mathrm{e})}[\wt(\vb*{\mu})]
                + \delta_{k1} \pi^{(\mathrm{o})}[\wt(\vb*{\mu})] \right\}
            - i \delta_{PY} \left\{ 
                \delta_{k0} \pi^{(\mathrm{o})}[\wt(\vb*{\mu})]
                + \delta_{k1} \pi^{(\mathrm{e})}[\wt(\vb*{\mu})] \right\}, \\
        \phi^Q
        &= \delta_{QX} \left\{ 
                \delta_{l0} \pi^{(\mathrm{e})}[\wt(\vb*{\mu})]
                + \delta_{l1} \pi^{(\mathrm{o})}[\wt(\vb*{\mu})] \right\}
            - i \delta_{QY} \left\{ 
                \delta_{l0} \pi^{(\mathrm{o})}[\wt(\vb*{\mu})]
                + \delta_{l1} \pi^{(\mathrm{e})}[\wt(\vb*{\mu})] \right\},
        \label{app-eq:product/psi-PQ-XX-XY-YY-2}
    \end{split}\end{equation}
    where $\pi^{(\mathrm{e})}(x)$ and $\pi^{(\mathrm{o})}(x) = 1 - \pi^{(\mathrm{e})}(x)$ are parity indicators of an input $x \in \mathbb{N}$, such that $\pi^{(\mathrm{e})}(x)$ has unit value when $x$ is even and is zero otherwise. Substituting \cref{app-eq:product/psi-PQ-XX-XY-YY-1,app-eq:product/psi-PQ-XX-XY-YY-2} into \cref{app-eq:product/xi-PQ-XX-XY-YY-initial}, and using \cref{eq:app/preliminaries/identities/trig-addsub-prod-expansion} for trigonometric simplifications, we find
    \begin{equation*}\begin{split}
        \xi^{XX}_{uv}(\vb*{\gamma})
        &= R^\bbslash_{uv}(\vb*{\gamma})
            R^\bbslash_{vu}(\vb*{\gamma})
            \left[
                \cos(2 h'_u \gamma_u)
                \cos(2 h'_v \gamma_v)
                \sum_{\substack{\vb*{\mu} \in \mathbb{F}_2^f \\ \wt(\vb*{\mu}) \, \text{even}}}
                + \sin(2 h'_u \gamma_u) \sin(2 h'_v \gamma_v)
                \sum_{\substack{\vb*{\mu} \in\mathbb{F}_2^f \\ \wt(\vb*{\mu}) \, \text{odd}}}
            \right] \\
            & \qquad \qquad \times 
            \prod_{j = 1}^f
            \cos[1 - \mu_j](2 J'_{u a_j} \gamma_{u a_j})
            \cos[1 - \mu_j](2 J'_{v a_j} \gamma_{v a_j})
            \sin[\mu_j](2 J'_{u a_j} \gamma_{u a_j})
            \sin[\mu_j](2 J'_{v a_j} \gamma_{v a_j}) \\
        &= \frac{1}{2}
            R^\bbslash_{uv}(\vb*{\gamma})
            R^\bbslash_{vu}(\vb*{\gamma})
            \left\{
                \cos(2 h'_u \gamma_u)
                \cos(2 h'_v \gamma_v)
                \left[ R^-_{uv}(\vb*{\gamma}) + R^+_{uv}(\vb*{\gamma}) \right]
                + \sin(2 h'_u \gamma_u)
                \sin(2 h'_v \gamma_v)
                \left[ R^-_{uv}(\vb*{\gamma}) - R^+_{uv}(\vb*{\gamma}) \right]
            \right\},
    \end{split}\end{equation*}
    \begin{equation*}\begin{split}
        \xi^{XY}_{uv}(\vb*{\gamma})
        &= R^\bbslash_{uv}(\vb*{\gamma})
            R^\bbslash_{vu}(\vb*{\gamma})
            \left[ 
                \cos(2 h'_u \gamma_u)
                \sin(2 h'_v \gamma_v)
                \sum_{\substack{\vb*{\mu} \in \mathbb{F}_2^f \\ \wt(\vb*{\mu}) \, \text{even}}}
                - \sin(2 h'_u \gamma_u) 
                \cos(2 h'_v \gamma_v)
                \sum_{\substack{\vb*{\mu} \in \mathbb{F}_2^f \\ \wt(\vb*{\mu}) \, \text{odd}}}
            \right] \\
            & \qquad \qquad \times 
            \prod_{j = 1}^f
            \cos[1 - \mu_j](2 J'_{u a_j} \gamma_{u a_j})
            \cos[1 - \mu_j](2 J'_{v a_j} \gamma_{v a_j})
            \sin[\mu_j](2 J'_{u a_j} \gamma_{u a_j})
            \sin[\mu_j](2 J'_{v a_j} \gamma_{v a_j}) \\
        &= \frac{1}{2}
            R^\bbslash_{uv}(\vb*{\gamma})
            R^\bbslash_{vu}(\vb*{\gamma})
            \left\{
                \cos(2 h'_u \gamma_u)
                \sin(2 h'_v \gamma_v)
                \left[ R^-_{uv}(\vb*{\gamma}) + R^+_{uv}(\vb*{\gamma}) \right]
                - \sin(2 h'_u \gamma_u)
                \cos(2 h'_v \gamma_v)
                \left[ R^-_{uv}(\vb*{\gamma}) - R^+_{uv}(\vb*{\gamma}) \right]
            \right\},
    \end{split}\end{equation*}
    \begin{equation}\begin{split}
        \xi^{YY}_{uv}(\vb*{\gamma}) 
        &= R^\bbslash_{uv}(\vb*{\gamma})
            R^\bbslash_{vu}(\vb*{\gamma})
            \left[
                \cos(2 h'_u \gamma_u)
                \cos(2 h'_v \gamma_v)
                \sum_{\substack{\vb*{\mu} \in \mathbb{F}_2^f \\ \wt(\vb*{\mu}) \, \text{odd}}}
                + \sin(2 h'_u \gamma_u)
                \sin(2 h'_v \gamma_v)
                \sum_{\substack{\vb*{\mu}\in\mathbb{F}_2^f \\ \wt(\vb*{\mu}) \, \text{even}}}
            \right] \\
            & \qquad \qquad \times
            \prod_{j = 1}^f
            \cos[1 - \mu_j](2 J'_{u a_j} \gamma_{u a_j})
            \cos[1 - \mu_j](2 J'_{v a_j} \gamma_{v a_j})
            \sin[\mu_j](2 J'_{u a_j} \gamma_{u a_j})
            \sin[\mu_j](2 J'_{v a_j} \gamma_{v a_j}) \\
        &= \frac{1}{2}
            R^\bbslash_{uv}(\vb*{\gamma})
            R^\bbslash_{vu}(\vb*{\gamma})
            \left\{
                \cos(2 h'_u \gamma_u)
                \cos(2 h'_v \gamma_v)
                \left[ R^-_{uv}(\vb*{\gamma}) - R^+_{uv}(\vb*{\gamma}) \right]
                + \sin(2 h'_u \gamma_u)
                \sin(2 h'_v \gamma_v)
                \left[ R^-_{uv}(\vb*{\gamma}) + R^+_{uv}(\vb*{\gamma}) \right]
            \right\}.
        \label{app-eq:product/xi-PQ-XX-XY-YY-final}
    \end{split}\end{equation}

    The coefficient $\xi^{YX}_{uv}(\vb*{\gamma})$ follows immediately from $\xi^{XY}_{uv}(\vb*{\gamma})$ by the symmetry property of \cref{app-eq:/product/symmetry/property}, but can also be worked out explicitly in the same manner as above.

    \item $PQ \in \{XZ, YZ\}$. Then invoking the expansion of the phase unitary in \cref{app-eq:product/expansion-A-u-A-v-1},
    \begin{equation}\begin{split}
        \xi^{PQ}_{uv}(\vb*{\gamma}) 
        &= \mel{s}{
            e^{2i \gamma_u h'_u Z_u}
            e^{2i \gamma_{uv} J'_{uv} Z_u Z_v}
            e^{2i \vb*{\gamma} \cdot \vb{A}^{(2)}_u}
            P_u Q_v}
            {s} \\
        &= \sum_{k \in \mathbb{F}_2}
            \sum_{s \in \mathbb{F}_2}
            \sum_{\vb*{x} \in \mathbb{F}_2^d}
            \cos[1 - k](2 h'_u \gamma_u)
            \cos[1 - s](2 J'_{uv} \gamma_{uv})
            \left[ i \sin(2 h'_u \gamma_u) \right]^k
            \left[ i \sin(2 J'_{uv} \gamma_{uv}) \right]^s \\
            & \qquad \times
            \left\{ \prod_{j = 1}^d 
                \cos[1 - x_j](2 J'_{u g_j} \gamma_{u g_j})
                \left[ i \sin(2 J'_{u g_j} \gamma_{u g_j}) \right]^{x_j}
            \right\}
            \underbrace{
                \tr( \ket{s} \bra{s}
                    Z_u^{k + s + \wt(\vb{x})}
                    P_u Z_v^{s + 1}
                    \prod_{j = 1}^d Z_{g_j}^{x_j} )
            }_{\psi^P},
        \label{app-eq:product/xi-PQ-XZ-YZ-initial}
    \end{split}\end{equation}
    where we have denoted the trace term as $\psi^P$. We expand
    \begin{equation}\begin{split}
        \psi^P 
        &= \tr(
                \frac{\mathbb{I} + X_u}{2}
                Z_u^{k + s + \wt(\vb*{x})}
                P_u)
            \tr(
                \frac{\mathbb{I} + X_v}{2}
                Z_v^{s + 1})
            \prod_{j = 1}^d \tr(
                \frac{\mathbb{I} + X_{g_j}}{2}
                Z_{g_j}^{x_j}) \\
        &= \tr(
                \frac{\mathbb{I} + X_u}{2}
                Z_u^{k + 1}
                P_u)
            \delta_{s1} \delta_{\vb{x} \vb{0}}
        = \left( 
                \delta_{PX} \delta_{k1} 
                - i \delta_{PY} \delta_{k0} 
            \right) \delta_{s1} \delta_{\vb{x} \vb{0}}.
        \label{app-eq:product/psi-PQ-XZ-YZ}
    \end{split}\end{equation}

    Substituting \cref{app-eq:product/psi-PQ-XZ-YZ} into \cref{app-eq:product/xi-PQ-XZ-YZ-initial}, we obtain
    \begin{equation}\begin{split}
        \xi^{XZ}_{uv}(\vb*{\gamma}) 
        &= -\sin(2 h'_u \gamma_u)
            \sin(2 J'_{uv} \gamma_{uv})
            R^\setminus_{uv}(\vb*{\gamma}), \\
        \xi^{YZ}_{uv}(\vb*{\gamma}) 
        &= \cos(2 h'_u \gamma_u) 
            \sin(2 J'_{uv} \gamma_{uv})
            R^\setminus_{uv}(\vb*{\gamma}).
        \label{app-eq:product/xi-PQ-XZ-YZ-final}
    \end{split}\end{equation}
    
    \item $PQ \in \{ZZ\}$. Then trivially
    \begin{equation}\begin{split}
        \xi^{ZZ}_{uv}(\vb*{\gamma}) 
        = \tr(\ket{s} \bra{s} Z_u Z_v)
        = \left[ \prod_{q \neq u, v}
                \tr(\frac{\mathbb{I} + X_q}{2}) \right] 
            \tr(\frac{\mathbb{I} + X_u}{2} Z_u)
            \tr(\frac{\mathbb{I} + X_v}{2} Z_v)
        = 0.
        \label{app-eq:product/xi-ZZ}
    \end{split}\end{equation}
    
\end{itemize}

The results for the coefficients $\xi^P_u(\vb*{\gamma})$ and $\xi^{PQ}_{uv}(\vb*{\gamma})$ derived above are also summarized in \cref{app-tab:product/summary-xi-expressions}. We remind that these expressions, and the shorthand in \cref{app-eq:product/PM/R}, are slightly dissimilar to those in \cref{eq:intro-example/PM/xi,eq:intro-example/PM/R} of the main text, which have been simplified for the particular problem setting of \textsc{MaxCut} discussed in \cref{sec:intro-example-maxcut}.

\clearpage 
\pagebreak

\subsection{Specializing to specific QAOA variants}
\label{app:product/specific}

In this subsection, we consider specialization of our general results of PM-QAOA to specific algorithm variants. The expressions for vanilla QAOA have been previously reported in~\cite{PhysRevA.97.022304,hadfield2018quantum,herrman2022multi,ozaeta2022expectation,vijendran2024expressive}. The expressions obtained for WS-QAOA and FAM-QAOA are new to the best of our knowledge. 

\subsubsection{Vanilla QAOA}
\label{app:product/specific/qaoa}

The vanilla QAOA mixer is a special case of our product mixer in \cref{eq:product_mixer} specified by the $\smash{\vu{r}_u}$ values listed in \cref{tab:product_mixer_Ham_coeff}. Explicitly, the vanilla QAOA mixer Hamiltonian is
\begin{equation}
    B^{\mathrm{QAOA}} = \sum_{u = 1}^n B^{\mathrm{QAOA}}_u,
    \qquad
    B^{\mathrm{QAOA}}_u = X_u,
    \qquad
    \vb{B}^{\mathrm{QAOA}}
    = \mqty(B^{\mathrm{QAOA}}_1, B^{\mathrm{QAOA}}_2, \ldots, B^{\mathrm{QAOA}}_n).
\end{equation}
As such, \cref{app-eq:conj-Zu-Zu-Zv} reduces to 
\begin{equation}\begin{split}
    e^{i \vb*{\beta} \cdot \vb{B}^{\mathrm{QAOA}}} 
        Z_u 
        e^{-i \vb*{\beta} \cdot \vb{B}^{\mathrm{QAOA}}} 
    &= \underbrace{
            \sin(2 \beta_u)
        }_{a^Y_u(\vb*{\beta})} Y_u 
        + \underbrace{
            \cos(2 \beta_u)
        }_{a^Z_u(\vb*{\beta})} Z_u, \\
    e^{i \vb*{\beta} \cdot \vb{B}^{\mathrm{QAOA}}}
        Z_u Z_v 
        e^{-i \vb*{\beta} \cdot \vb{B}^{\mathrm{QAOA}}} 
    &= \underbrace{
            \sin(2 \beta_u) \sin(2 \beta_v)
        }_{a^{YY}_{uv}(\vb*{\beta})} Y_u Y_v
        + \underbrace{
            \sin(2 \beta_u) \cos(2 \beta_v)
        }_{a^{YZ}_{uv}(\vb*{\beta})} Y_u Z_v \\
        & \qquad 
        + \underbrace{
            \cos(2 \beta_u) \sin(2 \beta_v)
        }_{a^{ZY}_{uv}(\vb*{\beta})} Z_u Y_v
        + \underbrace{
            \cos(2 \beta_u) \cos(2 \beta_v)
        }_{a^{ZZ}_{uv}(\vb*{\beta})} Z_u Z_v.
    \label{app-eq:product/specific/qaoa/a}
\end{split}\end{equation}

\subsubsection{Warm-start QAOA (WS-QAOA)}
\label{app:product/specific/ws-qaoa}

Likewise, the WS-QAOA mixer is another special case of our product mixer in \cref{eq:product_mixer} specified by the $\smash{\vu{r}_u}$ values listed in \cref{tab:product_mixer_Ham_coeff}. Explicitly,
\begin{equation}
    B^{\mathrm{WS}} = \sum_{u = 1}^n B^{\mathrm{WS}}_u,
    \qquad
    B^{\mathrm{WS}}_u = - \sin{\theta_u} X_u - \cos{\theta_u} Z_u,
    \qquad
    \vb{B}^{\mathrm{WS}}
    = \mqty(B^{\mathrm{WS}}_1, B^{\mathrm{WS}}_2, \ldots, B^{\mathrm{WS}}_n).
\end{equation}
Then, \cref{app-eq:conj-Zu-Zu-Zv} produces $\smash{a^P_u(\vb*{\beta})}$ and $\smash{a^{PQ}_{uv}(\vb*{\beta})}$ coefficients simpler than those for general PM-QAOA listed in \cref{app-tab:product/summary-a-expressions}. We give these simplified expressions in \cref{app-tab:product/summary-a-expressions-ws}.

\begin{table*}[!ht]
    \centering
    \setlength{\tabcolsep}{0pt}
    \renewcommand*{\arraystretch}{1.2}
    \newcommand\tstrut{\rule{0pt}{12pt}}
    \newcommand\bstrut{\rule[-8pt]{0pt}{0pt}}
    \begin{tabular}[t]{p{1.5cm} p{16cm}}
        \toprule
        $P$ & $a^P_u(\vb*{\beta})$ \\
        \midrule
        $X$ &
            $\begin{aligned}
                \tstrut
                \sin^2(\beta_u) \sin(2 \theta_u)
                \bstrut
            \end{aligned}$ \\
        \tabdashline & \tabdashline
        $Y$ &
            $\begin{aligned}
                \tstrut
                - \sin(2 \beta_u) \sin(\theta_u)
                \bstrut
            \end{aligned}$ \\
        \tabdashline & \tabdashline
        $Z$ &
            $\begin{aligned}
                \tstrut
                \cos^2(\beta_u) 
                + \sin^2(\beta_u) \cos(2 \theta_u) 
                \bstrut
            \end{aligned}$ \\
        \bottomrule
    \end{tabular} \\[1.5\baselineskip]
    \begin{tabular}[t]{p{1.5cm} p{16cm}}
        \toprule
        $PQ$ & $a^{PQ}_{uv}(\vb*{\beta})$ \\
        \midrule
        $XX$ &
            $\begin{aligned}
                \tstrut
                \sin^2(\beta_u)\sin^2(\beta_v)\sin(2\theta_u)\sin(2\theta_v)
                \bstrut
            \end{aligned}$ \\
        \tabdashline & \tabdashline
        $XY$ &
            $\begin{aligned}
                \tstrut
                -\sin^2(\beta_u)\sin(2\beta_v)\sin(2\theta_u)\sin(\theta_v)
                \bstrut
            \end{aligned}$ \\
        \tabdashline & \tabdashline
        $YY$ &
            $\begin{aligned}
                \tstrut
                \sin(2\beta_u)\sin(2\beta_v)\sin(\theta_u)\sin(\theta_v)
                \bstrut
            \end{aligned}$ \\
        \tabdashline & \tabdashline
        $XZ$ &
            $\begin{aligned}
                \tstrut
                \sin^2(\beta_u)\sin(2\theta_u)\left(\cos^2(\beta_v) + \sin^2(\beta_v)\cos(2\theta_v)\right)
                \bstrut
            \end{aligned}$ \\
        \tabdashline & \tabdashline
        $YZ$ &
            $\begin{aligned}
                \tstrut
                - \sin(2\beta_u)\sin(\theta_u)\left(\cos^2(\beta_v) + \cos(2\theta_v)\sin^2(\beta_v)\right)
                \bstrut
            \end{aligned}$ \\
        \tabdashline & \tabdashline
        $ZZ$ &
            $\begin{aligned}
                \tstrut
                \left(\cos^2(\beta_u) + \cos(2\theta_u)\sin^2(\beta_u)\right)\left(\cos^2(\beta_v) + \cos(2\theta_v)\sin^2(\beta_v)\right)
                \bstrut
            \end{aligned}$ \\
        \bottomrule
    \end{tabular}
    \caption{\textbf{List of $a^P_u(\vb*{\beta})$ and $a^{PQ}_{uv}(\vb*{\beta})$ expressions for WS-QAOA.} By the symmetry property in \cref{app-eq:/product/symmetry/property}, the coefficients $\smash{a^{YX}_{uv}(\vb*{\beta})}$, $\smash{a^{ZX}_{uv}(\vb*{\beta})}$, and $\smash{a^{ZY}_{uv}(\vb*{\beta})}$ can be obtained from $\smash{a^{XY}_{uv}(\vb*{\beta})}$, $\smash{a^{XZ}_{uv}(\vb*{\beta})}$, and $\smash{a^{YZ}_{uv}(\vb*{\beta})}$ respectively by swapping the $u$ and $v$ vertices.}
    \label{app-tab:product/summary-a-expressions-ws}
\end{table*}

\subsubsection{Free axis mixer QAOA (FAM-QAOA)}
\label{app:product/specific/fam-qaoa}

As a final example, the FAM-QAOA mixer is a third special case of our product mixer in \cref{eq:product_mixer} specified by the $\smash{\vu{r}_u}$ values listed in \cref{tab:product_mixer_Ham_coeff}. Explicitly,
\begin{equation}
    B^{\mathrm{FAM}} = \sum_{u = 1}^n B^{\mathrm{FAM}}_u,
    \qquad
    B^{\mathrm{FAM}}_u = \cos{\theta_u} X_u - \sin{\theta_u} Y_u,
    \qquad
    \vb{B}^{\mathrm{FAM}}
    = \mqty(B^{\mathrm{FAM}}_1, B^{\mathrm{FAM}}_2, \ldots, B^{\mathrm{FAM}}_n).
\end{equation}
Then, likewise, \cref{app-eq:conj-Zu-Zu-Zv} produces $\smash{a^P_u(\vb*{\beta})}$ and $\smash{a^{PQ}_{uv}(\vb*{\beta})}$ coefficients simpler than those for general PM-QAOA listed in \cref{app-tab:product/summary-a-expressions}. We give these simplified expressions in \cref{app-tab:product/summary-a-expressions-fam}.

\begin{table*}[!ht]
    \centering
    \setlength{\tabcolsep}{0pt}
    \renewcommand*{\arraystretch}{1.2}
    \newcommand\tstrut{\rule{0pt}{12pt}}
    \newcommand\bstrut{\rule[-8pt]{0pt}{0pt}}
    \begin{tabular}[t]{p{1.5cm} p{16cm}}
        \toprule
        $P$ & $a^P_u(\vb*{\beta})$ \\
        \midrule
        $X$ &
            $\begin{aligned}
                \tstrut
                \sin(2 \beta_u) \sin(\theta_u)
                \bstrut
            \end{aligned}$ \\
        \tabdashline & \tabdashline
        $Y$ &
            $\begin{aligned}
                \tstrut
                \sin(2 \beta_u) \cos(\theta_u)
                \bstrut
            \end{aligned}$ \\
        \tabdashline & \tabdashline
        $Z$ &
            $\begin{aligned}
                \tstrut
                \cos(2 \beta_u)
                \bstrut
            \end{aligned}$ \\
        \bottomrule
    \end{tabular} \\[1.5\baselineskip]
    \begin{tabular}[t]{p{1.5cm} p{16cm}}
        \toprule
        $PQ$ & $a^{PQ}_{uv}(\vb*{\beta})$ \\
        \midrule
        $XX$ &
            $\begin{aligned}
                \tstrut
                \sin(2\beta_u)\sin(2\beta_v)\sin(\theta_u)\sin(\theta_v)
                \bstrut
            \end{aligned}$ \\
        \tabdashline & \tabdashline
        $XY$ &
            $\begin{aligned}
                \tstrut
                \sin(2\beta_u)\sin(2\beta_v)\sin(\theta_u)\cos(\theta_v)
                \bstrut
            \end{aligned}$ \\
        \tabdashline & \tabdashline
        $YY$ &
            $\begin{aligned}
                \tstrut
                \sin(2\beta_u)\sin(2\beta_v)\cos(\theta_u)\cos(\theta_v)
                \bstrut
            \end{aligned}$ \\
        \tabdashline & \tabdashline
        $XZ$ &
            $\begin{aligned}
                \tstrut
                \sin(2\beta_u)\cos(2\beta_v)\sin(\theta_u)
                \bstrut
            \end{aligned}$ \\
        \tabdashline & \tabdashline
        $YZ$ &
            $\begin{aligned}
                \tstrut
                \sin(2\beta_u)\cos(2\beta_v)\cos(\theta_u)
                \bstrut
            \end{aligned}$ \\
        \tabdashline & \tabdashline
        $ZZ$ &
            $\begin{aligned}
                \tstrut
                \cos(2\beta_u)\cos(2\beta_v) + \sin^2(\beta_u)\sin^2(\beta_v)\cos(2\theta_u)\sin^2(\theta_v)
                \bstrut
            \end{aligned}$ \\
        \bottomrule
    \end{tabular}
    \caption{\textbf{List of $a^P_u(\vb*{\beta})$ and $a^{PQ}_{uv}(\vb*{\beta})$ expressions for FAM-QAOA.} By the symmetry property in \cref{app-eq:/product/symmetry/property}, the coefficients $\smash{a^{YX}_{uv}(\vb*{\beta})}$, $\smash{a^{ZX}_{uv}(\vb*{\beta})}$, and $\smash{a^{ZY}_{uv}(\vb*{\beta})}$ can be obtained from $\smash{a^{XY}_{uv}(\vb*{\beta})}$, $\smash{a^{XZ}_{uv}(\vb*{\beta})}$, and $\smash{a^{YZ}_{uv}(\vb*{\beta})}$ respectively by swapping the $u$ and $v$ vertices.}
    \label{app-tab:product/summary-a-expressions-fam}
\end{table*}

\clearpage 
\pagebreak

\subsection{Specializing to specific optimization problems}
\label{app:specialize}

Setting the cost Hamiltonian $C$ to specific optimization problems corresponds to choosing the problem graph and vertex and edge weights (\ie~coefficients of the Ising Hamiltonian). Many analytical expressions simplify drastically when specializing to certain classes of graphs. On triangle-free graphs, the number of triangles $f$ in our solutions for $\expval{C}$ can simply be zeroed. Simplifications in $\expval{C}$ are also straightforward on star graphs. Without loss of generality, one takes $u$ to be a leaf vertex and $v$ to be the center of the graph, in addition to setting $f = 0$~\cite{herrman2022multi}. In the remainder of this appendix, we elaborate on the \textsc{MaxCut} and \textsc{MaxIndependentSet} problem settings.

\subsubsection{\texorpdfstring{\textsc{MaxCut}}{MaxCut}: Proof of \texorpdfstring{\cref{thm:intro-example/PM}}{ref}}
\label{app:specialize/maxcut}

We re-iterate the definition of \textsc{MaxCut} from the main text:

\begin{center}\fbox{
\begin{minipage}{\linewidth}
\textbf{Definition:} Given a weighted undirected graph $G = (V_G, E_G, w)$ with vertex set $V_G = [n]$ and edge set $E_G = \{\{u, v\}: u, v \in V_G\}$ comprising $m$ edges, the \textsc{MaxCut} problem is to partition the vertices into $S_G \subset V_G$ and $V_G \setminus S_G$ such that the number of edges crossing the cut $(S_G, V_G \setminus S_G)$ is maximized. This optimization problem is formalized as
\begin{equation*}\begin{aligned}
    & \maximize_{\vb{x} \in \{0, 1\}^n} 
    \quad 
    \sum_{\{u, v\} \in E_G} w_{uv}(x_u \oplus x_v). 
\end{aligned}\end{equation*}
\end{minipage}}
\end{center}
\vspace{4pt}

Here, $\oplus$ denotes the modulo-2 sum or XOR operation. The cost function can be promoted to its corresponding diagonal cost Hamiltonian by the substitution $x_u = (\mathbb{I} - Z_u) / 2$. As in typical formulations of \textsc{MaxCut}, we consider the cost and phase Hamiltonians to be equal. From the general formulation of PM-QAOA (see \cref{sec:product} of the main text), by setting
\begin{equation}
    a = a' = \frac{1}{2} \sum_{u,v\in V_G} w_{uv},
    \qquad
    h_u = h_u' = 0,
    \qquad
    J_{uv} = J_{uv}' = -\frac{1}{2} w_{uv}
    \label{eq:specialize/maxcut/weights}
\end{equation}
in \cref{eq:product/definitions/ising-ham-cost}, we obtain the cost Hamiltonian of \textsc{MaxCut},
\begin{equation}
    C = A = \sum_{\{u, v\} \in E_G} C_{uv} 
    = \sum_{\{u, v\}\in E_G} \frac{w_{uv}}{2}
        \left(\mathbb{I} - Z_u Z_v\right).
\end{equation}
The general solution to the edge-resolved cost expectation $\expval{Z_uZ_v}$, from \cref{app-eq:Fu-Fuv}, is
\begin{equation}
    \expval{Z_uZ_v}
    = F_{uv}
    = \sum_{P, Q \in \mathcal{P}} a^{PQ}_{uv}(\vb*{\beta}) 
        \xi^{PQ}_{uv}(\vb*{\gamma}),
\end{equation}
where general forms for the coefficients $a^{PQ}_{uv}(\vb*{\beta})$ and $\xi^{PQ}_{uv}(\vb*{\gamma})$ are given in \cref{app-tab:product/summary-a-expressions,app-tab:product/summary-xi-expressions}. Note that, as $h_u = 0$, only $\xi^{XX}_{uv}(\vb*{\gamma})$, $\xi^{YY}_{uv}(\vb*{\gamma})$, $\xi^{YZ}_{uv}(\vb*{\gamma})$, and $\xi^{ZY}_{uv}(\vb*{\gamma})$ among the $\xi^{PQ}_{uv}(\vb*{\gamma})$ coefficients are non-zero. We produce the simplified versions of these coefficients after substitution of the $(a, h, J)$ weights in \cref{eq:specialize/maxcut/weights} defining \textsc{MaxCut}, in \cref{app-tab:summary-xi-expressions_maxcut}. 

For PM-QAOA in general, which subsumes variants such as WS-QAOA, FAM-QAOA, ma-QAOA, and XQAOA, we show the simplified edge-resolved cost expectation in the following cases:
\begin{itemize}
    
    \item \textit{Weighted problem graph, multi-angle circuit ansatz ($\vb*{\beta}$ and $\vb*{\gamma}$)}. Then
    \begin{equation}
        \expval{C_{uv}} = \frac{w_{uv}}{2}
            \left[
                1
                -a^{XX}_{uv}(\vbeta) \xi^{XX}_{uv}(\vb*{\gamma})
                -a^{YY}_{uv}(\vbeta) \xi^{YY}_{uv}(\vb*{\gamma})
                -a^{YZ}_{uv}(\vbeta) \xi^{YZ}_{uv}(\vb*{\gamma})
                -a^{ZY}_{uv}(\vbeta) \xi^{ZY}_{uv}(\vb*{\gamma})
            \right],
    \label{app-eq:maxcut_Cuv_pm}
    \end{equation}
    where the $\xi^{PQ}_{uv}(\vb*{\gamma})$ coefficients are given by the first part of \cref{app-tab:summary-xi-expressions_maxcut}. This is the same setting, and the same solution, as given in \cref{thm:intro-example/PM} of the main text.

    \item \textit{Unweighted problem graph, single-angle circuit ansatz ($\vb*{\beta} = \beta$, $\vb*{\gamma} = \gamma$, and $w_{uv} = 1$)}. Then additional simplifications can be made to the $\xi^{PQ}_{uv}(\vb*{\gamma})$ coefficients, which we provide in the second part of \cref{app-tab:summary-xi-expressions_maxcut}. Explicitly,
    \begin{equation}\begin{aligned}
        \expval{C_{uv}} 
        = \frac{1}{2} \bigg[ &
            1
            -\frac{a^{XX}_{uv}(\vbeta)}{2} \cos^{d + e - 2f}(\gamma)
                \left[1 + \cos^f(2\gamma) \right] 
            - \frac{a^{YY}_{uv}(\vbeta)}{2} \cos^{d + e -2f}(\gamma)
                \left[1 - \cos^f(2\gamma) \right] \\
            &\qquad 
            + a^{YZ}_{uv}(\vbeta) \sin(\gamma) \cos^d(\gamma) 
            + a^{ZY}_{uv}(\vbeta) \sin(\gamma) \cos^e(\gamma)
        \bigg].
    \label{app-eq:maxcut_Cuv_pm_simplified}
    \end{aligned}\end{equation}
    
\end{itemize}
The specific variant of QAOA used defines the mixer unit vectors $\vu{r}_u$ and thereby the $a^{PQ}_{uv}(\vbeta)$ coefficients. These coefficients are given in \cref{app-tab:product/summary-a-expressions,app-tab:product/summary-a-expressions-ws,app-tab:product/summary-a-expressions-fam} for PM-QAOA in general, WS-QAOA, and FAM-QAOA respectively. 

Lastly, for the vanilla QAOA variant, the simplicity of the $\vu{r}_u$ settings allows yet further simplifications to be made on $\expval{C_{uv}}$. The $a^{PQ}_{uv}(\vbeta)$ coefficients have been worked out in \cref{app-eq:product/specific/qaoa/a}. As before, we show the edge-resolved cost expectation in the following cases:
\begin{itemize}

    \item \textit{Weighted problem graph, multi-angle circuit ansatz ($\vb*{\beta}$ and $\vb*{\gamma}$)}. Then, explicitly,
    \begin{equation}\begin{aligned}
        \expval{C_{uv}} 
        &= \frac{w_{uv}}{2}
        \left(
            1 
            - \sin(2\beta_u) \sin(2\beta_v) \xi^{YY}_{uv}(\vb*{\gamma})
            - \sin(2\beta_u) \cos(2\beta_v) \xi^{YZ}_{uv}(\vb*{\gamma})
            - \cos(2\beta_u) \sin(2\beta_v) \xi^{ZY}_{uv}(\vb*{\gamma})
        \right),
        \label{app-eq:maxcut_Cuv_qaoa}
    \end{aligned}\end{equation}
    where the $\xi^{PQ}_{uv}(\vb*{\gamma})$ coefficients are given by the second part of \cref{app-tab:summary-xi-expressions_maxcut}. 

    \item \textit{Unweighted problem graph, single-angle circuit ansatz ($\vb*{\beta} = \beta$, $\vb*{\gamma} = \gamma$, and $w_{uv} = 1$)}. Then additional simplifications give
    \begin{equation}
    \begin{aligned}
        \expval{C_{uv}} &= 
        \frac{1}{2} \left[
            1
            - \frac{1}{2} \sin^2(2\beta) \cos^{d+e-2f}(\gamma) 
                \left(1 - \cos^f(2\gamma)\right)
            +\frac{1}{2} \sin(4\beta) \sin(\gamma) 
                \left(\cos^d(\gamma)+\cos^e(\gamma)\right)
        \right].
    \label{app-eq:maxcut_Cuv_qaoa_simplified}
    \end{aligned}
    \end{equation}

    We remark that \cref{app-eq:maxcut_Cuv_qaoa_simplified} was first derived in Refs.~\onlinecite{PhysRevA.97.022304,hadfield2018quantum}.
    
\end{itemize}

\begin{table*}[!ht]
    \centering
    \setlength{\tabcolsep}{0pt}
    \renewcommand*{\arraystretch}{1.2}
    \newcommand\tstrut{\rule{0pt}{12pt}}
    \newcommand\bstrut{\rule[-8pt]{0pt}{0pt}}
    \begin{tabular}[t]{p{1.5cm} p{16cm}}
        \toprule
        $PQ$ & $\xi^{PQ}_{uv}(\vb*{\gamma})$ \\
        \midrule
        $XX$ &
            $\begin{aligned}
                \tstrut
                \frac{1}{2}R^\bbslash_{uv}(\vb*{\gamma})R^\bbslash_{vu}(\vb*{\gamma})\big[R^-_{uv}(\vb*{\gamma}) + R^+_{uv}(\vb*{\gamma})\big]
                \bstrut
            \end{aligned}$ \\
        \tabdashline & \tabdashline
        $YY$ &
            $\begin{aligned}
                \tstrut
                \frac{1}{2}R^\bbslash_{uv}(\vb*{\gamma})R^\bbslash_{vu}(\vb*{\gamma})\big[R^-_{uv}(\vb*{\gamma}) - R^+_{uv}(\vb*{\gamma})\big]
                \bstrut
            \end{aligned}$ \\
        \tabdashline & \tabdashline
        $YZ$ &
            $\begin{aligned}
                \tstrut
                -\sin(w_{uv}\gamma_{uv})R^\setminus_{uv}(\vb*{\gamma})
                \bstrut
            \end{aligned}$ \\
        \midrule
        $XX$ &
            $\begin{aligned}
                \tstrut
                \frac{1}{2}\cos^{d+e-2f}(\gamma)\left(1+\cos^f(2\gamma)\right)
                \bstrut
            \end{aligned}$ \\
        \tabdashline & \tabdashline
        $YY$ &
            $\begin{aligned}
                \tstrut
                \frac{1}{2}\cos^{d+e-2f}(\gamma)\left(1-\cos^f(2\gamma)\right)
                \bstrut
            \end{aligned}$ \\
        \tabdashline & \tabdashline
        $YZ$ &
            $\begin{aligned}
                \tstrut
                -\sin(\gamma)\cos^d(\gamma)
                \bstrut
            \end{aligned}$ \\
        \bottomrule
    \end{tabular}
    \caption{\textbf{Reduced $\xi^P_u(\vb*{\gamma})$ and $\xi^{PQ}_{uv}(\vb*{\gamma})$ expressions for \textsc{MaxCut}}. Expressions for general optimization problems are provided in \cref{app-tab:product/summary-xi-expressions}, specialized here for the \textsc{MaxCut} setting. The first three rows are for weighted problem graphs and multi-angle circuit ansatzes; the last three rows are for unweighted graphs and single-angle circuit ansatzes ($\vb*{\beta} = \beta, \vb*{\gamma} = \gamma, w_{uv} = 1$). Here, following notation in \cref{app:product/expval-C}, $d = \deg(u) - 1$, $e = \deg(v) - 1$, and $f$ is the number of triangles that has an edge $\{u, v\}$. By the symmetry property in \cref{app-eq:/product/symmetry/property}, the coefficient $\smash{a^{ZY}_{uv}(\vb*{\beta})}$ can be obtained from $\smash{a^{YZ}_{uv}(\vb*{\beta})}$ by swapping the $u$ and $v$ vertices. The functions $\smash{R^\setminus_{uv}(\vb*{\gamma})}$, $\smash{R^\bbslash_{uv}(\vb*{\gamma})}$, and $\smash{R^\pm_{uv}(\vb*{\gamma})}$ are given in \cref{eq:intro-example/PM/R}.}
    \label{app-tab:summary-xi-expressions_maxcut}
\end{table*}

\subsubsection{\textsc{MaxIndependentSet} (MIS)}
\label{app:specialize/mis}

\begin{center}\fbox{\begin{minipage}{\linewidth}
\textbf{Definition:} Given a weighted undirected graph $G = (V_G, E_G, r)$ with vertex set $V_G = [n]$ and edge set $E_G = \{\{u, v\}: u, v \in V_G\}$ comprising $m$ edges, the \textsc{MaxIndependentSet} problem is to find the largest subset of vertices $S_G\subset V_G$ such that any two vertices in $S$ do not form an edge. This optimization problem is formalized as
\begin{equation*}
\begin{aligned}
    &\maximize_{\vb{x} \in \{0, 1\}^n} 
    \quad 
    \sum_{u = 1}^n x_u 
    \quad 
    \text{subject to} 
    \quad x_u x_v \neq 1 \,\, \forall \,\, u, v\in V_G. 
\end{aligned}
\end{equation*}
\end{minipage}}
\end{center}
\vspace{4pt}

Generically, there are two ways to treat constrained problems. One could consider a mixer that encodes the constraints such that the evolution of the initial state is restricted to a subspace of feasible states \cite{hadfield2019quantum}. Alternatively, the QUBO formalism encodes the constraints as an additional term in the cost function with an associated tuning parameter and there are no restrictions on the mixer type \cite{glover2022quantum}. We treat the latter case in this appendix. 

We can express the associated cost function in QUBO form \cite{glover2022quantum,brady2023iterative} as
\begin{equation}
    C = \lambda_1 \sum_{u = 1}^n s_u x_u 
    - \lambda_2 \sum_{\{u,v\}\in E_G} x_u x_v,
\end{equation}
where $s_u$ are the MIS weights on vertices and $\lambda_1, \lambda_2 \in \mathbb{R}$ are adjustable parameters that determine the relative importance of the objective and constraint terms in $C$. Likewise, the phase Hamiltonian $A$ has analogous coefficients as $C$ but with primes to allow for flexibility in defining the phase unitary. In general, for constrained problems (such as MIS), the phase and cost Hamiltonians may be taken to be different in the QAOA---that is, the graph weights $a \neq a'$, $h_u \neq h_u'$, and $J_{uv} \neq J_{uv}'$. Here, we generically allow the MIS weights to be different, $s_u \neq s_u'$, and we also allow the adjustable parameters $\lambda_1 \neq \lambda_1'$ and $\lambda_2 \neq \lambda_2'$. 

As before, the cost function can be promoted to its corresponding cost Hamiltonian by the substitution $x_u = (\mathbb{I} - Z_u) / 2$, and prefactors (\eg~factors of $1 / 2$) can be absorbed into $\lambda_1$ and $\lambda_2$. Making the promotions, the cost and phase Hamiltonians for MIS are~\cite{brady2023iterative,saleem2023approaches}
\begin{equation}\begin{aligned}
    C &= \lambda_1 \sum_{u = 1}^n 
            s_u(\mathbb{I} - Z_u) 
        - \lambda_2 \sum_{\{u, v\} \in E_G}
            (\mathbb{I} - Z_u)
            (\mathbb{I} - Z_v) \\
    &= \left( \lambda_1 \sum_{u = 1}^n s_u 
            - \lambda_2 m \right) \mathbb{I}
        + \sum_{u = 1}^n \left[ \lambda_2 D_u - \lambda_1 s_u \right] Z_u
        - \lambda_2 \sum_{\{u, v\} \in E_G} Z_u Z_v, \\
    A 
    &= \left( \lambda_1' \sum_{u = 1}^n s_u' 
            - \lambda_2' m \right) \mathbb{I}
        + \sum_{u = 1}^n \left[ \lambda_2' D_u - \lambda_1' s_u \right] Z_u
        - \lambda_2' \sum_{\{u, v\} \in E_G} Z_u Z_v,
    \label{app-eq:mis_C_A}
\end{aligned}\end{equation}
where $D_u = \deg(u)$. 

To reduce the general setting of PM-QAOA, which treats arbitrary problem graphs with vertex and edge weights, to the present MIS problem, we set the graph weights
\begin{equation}\begin{aligned}
    a &= \lambda_1 \sum_{u = 1}^n s_u - \lambda_2 m,
    \qquad
    h_u = \lambda_2 D_u - \lambda_1 s_u,
    \qquad
    J_{uv} = -\lambda_2, \\
    a' &= \lambda_1' \sum_{u = 1}^n s_u' - \lambda_2' m,
    \qquad
    h_u' = \lambda_2' D_u - \lambda_1' s_u',
    \qquad
    J_{uv}' = -\lambda_2'.
\end{aligned}\end{equation}
Then the PM-QAOA cost and phase Hamiltonians, as defined in \cref{eq:product/definitions/ising-ham-cost,eq:product/definitions/ising-ham-phase}, reduce to those of MIS as written in \cref{app-eq:mis_C_A}. Cost expectations are then given by \cref{thm:PM}. We provide $\xi^P_u(\vb*{\gamma})$ and $\xi^{PQ}_{uv}(\vb*{\gamma})$ coefficients specialized for the present MIS setting in \cref{app-tab:mis_xi_eq}, wherein we use the shorthand notation
\begin{equation}\begin{split}
    R^\setminus_{uv}(\vb*{\gamma}) 
    &= \prod_{g \in \mathcal{N}_{u \setminus v}} 
        \cos(2\lambda'_2 \gamma_{ug}), \\
    R^\bbslash_{uv}(\vb*{\gamma}) 
    &= \prod_{g \in \mathcal{N}_{u \bbslash v}} 
        \cos(2\lambda'_2 \gamma_{ug}), \\
    R^\pm_{uv}(\vb*{\gamma}) 
    &= \prod_{a \in \mathcal{N}_{uv}}
        \cos[2\lambda'_2(\gamma_{ua} \pm \gamma_{va})],
    \label{eq:specialize/mis/R}
\end{split}\end{equation}
which is the analogue of \cref{app-eq:product/PM/R} or \cref{eq:intro-example/PM/R} in the main text but specialized here for MIS. We remind that $a^P_u(\vbeta)$ and $a^{PQ}_{uv}(\vbeta)$ coefficients are listed in \cref{app-tab:product/summary-a-expressions,app-eq:product/specific/qaoa/a,app-tab:product/summary-a-expressions-ws,app-tab:product/summary-a-expressions-fam} for general PM-QAOA and the various variants subsumed as special cases.

We remark that there are different formulations of MIS in existing analytical literature. For example, Ref.~\onlinecite{brady2023iterative} neglects the constant and quadratic terms and only evaluates the expectation of the linear terms ($Z_u$ terms) in the cost Hamiltonian $C$. In comparison, Ref.~\onlinecite{saleem2023approaches} considers phase unitaries comprising only the quadratic (penalty) terms, and cost expectations with respect to linear and quadratic terms of the cost Hamiltonian are evaluated.

\begin{table*}[!ht]
    \centering
    \setlength{\tabcolsep}{0pt}
    \renewcommand*{\arraystretch}{1.2}
    \newcommand\tstrut{\rule{0pt}{12pt}}
    \newcommand\bstrut{\rule[-8pt]{0pt}{0pt}}
    \begin{tabular}[t]{p{1.5cm} p{16cm}}
        \toprule
        $P$ & $\xi^P_u(\vb*{\gamma})$ \\
        \midrule
        $X$ &
            $\begin{aligned}
                \tstrut
                \cos[2\gamma_u(\lambda'_2D_u-\lambda'_1s_u)]\cos(2\lambda'_2\gamma_{uv})R^\setminus_{uv}(\vb*{\gamma})
                \bstrut
            \end{aligned}$ \\
        \tabdashline & \tabdashline
        $Y$ &
            $\begin{aligned}
                \tstrut
                \sin[2\gamma_u(\lambda'_2D_u-\lambda'_1s_u)]\cos(2\lambda'_2\gamma_{uv})R^\setminus_{uv}(\vb*{\gamma})
                \bstrut
            \end{aligned}$ \\
        \midrule
        $X$ &
            $\begin{aligned}
                \tstrut
                \cos[2\gamma(\lambda'_2D_u-\lambda'_1)]\cos^{d+1}(2\lambda'_2\gamma)
                \bstrut
            \end{aligned}$ \\
        \tabdashline & \tabdashline
        $Y$ &
            $\begin{aligned}
                \tstrut
                \sin[2\gamma(\lambda'_2D_u-\lambda'_1)]\cos^{d+1}(2\lambda'_2\gamma)
                \bstrut
            \end{aligned}$ \\
        \bottomrule
    \end{tabular} \\[1.5\baselineskip]
    \begin{tabular}[t]{p{1.5cm} p{16cm}}
        \toprule
        $PQ$ & $\xi^{PQ}_{uv}(\vb*{\gamma})$ \\
        \midrule
        $XX$ &
            $\begin{aligned}
                \tstrut
                &\frac{1}{2}R^\bbslash_{uv}(\vb*{\gamma})R^\bbslash_{vu}(\vb*{\gamma})\bigg[\cos[2\gamma_u(\lambda'_2D_u-\lambda'_1s_u)]\cos[2\gamma_v(\lambda'_2D_v-\lambda'_1s_v)]\big(R^-_{uv}(\vb*{\gamma})+R^+_{uv}(\vb*{\gamma})\big) \\
                &\qquad +\sin[2\gamma_u(\lambda'_2D_u-\lambda'_1s_u)]\sin[2\gamma_v(\lambda'_2D_v-\lambda'_1s_v)]\big(R^-_{uv}(\vb*{\gamma})-R^+_{uv}(\vb*{\gamma})\big)\bigg]
                \bstrut
            \end{aligned}$ \\
        \tabdashline & \tabdashline
        $XY$ &
            $\begin{aligned}
                \tstrut
                &\frac{1}{2}R^\bbslash_{uv}(\vb*{\gamma})R^\bbslash_{vu}(\vb*{\gamma})\bigg[\cos[2\gamma_u(\lambda'_2D_u-\lambda'_1s_u)]\sin[2\gamma_v(\lambda'_2D_v-\lambda_1's_v)]\big(R^-_{uv}(\vb*{\gamma})+R^+_{uv}(\vb*{\gamma})\big) \\
                &\qquad -\sin[2\gamma_u(\lambda'_2D_u-\lambda'_1s_u)]\cos[2\gamma_v(\lambda'_2D_v-\lambda'_1s_v)]\big(R^-_{uv}(\vb*{\gamma})-R^+_{uv}(\vb*{\gamma})\big)\bigg]
                \bstrut
            \end{aligned}$ \\
        \tabdashline & \tabdashline
        $YY$ &
            $\begin{aligned}
                \tstrut
                &\frac{1}{2}R^\bbslash_{uv}(\vb*{\gamma})R^\bbslash_{vu}(\vb*{\gamma})\bigg[\cos[2\gamma_u(\lambda'_2D_u-\lambda'_1s_u)]\cos[2\gamma_v(\lambda'_2D_v-\lambda'_1s_v)]\big(R^-_{uv}(\vb*{\gamma})-R^+_{uv}(\vb*{\gamma})\big) \\
                &\qquad +\sin[2\gamma_u(\lambda'_2D_u-\lambda'_1s_u)]\sin[2\gamma_v(\lambda'_2D_v-\lambda'_1s_v)]\big(R^-_{uv}(\vb*{\gamma})+R^+_{uv}(\vb*{\gamma})\big)\bigg]
                \bstrut
            \end{aligned}$ \\
        \tabdashline & \tabdashline
        $XZ$ &
            $\begin{aligned}
                \tstrut
                \sin[2\gamma_u(\lambda'_2D_u-\lambda'_1s_u)]\sin(2\lambda'_2\gamma_{uv})R^\setminus_{uv}(\vb*{\gamma})
                \bstrut
            \end{aligned}$ \\
        \tabdashline & \tabdashline
        $YZ$ &
            $\begin{aligned}
                \tstrut
                -\cos[2\gamma_u(\lambda'_2D_u-\lambda'_1s_u)]\sin(2\lambda'_2\gamma_{uv})R^\setminus_{uv}(\vb*{\gamma})
                \bstrut
            \end{aligned}$ \\
        \midrule
        $XX$ &
            $\begin{aligned}
                \tstrut
                &\frac{1}{2}\cos^{d+e-2f}(2\lambda'_2\gamma)\bigg[\cos[2\gamma(\lambda'_2D_u-\lambda'_1)]\cos[2\gamma(\lambda'_2D_v-\lambda'_1)]\big(1+\cos^f(4\lambda'_2\gamma)\big) \\
                &\qquad +\sin[2\gamma(\lambda'_2D_u-\lambda'_1)]\sin[2\gamma(\lambda'_2D_v-\lambda'_1)]\big(1-\cos^f(4\lambda'_2\gamma)\big)\bigg]
                \bstrut
            \end{aligned}$ \\
            \tabdashline & \tabdashline
            $XY$ &
            $\begin{aligned}
                \tstrut
                &\frac{1}{2}\cos^{d+e-2f}(2\lambda'_2\gamma)\bigg[\cos[2\gamma(\lambda'_2D_u-\lambda'_1)]\sin[2\gamma(\lambda'_2D_v-\lambda'_1)]\big(1+\cos^f(4\lambda'_2\gamma)\big) \\
                &\qquad -\sin[2\gamma(\lambda'_2D_u-\lambda'_1)]\cos[2\gamma(\lambda'_2D_v-\lambda'_1)]\big(1-\cos^f(4\lambda'_2\gamma)\big)\bigg]
                \bstrut
            \end{aligned}$ \\
        \tabdashline & \tabdashline
        $YY$ &
            $\begin{aligned}
                \tstrut
                &\frac{1}{2}\cos^{d+e-2f}(2\lambda'_2\gamma)\bigg[\cos[2\gamma(\lambda'_2D_u-\lambda'_1)]\cos[2\gamma(\lambda'_2D_v-\lambda'_1)]\big(1-\cos^f(4\lambda'_2\gamma)\big) \\
                &\qquad +\sin[2\gamma(\lambda'_2D_u-\lambda'_1)]\sin[2\gamma(\lambda'_2D_v-\lambda'_1)]\big(1+\cos^f(4\lambda'_2\gamma)\big)\bigg]
                \bstrut
            \end{aligned}$ \\
        \tabdashline & \tabdashline
        $XZ$ &
            $\begin{aligned}
                \tstrut
                \sin[2\gamma(\lambda'_2D_u-\lambda'_1)]\sin(2\lambda'_2\gamma)\cos^d(2\lambda'_2\gamma)
                \bstrut
            \end{aligned}$ \\
        \tabdashline & \tabdashline
        $YZ$ &
            $\begin{aligned}
                \tstrut
                -\cos[2\gamma(\lambda'_2D_u-\lambda'_1)]\sin(2\lambda'_2\gamma)\cos^d(2\lambda'_2\gamma)
                \bstrut
            \end{aligned}$ \\
        \bottomrule
    \end{tabular}
    \caption{\textbf{Reduced $\xi^P_u(\vb*{\gamma})$ and $\xi^{PQ}_{uv}(\vb*{\gamma})$ expressions for MIS}. Expressions for general optimization problems are provided in \cref{app-tab:product/summary-xi-expressions}, specialized here for the MIS setting. Here, $s_u$ are vertex weights on the MIS graph and $\lambda_1'$, $\lambda_2'$ are adjustable parameters of the optimization problem. The first three rows are for weighted problem graphs and multi-angle circuit ansatzes; the last three rows are for unweighted graphs and single-angle circuit ansatzes ($\vb*{\beta} = \beta, \vb*{\gamma} = \gamma, s_u = 1$). By the symmetry property in \cref{app-eq:/product/symmetry/property}, the coefficients $\smash{a^{YX}_{uv}(\vb*{\beta})}$, $\smash{a^{ZX}_{uv}(\vb*{\beta})}$, and $\smash{a^{ZY}_{uv}(\vb*{\beta})}$ can be obtained from $\smash{a^{XY}_{uv}(\vb*{\beta})}$, $\smash{a^{XZ}_{uv}(\vb*{\beta})}$, and $\smash{a^{YZ}_{uv}(\vb*{\beta})}$ respectively by swapping the $u$ and $v$ vertices. The functions $\smash{R^\setminus_{uv}(\vb*{\gamma})}$, $\smash{R^\bbslash_{uv}(\vb*{\gamma})}$, and $\smash{R^\pm_{uv}(\vb*{\gamma})}$ are given in \cref{eq:specialize/mis/R}. }
    \label{app-tab:mis_xi_eq}
\end{table*}

\clearpage
\pagebreak

\section{Derivation for QAOA for Grover-Type Mixers}
\label{app:grover}

In this section, we provide a comprehensive walk-through of derivations leading to our results reported in \cref{sec:grover}.

\subsection{General weighted hypergraphs}
\label{app:grover/weighted-hypergraph}

We adopt the same general setting and notation as in \cref{sec:grover/definitions}, which we reproduce here for convenience. We consider a weighted undirected hypergraph $G = (V_G, E_G, w)$ with vertex set $V_G = [n]$, edge set $\smash{E_G \subseteq 2^{V_G}}$ of size $m$, and weight function $w: E_G \to \mathbb{R}$, which defines the cost Hamiltonian $C$. We allow a generically different weight function $w': E_G \to \mathbb{R}$ which defines the phase Hamiltonian $A$. The cost and phase Hamiltonians and their terms are then 
\begin{equation}\begin{split}
    C &= \sum_{e \in E_G} C_e ,
    \qquad
    C_e = w_e Z_e,
    \qquad
    \vb{C} = \left[C_e\right]_{e \in E_G},
    \\
    A &= \sum_{e \in E_G} A_e,
    \qquad
    A_e = w'_e Z_e,
    \qquad
    \vb{A} = \left[A_e\right]_{e \in E_G},
    \label{eq:grover/definitions/general-hypergraph-cost-ham}
\end{split}\end{equation}
where $C_e$ and $A_e$ are the cost and phase contributions by edge $e$ respectively, and $\vb{C}$ and $\vb{A}$ are vectors containing these contributions over all edges.

\subsubsection{\texorpdfstring{Calculating $\mel{\Omega}{e^{i 
\vb*{\gamma} \cdot \vb{A}}}{\Omega}$ and $\mel{\Omega}{e^{i 
\vb*{\gamma} \cdot \vb{A}} Z_e}{\Omega}$ for an edge $e$, and analogues for $\ket{s}$}{}}
\label{app:grover/weighted-hypergraph/mel-Omega-exp-C-A-Omega}

For fixed angles $\vb*{\lambda} = (\lambda_1, \lambda_2, \ldots, \lambda_n) \in [0, 2 \pi)^n$ and $\vb*{\omega} = (\omega_1, \omega_2, \ldots, \omega_n) \in [0, \pi)^n$, we observe
\begin{equation}\begin{split}
    \bra{\Omega(\vb*{\lambda}, \vb*{\omega})} 
        \prod_{j = 1}^n Z_j^{r_j}
        \ket{\Omega(\vb*{\lambda}, \vb*{\omega})}
    = \prod_{j = 1}^n 
        \bra{\Omega(\lambda_j, \omega_j)}
        Z^{r_j}
        \ket{\Omega(\lambda_j, \omega_j)}
    = \prod_{\substack{j = 1 \\ r_j \,\, \mathrm{odd}}}^n
        \cos{\omega_j},
    \label{app:grover/weighted-hypergraph/mel-Omega-exp-C-A-Omega/mel-Omega-Z-Omega}
\end{split}\end{equation}
for arbitrary powers $\vb{r} = (r_1, r_2, \ldots r_n) \in \mathbb{Z}^n$. Thus, whenever $\cos{\omega_j} = 0$ and $r_j$ is odd for some $j \in [n]$, the inner product above vanishes. Since setting $\vb*{\lambda} = \vb{0}$ and $\vb*{\omega} = (\pi / 2, \pi / 2, \ldots, \pi / 2)$ recovers the special case $\ket{\Omega(\vb*{\lambda}, \vb*{\omega})} = \ket{s}$, the inner product above with respect to $\ket{s}$ is nonzero only when all $r_j$ is even. This is summarized as
\begin{equation}\begin{split}
    \bra{\Omega(\vb*{\lambda}, \vb*{\omega})} 
        \prod_{j = 1}^n Z_j^{r_j}
        \ket{\Omega(\vb*{\lambda}, \vb*{\omega})}
    &= 0
    \qquad \text{when} \qquad
    \exists j \in [n]: 
        \left( \omega_j \in \left\{\frac{\pi}{2}, \frac{3\pi}{2}\right\} \right) 
        \land 
        \left( \text{$r_j$ odd} \right), \\
    \bra{s} \prod_{j = 1}^n Z_j^{r_j} \ket{s}
    &= \begin{dcases}
        1 & \text{all $r_j$ even}, \\
        0 & \text{otherwise}.
    \end{dcases}
    \label{app:grover/weighted-hypergraph/mel-Omega-exp-C-A-Omega/mel-s-Z-s}
\end{split}\end{equation}

For the sake of notational simplicity, henceforth whenever unambiguous we shall suppress the fixed angles in $\ket{\Omega(\vb*{\lambda}, \vb*{\omega})}$ and simply write $\ket{\Omega}$. For cost angles $\vb*{\gamma} \in \mathbb{R}^m$ and phase Hamiltonian terms $\vb{A}$, and an arbitrary operator $O$, we have
\begin{equation}\begin{split}
    \bra{\Omega} e^{i \vb*{\gamma} \cdot \vb{A}} O \ket{\Omega}
    = \bra{\Omega} 
        \exp( i \sum_{e \in E_G} \gamma_e w'_e Z_e )
        O \ket{\Omega}
    &= \bra{\Omega} 
        \left[ \prod_{e \in E_G}
        \exp( i \gamma_e w'_e Z_e ) \right]
        O \ket{\Omega},
    \label{app:grover/weighted-hypergraph/mel-Omega-exp-C-A-Omega/main-derive-A-1}
\end{split}\end{equation}
since all the terms in $\vb{A}$ commute. Now note that $Z_e$ is involutory. Hence, using the identity in \cref{eq:app/preliminaries/identities/exp-involutory-cos-sin} and rewriting with \cref{eq:app/preliminaries/identities/prod-sum-rewriting-binary}, we obtain
\begin{equation}\begin{split}
    \prod_{e \in E_G}
        \exp( i \gamma_e w'_e Z_e )
    &= \prod_{e \in E_G}
        \left[ \cos(\gamma_e w'_e) \mathbb{I}
            + i \sin(\gamma_e w'_e) Z_e \right] \\
    &= \sum_{\vb{f} \in \mathbb{F}_2^m}
        \prod_{k = 1}^m
        \left[ \cos(\gamma_{e_k} w'_{e_k}) \right]^{1 - f_k}
        \left[ i \sin(\gamma_{e_k} w'_{e_k}) Z_{e_k} \right]^{f_k},
    \label{app:grover/weighted-hypergraph/mel-Omega-exp-C-A-Omega/main-derive-A-2}
\end{split}\end{equation}
where $e_k \in E_G$ denotes the $k^{\mathrm{th}}$ edge in the edge set assuming an arbitrary fixed ordering of edges, and the binary variable $f_k$ selects between the cosine and sine terms for that edge in the expansion of the product. But $\vb{f}$ can equally be interpreted as selecting a subset of edges of $G$ to form a subhypergraph. The edges in the subhypergraph contribute their sine terms, and remaining edges contribute their cosine terms. Let $\mathcal{H}_G$ be the set of all subhypergraphs of $G$---such that distinct elements of $\mathcal{H}_G$ have distinct edge sets. Then we can write
\begin{equation}\begin{split}
    \bra{\Omega} 
        \left[ \prod_{e \in E_G}
        \exp( i \gamma_e w'_e Z_e ) \right]
        O \ket{\Omega}
    &= \bra{\Omega} \sum_{H \in \mathcal{H}_G}
        \left[ \prod_{e \in E_H}
            i \sin(\gamma_e w'_e) Z_e \right]
        \left[ \prod_{e \in E_G \setminus E_H}
            \cos(\gamma_e w'_e) \right]
        O \ket{\Omega} \\
    &= \sum_{H \in \mathcal{H}_G}
        \bra{\Omega}
        \left[ \prod_{e \in E_H}
            i \sin(\gamma_e w'_e) \prod_{u \in e} Z_u \right]
        \left[ \prod_{e \in E_{G'} \setminus E_H}
            \cos(\gamma_e w'_e) \right]
        O \ket{\Omega} \\
    &= \sum_{H \in \mathcal{H}_G}
        \left( i^{m_H} \right)
        \left[ \prod_{e \in E_H}
            \sin(\gamma_e w'_e) \right]
        \left[ \prod_{e \in E_{G'} \setminus E_H}
            \cos(\gamma_e w'_e) \right]
        \bra{\Omega}
        \left[ \prod_{u \in V_H} Z_u^{\deg_H(u)} \right]
        O \ket{\Omega}.
    \label{app:grover/weighted-hypergraph/mel-Omega-exp-C-A-Omega/main-derive-A-3}
\end{split}\end{equation}

In the last equality above, we have collated the powers of $Z_u$ for each vertex $u \in V_H$, each accumulating to $\deg_H(u)$, the number of times $u$ appears in edges of the subhypergraph $H \in \mathcal{H}_G$. For notational ease, we define the odd-degree subset of vertices for an arbitrary hypergraph $H'$,
\begin{equation}\begin{split}
    V_{H'}^{\mathrm{odd}} &= \left\{ u \in V_{H'} : \deg_{H'}(u) \,\, \mathrm{odd} \right\}.
\end{split}\end{equation}

Examining $O = Z_e$ for an edge $e \in E_G$ and applying \cref{app:grover/weighted-hypergraph/mel-Omega-exp-C-A-Omega/mel-Omega-Z-Omega} gives
\begin{equation}\begin{split}
    \bra{\Omega}
    \left[ \prod_{u \in V_H} Z_u^{\deg_H(u)} \right]
    Z_e
    \ket{\Omega}
    = \bra{\Omega}
    \left[ \prod_{u \in V_H} Z_u^{\deg_H(u)} \right]
    \left( \prod_{v \in e} Z_v \right)
    \ket{\Omega}
    = \prod_{u \in V_{H \symdiff \{e\}}^{\mathrm{odd}}}
        \cos{\omega_u},
    \label{app:grover/weighted-hypergraph/mel-Omega-exp-C-A-Omega/mel-deg-Omega}
\end{split}\end{equation}
where $H \symdiff \{e\}$ is the symmetric difference between subhypergraph $H$ and the edge set $\{e\}$, that is, the edge $e$ is added to $H$ if absent in $H$ and removed otherwise. Explicitly, the following relations hold,
\begin{equation}\begin{split}
    E_{H \symdiff \{e\}} = E_H \symdiff \{e\},
    \qquad
    V_{H \symdiff \{e\}}^{\mathrm{odd}} 
        = V_H^{\mathrm{odd}} \symdiff e,
    \label{app:grover/weighted-hypergraph/mel-Omega-exp-C-A-Omega/V-H-symdiff-e}
\end{split}\end{equation}
and trivially $H \symdiff \{\varnothing\} = H$. Our results above motivate the definition of the structural factor
\begin{equation}\begin{split}
    T_G^e(\vb*{\gamma})
    &= \sum_{H \in \mathcal{H}_G}
        \left( i^{m_H} \right)
        \left[ \prod_{f \in E_H}
            \sin(\gamma_f w'_f) \right]
        \left[ \prod_{f \in E_G \setminus E_H}
            \cos(\gamma_f w'_f) \right]
        \left[ \prod_{u \in V_{H \symdiff \{e\}}^{\mathrm{odd}}}
            \cos{\omega_u} \right] \\
    &= \left[ \prod_{f \in E_G}
            \cos(\gamma_f w'_f) \right]
        \times \sum_{H \in \mathcal{H}_G}
        \left( i^{m_H} \right)
        \left[ \prod_{f \in E_H}
            \tan(\gamma_f w'_f) \right]
        \left[ \prod_{u \in V_{H \symdiff \{e\}}^{\mathrm{odd}}}
            \cos{\omega_u} \right],
    \label{app:grover/weighted-hypergraph/mel-Omega-exp-C-A-Omega/definition-loop-factors-T}
\end{split}\end{equation}
such that putting \cref{app:grover/weighted-hypergraph/mel-Omega-exp-C-A-Omega/mel-deg-Omega,app:grover/weighted-hypergraph/mel-Omega-exp-C-A-Omega/main-derive-A-3,app:grover/weighted-hypergraph/mel-Omega-exp-C-A-Omega/main-derive-A-1} together, we have
\begin{equation}\begin{split}
    \bra{\Omega} e^{i \vb*{\gamma} \cdot \vb{A}} Z_e \ket{\Omega}
    &= T_G^e(\vb*{\gamma}).
    \label{app:grover/weighted-hypergraph/mel-Omega-exp-C-A-Omega/final-result-Omega}
\end{split}\end{equation}

As $Z_\varnothing = \mathbb{I}$, we have $\smash{\bra{\Omega} e^{i \vb*{\gamma} \cdot \vb{A}} \ket{\Omega} = T_G^\varnothing(\vb*{\gamma})}$ as a special case. We point out the symmetry
\begin{equation}\begin{split}
    T_G^e(-\vb*{\gamma}) = T_G^e(\vb*{\gamma})^*,
    \label{app:grover/weighted-hypergraph/mel-Omega-exp-C-A-Omega/T-angle-sign-symmetry}
\end{split}\end{equation}
as can be observed from the form in \cref{app:grover/weighted-hypergraph/mel-Omega-exp-C-A-Omega/definition-loop-factors-T} above. Consider also that for $\vb*{\gamma} = \vb{0}$, the contributions from all non-empty $H \in \mathcal{H}_G$ vanish because of the sine term in the products. Only the empty subhypergraph contributes,
\begin{equation}\begin{split}
    T_G^e(\vb{0}) = \prod_{u \in e} \cos{\omega_u},
    \qquad
    T_G^\varnothing(\vb{0}) = 1.
    \label{app:grover/weighted-hypergraph/mel-Omega-exp-C-A-Omega/T-special-case-gamma-0}
\end{split}\end{equation}

Next, we consider specializing to the $\ket{s}$ state from the more general $\ket{\Omega}$. Substituting $\vb*{\omega} = (\pi / 2, \pi / 2, \ldots, \pi / 2)$ into \cref{app:grover/weighted-hypergraph/mel-Omega-exp-C-A-Omega/mel-deg-Omega}, or equivalently using \cref{app:grover/weighted-hypergraph/mel-Omega-exp-C-A-Omega/mel-s-Z-s} instead of \cref{app:grover/weighted-hypergraph/mel-Omega-exp-C-A-Omega/mel-Omega-Z-Omega} above, we find
\begin{equation}\begin{split}
    \bra{s}
    \left[ \prod_{u \in V_H}
        Z_u^{\deg_H(u)} \right]
    Z_e
    \ket{s} 
    &= \begin{dcases}
        1 & \text{$\deg_H(u)$ is even $\forall u \in V_H \setminus e$ but odd $\forall u \in e$} \\
        0 & \text{otherwise}
    \end{dcases} \\
    &= \begin{dcases}
        1 & \text{$H \symdiff \{e\}$ is even-regular} \\
        0 & \text{otherwise},
    \end{dcases}
    \label{app:grover/weighted-hypergraph/mel-Omega-exp-C-A-Omega/mel-deg-s}
\end{split}\end{equation}
where we say a hypergraph is even-regular (or even) if all of its vertices have even degree; that is, an even number of edges is incident to each vertex. This motivates the analogous definition of the structural factor
\begin{equation}\begin{split}
    L_G^e(\vb*{\gamma})
    &= \sum_{H \in \mathcal{C}_G^e}
        \left( i^{m_H} \right)
        \left[ \prod_{f \in E_H}
            \sin(\gamma_f w'_f) \right]
        \left[ \prod_{f \in E_G \setminus E_H}
            \cos(\gamma_f w'_f) \right] \\
    &= \left[ \prod_{f \in E_G}
            \cos(\gamma_f w'_f) \right]
        \sum_{H \in \mathcal{C}_G^e}
        \left( i^{m_H} \right)
        \left[ \prod_{f \in E_H}
            \tan(\gamma_f w'_f) \right],
    \label{app:grover/weighted-hypergraph/mel-Omega-exp-C-A-Omega/definition-loop-factors-L}
\end{split}\end{equation}
where $\mathcal{C}_G^e$ is the set of subhypergraphs $H$ of $G$ such that the symmetric difference $H \symdiff \{e\}$ is even-regular. Explicitly, defining $\mathcal{C}_G$ to be the set of even subhypergraphs of $G$, we write
\begin{equation}\begin{split}
    \mathcal{C}_G^e
    &= \{H \in \mathcal{H}_G: H \symdiff \{e\} \,\, \text{is even-regular}\} \\
    &= \{H \in \mathcal{H}_G: H \symdiff \{e\} \in \mathcal{C}_G\} \\
    &= \{H \symdiff \{e\}: H \in \mathcal{C}_G\},
    \label{app:grover/weighted-hypergraph/mel-Omega-exp-C-A-Omega/definition-C-G-e}
\end{split}\end{equation}
as also reported in \cref{eq:grover/single-layer/definition-C-G-e} of the main text, and we note trivially $\mathcal{C}_G^\varnothing = \mathcal{C}_G$. The last line of \cref{app:grover/weighted-hypergraph/mel-Omega-exp-C-A-Omega/definition-C-G-e} arises from the identity $H \symdiff \{e\} \symdiff \{e\} = H$, and establishes a bijection between $\mathcal{C}_G^e$ and $\mathcal{C}_G$, namely, that each subhypergraph in $\mathcal{C}_G^e$ can be produced by selecting a subhypergraph in $\mathcal{C}_G$ and taking the symmetric difference with $e$. With the loop factor as written,
\begin{equation}\begin{split}
    \bra{s} e^{i \vb*{\gamma} \cdot \vb{A}} Z_e \ket{s}
    &= L_G^e(\vb*{\gamma}),
    \label{app:grover/weighted-hypergraph/mel-Omega-exp-C-A-Omega/final-result-s}
\end{split}\end{equation}
and likewise we have $\smash{\bra{s} e^{i \vb*{\gamma} \cdot \vb{A}} \ket{s} = L_G^\varnothing(\vb*{\gamma})}$ as a special case. Like before, we point out the symmetry
\begin{equation}\begin{split}
    L_G^e(-\vb*{\gamma}) = L_G^e(\vb*{\gamma})^*,
    \label{app:grover/weighted-hypergraph/mel-Omega-exp-C-A-Omega/L-angle-sign-symmetry}
\end{split}\end{equation}
as can be observed from the form in \cref{app:grover/weighted-hypergraph/mel-Omega-exp-C-A-Omega/definition-loop-factors-L} above. Moreover at $\vb*{\gamma} = \vb{0}$, the contributions from all non-empty subhypergraphs $H$ vanish because of the sine term in the products, and only the empty subhypergraph can contribute. While $\mathcal{C}_G = \mathcal{C}_G^\varnothing$ contains the empty subhypergraph, $\mathcal{C}_G^e$ for $e \neq \varnothing$ does not, as the symmetric difference between the empty subhypergraph and the single edge non-empty $e$ cannot be even-regular. Thus
\begin{equation}\begin{split}
    L_G^e(\vb{0}) = \delta_{e \varnothing}.
    \label{app:grover/weighted-hypergraph/mel-Omega-exp-C-A-Omega/L-special-case-gamma-0}
\end{split}\end{equation}

In summary, to specialize to $\ket{s}$ from the more general $\ket{\Omega}$, one simply replaces $T_G^e \to L_G^e$, as in going from \cref{app:grover/weighted-hypergraph/mel-Omega-exp-C-A-Omega/final-result-Omega} to \cref{app:grover/weighted-hypergraph/mel-Omega-exp-C-A-Omega/final-result-s}. We remark that besides the circuit angles $\vb*{\gamma}$ and edge $e$ given as arguments for $T_G^e(\vb*{\gamma})$, the structural factor itself depends only on the structure of the hypergraph $G$---in particular it comprises sums of trigonometric factors multiplied over the edges and vertices of subhypergraphs of $G$. The same is true for $L_G^e(\vb*{\gamma})$, only that it is simpler and involves only the edges but not the vertices. The structural factors $T_G^e(\vb*{\gamma}), L_G^e(\vb*{\gamma})$ are used in $\smash{\expval{Z_e}}$ calculations, as described next in \cref{app:grover/weighted-hypergraph/p-layer-expval-Z-e}.

For future use, we define also super-structural factors $\overline{T}_G(\vb*{\gamma}), \overline{L}_G(\vb*{\gamma})$, which are $T_G^e(\vb*{\gamma}), L_G^e(\vb*{\gamma})$ summed over all edges $e \in E_G$ and weighted by the cost edge weights. We have
\begin{equation}\begin{split}
    \overline{T}_G(\vb*{\gamma})
    &= \sum_{e \in E_G} w_e T_G^e(\vb*{\gamma}),
    \qquad
    \overline{L}_G(\vb*{\gamma})
    = \sum_{e \in E_G} w_e L_G^e(\vb*{\gamma}),
    \label{app:grover/weighted-hypergraph/mel-Omega-exp-C-A-Omega/definition-loop-factors-T-L-super}
\end{split}\end{equation}
where $T_G^e(\vb*{\gamma}), L_G^e(\vb*{\gamma})$ are given in \cref{app:grover/weighted-hypergraph/mel-Omega-exp-C-A-Omega/definition-loop-factors-T,app:grover/weighted-hypergraph/mel-Omega-exp-C-A-Omega/definition-loop-factors-L}.

An alternative expression for $\overline{L}_G(\vb*{\gamma})$ can be obtained by rewriting $L_G^e(\vb*{\gamma})$ as a sum over subhypergraphs in $\mathcal{C}_G$ rather than $\mathcal{C}_G^e$, utilizing the bijection between $\mathcal{C}_G$ rather than $\mathcal{C}_G^e$ as established in the last line of \cref{app:grover/weighted-hypergraph/mel-Omega-exp-C-A-Omega/definition-C-G-e}. Expanding the symmetric difference, we have
\begin{equation}\begin{split}
    L_G^e(\vb*{\gamma})
    &= \sum_{\substack{H \in \mathcal{C}_G \\ e \in E_H}}
        \left( i^{m_H - 1} \right)
        \left[ \prod_{f \in E_H \setminus \{e\}}
            \sin(\gamma_f w'_f) \right]
        \left[ \prod_{f \in \left(E_G \setminus E_H\right) \cup \{e\}}
            \cos(\gamma_f w'_f) \right] \\
        & \qquad + \sum_{\substack{H \in \mathcal{C}_G \\ e \notin E_H}}
        \left( i^{m_H + 1} \right)
        \left[ \prod_{f \in E_H \cup \{e\}}
            \sin(\gamma_f w'_f) \right]
        \left[ \prod_{f \in \left(E_G \setminus E_H\right) \setminus \{e\}}
            \cos(\gamma_f w'_f) \right],
    \label{app:grover/weighted-hypergraph/mel-Omega-exp-C-A-Omega/definition-loop-factors-L-altform}
\end{split}\end{equation}
and thus
\begin{equation}\begin{split}
    \overline{L}_G(\vb*{\gamma})
    &= \sum_{H \in \mathcal{C}_G} 
        \left\{ \sum_{e \in E_H}
        w_e \left( i^{m_H - 1} \right)
        \left[ \prod_{f \in E_H \setminus \{e\}}
            \sin(\gamma_f w'_f) \right]
        \left[ \prod_{f \in \left(E_G \setminus E_H\right) \cup \{e\}}
            \cos(\gamma_f w'_f) \right] \right. \\
        & \qquad \qquad\quad + 
        \left. \sum_{e \in E_G \setminus E_H}
        w_e \left( i^{m_H + 1} \right)
        \left[ \prod_{f \in E_H \cup \{e\}}
            \sin(\gamma_f w'_f) \right]
        \left[ \prod_{f \in \left(E_G \setminus E_H\right) \setminus \{e\}}
            \cos(\gamma_f w'_f) \right] \right\}.
    \label{app:grover/weighted-hypergraph/mel-Omega-exp-C-A-Omega/definition-loop-factors-L-super-altform}
\end{split}\end{equation}

Similar to \cref{app:grover/weighted-hypergraph/mel-Omega-exp-C-A-Omega/T-angle-sign-symmetry,app:grover/weighted-hypergraph/mel-Omega-exp-C-A-Omega/L-angle-sign-symmetry}, we point out the symmetries
\begin{equation}\begin{split}
    & \overline{T}_G(-\vb*{\gamma}) = \overline{T}_G(\vb*{\gamma})^*,
    \qquad 
    \overline{L}_G(-\vb*{\gamma}) = \overline{L}_G(\vb*{\gamma})^*,
    \label{app:grover/weighted-hypergraph/mel-Omega-exp-C-A-Omega/T-L-super-angle-sign-symmetry}
\end{split}\end{equation}
and similar to \cref{app:grover/weighted-hypergraph/mel-Omega-exp-C-A-Omega/T-special-case-gamma-0,app:grover/weighted-hypergraph/mel-Omega-exp-C-A-Omega/L-special-case-gamma-0} we have
\begin{equation}\begin{split}
    \overline{T}_G(\vb{0}) 
    = \sum_{e \in E_G} 
        w_e
        \prod_{u \in e}
        \cos{\omega_u},
    \qquad
    \overline{L}_G(\vb{0})
    = w_\varnothing.
    \label{app:grover/weighted-hypergraph/mel-Omega-exp-C-A-Omega/T-L-super-special-case-gamma-0}
\end{split}\end{equation}

We remark that when the hypergraph $G$ is unweighted---more precisely without phase weights---further simplifications become available for the structural and super-structural factors, as discussed in \cref{app:grover/unweighted-hypergraph}.

\subsubsection{Calculating \texorpdfstring{$\expval{Z_e}$}{<Ze>} for an edge \texorpdfstring{$e$}{e} on \texorpdfstring{$p$}{p}-layer ansatz}
\label{app:grover/weighted-hypergraph/p-layer-expval-Z-e}

For a $p$-layer QAOA ansatz with cost angles $\smash{\left(\vb*{\gamma}^{(1)}, \vb*{\gamma}^{(2)}, \ldots, \vb*{\gamma}^{(p)}\right)}$ and mixer angles $\vb*{\beta} = (\beta_1, \beta_2, \ldots, \beta_p) \in \mathbb{R}^{p}$ for the layers, we are concerned with calculating the expectation value of $Z_e$ as measured at the end of the QAOA circuit. Trivially, for the empty edge $e = \varnothing$, as $Z_\varnothing = \mathbb{I}$ we straightforwardly have $\smash{\expval{Z_e}_{\ket{\Omega}}} = \smash{\expval{Z_e}_{\ket{s}}} = 1$; thus we are interested in the non-empty edges $e \neq \varnothing$ (though our analysis below nonetheless holds for the empty case also). We begin by expressing
\begin{equation}\begin{split}
    \expval{Z_e}_{\ket{\Omega}}
    &= \bra{\Omega} 
        e^{i \vb*{\gamma}^{(1)} \cdot \vb{A}} e^{i \beta_1 \ket{\Omega} \bra{\Omega}} 
        \ldots
        e^{i \vb*{\gamma}^{(p)} \cdot \vb{A}} e^{i \beta_p \ket{\Omega} \bra{\Omega}} 
        \, Z_e \,
        e^{-i \beta_p \ket{\Omega} \bra{\Omega}} e^{-i \vb*{\gamma}^{(p)} \cdot \vb{A}}
        \ldots
        e^{-i \beta_1 \ket{\Omega} \bra{\Omega}} e^{-i \vb*{\gamma}^{(1)} \cdot \vb{A}}
        \ket{\Omega} \\
    &= \bra{\Omega} 
        \left( \prodr_{l = 1}^p
            e^{i \vb*{\gamma}^{(l)} \cdot \vb{A}} e^{i \beta_l \ket{\Omega} \bra{\Omega}} \right)
        Z_e
        \left( \prodl_{l = 1}^p
            e^{-i \beta_l \ket{\Omega} \bra{\Omega}} e^{-i \vb*{\gamma}^{(l)} \cdot \vb{A}} \right)
        \ket{\Omega}.
\end{split}\end{equation}

As $\ket{\Omega} \bra{\Omega}$ is a projector, \cref{eq:app/preliminaries/identities/exp-projector} followed by \cref{eq:app/preliminaries/identities/prod-sum-rewriting-binary} can be invoked to obtain
\begin{equation}\begin{split}
    \expval{Z_e}_{\ket{\Omega}}
    &= \bra{\Omega} 
        \left\{ \prodr_{l = 1}^p
            e^{i \vb*{\gamma}^{(l)} \cdot \vb{A}} 
            \left[ \mathbb{I} + \left( e^{i \beta_l} - 1 \right) 
                \ket{\Omega} \bra{\Omega} \right]
            \right\}
        Z_e
        \left\{ \prodl_{l = 1}^p
            \left[ \mathbb{I} + \left( e^{-i \beta_l} - 1 \right) 
                \ket{\Omega} \bra{\Omega} \right]
            e^{-i \vb*{\gamma}^{(l)} \cdot \vb{A}} \right\}
        \ket{\Omega} \\
    &= \bra{\Omega}
        \left\{ \sum_{\vb{f} \in \mathbb{F}_2^p}
            \prodr_{l = 1}^p 
                e^{i \vb*{\gamma}^{(l)} \cdot \vb{A}} 
                \left[ \left( e^{i \beta_l} - 1 \right) 
                    \ket{\Omega} \bra{\Omega} \right]^{f_l}
        \right\} 
        Z_e
        \left\{ \sum_{\vb{g} \in \mathbb{F}_2^p}
            \prodl_{l = 1}^p
                \left[ \left( e^{-i \beta_l} - 1 \right) 
                    \ket{\Omega} \bra{\Omega} \right]^{g_l}
                e^{-i \vb*{\gamma}^{(l)} \cdot \vb{A}} 
        \right\}
        \ket{\Omega} \\
    &= \sum_{\vb{f} \in \mathbb{F}_2^p}
        \sum_{\vb{g} \in \mathbb{F}_2^p}
        \underbrace{
            \bra{\Omega}
            \left\{ 
                \prodr_{l = 1}^p 
                    e^{i \vb*{\gamma}^{(l)} \cdot \vb{A}} 
                    \left[ \left( e^{i \beta_l} - 1 \right) 
                        \ket{\Omega} \bra{\Omega} \right]^{f_l}
            \right\} 
            Z_e
            \left\{
                \prodl_{l = 1}^p
                    \left[ \left( e^{-i \beta_l} - 1 \right) 
                        \ket{\Omega} \bra{\Omega} \right]^{g_l}
                    e^{-i \vb*{\gamma}^{(l)} \cdot \vb{A}} 
            \right\}
            \ket{\Omega}
        }_{Q^{(e)}_{\vb{f} \vb{g}}}.
\end{split}\end{equation}

First, note that the $(e^{\pm i \beta_l} - 1)$ scalar factors can be moved out of the $\smash{Q^{(e)}_{\vb{f} \vb{g}}}$ inner product. Next, observe that each occurrence of the $\ket{\Omega} \bra{\Omega}$ projector, decided by $f_l$ or $g_l$ binary variables, splits the large inner product $\bra{\Omega} \ldots \ket{\Omega}$ into two. As a whole, $\vb{f}$ and $\vb{g}$ can thus be thought of as partitioning the entire $\bra{\Omega} \ldots \ket{\Omega}$ inner product into multiple shorter ones, with the partitions inserted at locations of nonzero $f_l$ and $g_l$ entries. To recast into this picture, we examine
\begin{equation}\begin{split}
    \vb{I}(\vb{f}) 
    &= \left( \text{indices where $\vb{f}$ is nonzero} \right)
    = \left( k = 1, 2, \ldots, \abs{\vb{f}}: f_k = 1 \right).
\end{split}\end{equation}
As defined, $\vb{I}(\vb{0})$ is the empty vector containing no entries, and $\abs{\vb{I}(\vb{f})} = \wt(\vb{f})$ in general. One can then rewrite
\begin{equation}\begin{split}
    Q^{(e)}_{\vb{f} \vb{g}} =
    \left[ \prod_{l = 1}^p 
        \left( e^{i \beta_l} - 1 \right)^{f_l}
        \left( e^{-i \beta_l} - 1 \right)^{g_l} 
    \right]
    & \left( 
        \bra{\Omega}
        \prodr_{l = 1}^{I(\vb{f})_1}
            e^{i \vb*{\gamma}^{(l)} \cdot \vb{A}} 
        \ket{\Omega} 
    \right)
    \left( 
        \prod_{k = 1}^{\abs{\vb{I}(\vb{f})} - 1}
        \bra{\Omega}
        \prodr_{l = I(\vb{f})_k + 1}^{I(\vb{f})_{k + 1}}
            e^{i \vb*{\gamma}^{(l)} \cdot \vb{A}}
        \ket{\Omega} 
    \right) \\
    & \times 
    \left[ 
        \bra{\Omega}
        \left(
        \prodr_{l = I(\vb{f})_{-1} + 1}^{p}
            e^{i \vb*{\gamma}^{(l)} \cdot \vb{A}}
        \right)
        Z_e
        \left(
        \prodl_{l = I(\vb{g})_{-1} + 1}^{p}
            e^{-i \vb*{\gamma}^{(l)} \cdot \vb{A}}
        \right) \ket{\Omega}
    \right] \\
    & \times 
    \left( 
        \prod_{k = 1}^{\abs{\vb{I}(\vb{g})} - 1}
        \bra{\Omega}
        \prodl_{l = I(\vb{g})_k + 1}^{I(\vb{g})_{k + 1}}
            e^{-i \vb*{\gamma}^{(l)} \cdot \vb{A}}
        \ket{\Omega} 
    \right)
    \left( 
        \bra{\Omega}
        \prodl_{l = 1}^{I(\vb{g})_1}
            e^{-i \vb*{\gamma}^{(l)} \cdot \vb{A}}
        \ket{\Omega} 
    \right).
\end{split}\end{equation}

Products above that are invalid---either because the upper limit is smaller than the lower limit, or because referenced entries in the limits do not exist---are to be regarded as being unit-valued. For notational ease, we define $I(\vb{f})_{-1} = I(\vb{f})_{\abs{\vb{I}(\vb{f})}}$ to be the last entry of $\vb{I}(\vb{f})$ when $\vb{I}(\vb{f})$ is non-empty and zero otherwise. Noting that $Z_e$ and the terms in $\vb{A}$ commute, and combining $\smash{e^{\pm i \vb*{\gamma}^{(l)} \cdot \vb{A}}}$ exponentials,
\begin{equation}\begin{split}
    Q^{(e)}_{\vb{f} \vb{g}} =
    \left[ \prod_{l = 1}^p 
        \left( e^{i \beta_l} - 1 \right)^{f_l}
        \left( e^{-i \beta_l} - 1 \right)^{g_l} 
    \right]
    & 
    \bra{\Omega}
    e^{i \sum_{l = 1}^{I(\vb{f})_1} \vb*{\gamma}^{(l)} \cdot \vb{A}}
    \ket{\Omega} 
    \left( 
        \prod_{k = 1}^{\abs{\vb{I}(\vb{f})} - 1}
        \bra{\Omega}
        e^{i \sum_{l = I(\vb{f})_k + 1}^{I(\vb{f})_{k + 1}} \vb*{\gamma}^{(l)} \cdot \vb{A}}
        \ket{\Omega} 
    \right) \\
    & \times 
    \bra{\Omega}
    e^{i \left( \sum_{l = I(\vb{f})_{-1} + 1}^p \vb*{\gamma}^{(l)}
        - \sum_{l = I(\vb{g})_{-1} + 1}^p \vb*{\gamma}^{(l)} \right) \cdot \vb{A}}
    Z_e
    \ket{\Omega} \\
    & \times 
    \left( 
        \prod_{k = 1}^{\abs{\vb{I}(\vb{g})} - 1}
        \bra{\Omega}
        e^{-i \sum_{l = I(\vb{g})_k + 1}^{I(\vb{g})_{k + 1}} \vb*{\gamma}^{(l)} \cdot \vb{A}}
        \ket{\Omega} 
    \right)
    \bra{\Omega}
    e^{-i \sum_{l = 1}^{I(\vb{g})_1} \vb*{\gamma}^{(l)} \cdot \vb{A}}
    \ket{\Omega}.
\end{split}\end{equation}

The inner product terms above depend only on the sum of the $\smash{\vb*{\gamma}^{(l)}}$ angles over layer intervals specified by $\vb{I}(\vb{f}), \vb{I}(\vb{g})$. Thus, for further simplification, we define
\begin{equation}\begin{split}
    \vb*{\Gamma}(\vb{f})
    = \left( 
        \sum_{l = 1}^{I(\vb{f})_1} \vb*{\gamma}^{(l)},
        \sum_{l = I(\vb{f})_1 + 1}^{I(\vb{f})_2} \vb*{\gamma}^{(l)},
        \ldots,
        \sum_{l = I(\vb{f})_{-1} + 1}^{p} \vb*{\gamma}^{(l)}
    \right).
    \label{eq:app/grover/weighted-hypergraph/expval-Z-e/definition-Gamma}
\end{split}\end{equation}

Sums above that are invalid---because the referenced entries in their limits do not exist---are to be regarded as absent in the $\vb*{\Gamma}(\vb{f})$ vector. In general $\abs{\vb*{\Gamma}(\vb{f})} = \wt(\vb{f}) + 1$, and since $\vb*{\Gamma}(\vb{f})$ is never empty, $\vb*{\Gamma}(\vb{f})_{-1} = \vb*{\Gamma}(\vb{f})_{\abs{\vb*{\Gamma}(\vb{f})}}$ always refers to the last entry of $\vb*{\Gamma}(\vb{f})$ by our notational convention. Then
\begin{equation}\begin{split}
    Q^{(e)}_{\vb{f} \vb{g}} =
    & 
    \left[ \prod_{l = 1}^p 
        \left( e^{i \beta_l} - 1 \right)^{f_l}
        \left( e^{-i \beta_l} - 1 \right)^{g_l} 
    \right] \\
    & \times 
    \left( 
        \prod_{k = 1}^{\wt(\vb{f})}
        \bra{\Omega}
        e^{i \vb*{\Gamma}(\vb{f})_k \cdot \vb{A}}
        \ket{\Omega} 
    \right)
    \left( 
        \prod_{k = 1}^{\wt(\vb{g})}
        \bra{\Omega}
        e^{-i \vb*{\Gamma}(\vb{g})_k \cdot \vb{A}}
        \ket{\Omega} 
    \right)
    \bra{\Omega}
    e^{i \left[ \vb*{\Gamma}(\vb{f})_{-1} - \vb*{\Gamma}(\vb{g})_{-1} \right] \cdot \vb{A}}
    Z_e
    \ket{\Omega}.
\end{split}\end{equation}

All inner product terms above are of the form analyzed in \cref{app:grover/weighted-hypergraph/mel-Omega-exp-C-A-Omega}. Substituting \cref{app:grover/weighted-hypergraph/mel-Omega-exp-C-A-Omega/final-result-Omega},
\begin{equation}\begin{split}
    Q^{(e)}_{\vb{f} \vb{g}} =
    \left[ \prod_{l = 1}^p 
        \left( e^{i \beta_l} - 1 \right)^{f_l}
        \left( e^{-i \beta_l} - 1 \right)^{g_l} 
    \right]
    \left(
        \prod_{k = 1}^{\wt(\vb{f})}
        T_G^\varnothing\left[\vb*{\Gamma}(\vb{f})_k\right] 
    \right)
    \left(
        \prod_{k = 1}^{\wt(\vb{g})}
        T_G^\varnothing\left[-\vb*{\Gamma}(\vb{g})_k\right] 
    \right)
    T_G^e\left[\vb*{\Gamma}(\vb{f})_{-1} - \vb*{\Gamma}(\vb{g})_{-1}\right].
    \label{eq:app/grover/weighted-hypergraph/expval-Z-e/expval-Omega-final-T-1}
\end{split}\end{equation}

Using the conjugation symmetry of $T_G^e$ in \cref{app:grover/weighted-hypergraph/mel-Omega-exp-C-A-Omega/T-angle-sign-symmetry}, we note the property
\begin{equation}\begin{split}
    Q^{(e)}_{\vb{g} \vb{f}} = \left[ Q^{(e)}_{\vb{f} \vb{g}} \right]^*,
    \label{eq:app/grover/weighted-hypergraph/expval-Z-e/Q-f-g-transpose-symmetry}
\end{split}\end{equation}
which is a statement about the Hermiticity of $\smash{Q^{(e)}}$ when interpreted as a matrix and implies that the diagonal elements are real, $\smash{Q^{(e)}_{\vb{f} \vb{f}} \in \mathbb{R}}$. Then the summation over $\vb{f}, \vb{g}$ could essentially be halved,
\begin{equation}\begin{split}
    \expval{Z_e}_{\ket{\Omega}}
    = \sum_{\vb{f} \in \mathbb{F}_2^p}
        \sum_{\vb{g} \in \mathbb{F}_2^p}
        Q^{(e)}_{\vb{f} \vb{g}}
    = \sum_{\vb{f} \in \mathbb{F}_2^p} 
        Q^{(e)}_{\vb{f} \vb{f}}
        + 2 \Re{ 
        \sum_{\vb{f} \in \mathbb{F}_2^p}
        \sum_{\substack{\vb{g} \in \mathbb{F}_2^p \\ \vb{g} < \vb{f}}}
        Q^{(e)}_{\vb{f} \vb{g}} },
    \label{eq:app/grover/weighted-hypergraph/expval-Z-e/expval-Omega-final-T-2}
\end{split}\end{equation}
where the comparison $\vb{g} < \vb{f}$ is lexicographic, or equivalently, performed by treating the bitstrings $\vb{f}, \vb{g}$ as binary representations of integers.

To specialize our result in \cref{eq:app/grover/weighted-hypergraph/expval-Z-e/expval-Omega-final-T-1,eq:app/grover/weighted-hypergraph/expval-Z-e/expval-Omega-final-T-2} to the state $\ket{s}$ from the more general $\ket{\Omega}$, one simply replaces $T_G^e \to L_G^e$. For a non-empty edge $e \neq \varnothing$, a small simplification can be had by recalling that $L_G^e(\vb{0}) = \delta_{e \varnothing}$, as in \cref{app:grover/weighted-hypergraph/mel-Omega-exp-C-A-Omega/L-special-case-gamma-0}. That is, $L_G^e\left[\Gamma(\vb{f})_{-1} - \Gamma(\vb{g})_{-1}\right]$ vanishes when $\Gamma(\vb{f})_{-1} = \Gamma(\vb{g})_{-1}$, which by the definition of $\vb*{\Gamma}$ in \cref{eq:app/grover/weighted-hypergraph/expval-Z-e/definition-Gamma} occurs whenever $\vb{f}$ and $\vb{g}$ share the same number of trailing zeros. Thus, $\smash{Q^{(e)}_{\vb{f} \vb{g}} = 0}$ whenever $\vb{f}$ and $\vb{g}$ share the same number of trailing zeros. Consequently, in \cref{eq:app/grover/weighted-hypergraph/expval-Z-e/expval-Omega-final-T-2} the $\smash{Q^{(e)}_{\vb{f} \vb{f}}}$ diagonal terms in the first sum all vanish, and there is a reduced number of terms in the second sum.

\subsubsection{Calculating \texorpdfstring{$\expval{C_e}$}{<Ce>} for an edge \texorpdfstring{$e$}{e} on \texorpdfstring{$p$}{p}-layer ansatz}
\label{app:grover/weighted-hypergraph/p-layer-expval-C-e}

Straightforwardly,
\begin{equation}\begin{split}
    \expval{C_e}_{\ket{\Omega}}
    = w_e \expval{Z_e}_{\ket{\Omega}},
    \label{app:grover/weighted-hypergraph/p-layer-expval-C-e/result}
\end{split}\end{equation}
where $\expval{Z_e}$ is given in \cref{eq:app/grover/weighted-hypergraph/expval-Z-e/expval-Omega-final-T-2} above, and likewise for $\expval{C_e}$ with respect to the $\ket{s}$ state instead of $\ket{\Omega}$. In the case of an empty edge $e = \varnothing$, trivially $\smash{\expval{C_\varnothing}_{\ket{\Omega}} = \expval{C_\varnothing}_{\ket{s}} = w_\varnothing}$ as $Z_\varnothing = \mathbb{I}$.

\subsubsection{Calculating \texorpdfstring{$\expval{C}$}{C} on \texorpdfstring{$p$}{p}-layer ansatz}
\label{app:grover/weighted-hypergraph/p-layer-expval-C}

Using \cref{eq:app/grover/weighted-hypergraph/expval-Z-e/expval-Omega-final-T-1} and the definition of the super-structural factor $\overline{T}_G$ in \cref{app:grover/weighted-hypergraph/mel-Omega-exp-C-A-Omega/definition-loop-factors-T-L-super}, we define analogously
\begin{equation}\begin{split}
    \overline{Q}_{\vb{f} \vb{g}}
    &= \sum_{e \in E_G} w_e Q^{(e)}_{\vb{f} \vb{g}} \\
    &= 
    \left[ \prod_{l = 1}^p 
        \left( e^{i \beta_l} - 1 \right)^{f_l}
        \left( e^{-i \beta_l} - 1 \right)^{g_l} 
    \right]
    \left(
        \prod_{k = 1}^{\wt(\vb{f})}
        T_G^\varnothing\left[\vb*{\Gamma}(\vb{f})_k\right] 
    \right)
    \left(
        \prod_{k = 1}^{\wt(\vb{g})}
        T_G^\varnothing\left[-\vb*{\Gamma}(\vb{g})_k\right] 
    \right)
    \overline{T}_G\left[\vb*{\Gamma}(\vb{f})_{-1} - \vb*{\Gamma}(\vb{g})_{-1}\right],
    \label{eq:app/grover/weighted-hypergraph/p-layer-expval-C/Q-f-g-super}
\end{split}\end{equation}
that is, $\smash{\overline{Q}_{\vb{f} \vb{g}}}$ has the exact same form as $\smash{Q^{(e)}_{\vb{f} \vb{g}}}$ but with $\smash{T_G^e}$ replaced by $\smash{\overline{T}_G}$. The conjugation symmetries of $\smash{T_G^\varnothing, \overline{T}_G}$ in \cref{app:grover/weighted-hypergraph/mel-Omega-exp-C-A-Omega/T-angle-sign-symmetry,app:grover/weighted-hypergraph/mel-Omega-exp-C-A-Omega/T-L-super-angle-sign-symmetry} yield likewise
\begin{equation}\begin{split}
    \overline{Q}_{\vb{g} \vb{f}} = \left[ \overline{Q}_{\vb{f} \vb{g}} \right]^*,
\end{split}\end{equation}
which is a statement about the Hermiticity of $\smash{\overline{Q}}$ when interpreted as a matrix, and also implies $\smash{\overline{Q}_{\vb{f} \vb{f}} \in \mathbb{R}}$. Then, with reference to \cref{eq:app/grover/weighted-hypergraph/expval-Z-e/expval-Omega-final-T-2,app:grover/weighted-hypergraph/p-layer-expval-C-e/result},
\begin{equation}\begin{split}
    \expval{C}_{\ket{\Omega}}
    = \sum_{e \in E_G} \expval{C_e}_{\ket{\Omega}}
    = \sum_{e \in E_G} w_e \expval{Z_e}_{\ket{\Omega}}
    &= \sum_{\vb{f} \in \mathbb{F}_2^p}
        \overline{Q}_{\vb{f} \vb{f}}
        + 2 \Re{ 
            \sum_{\vb{f} \in \mathbb{F}_2^p}
            \sum_{\substack{\vb{g} \in \mathbb{F}_2^p \\ \vb{g} < \vb{f}}}
            \overline{Q}_{\vb{f} \vb{g}} }.            
    \label{eq:app/grover/weighted-hypergraph/p-layer-expval-C/result}
\end{split}\end{equation}

Specializing from $\ket{\Omega}$ to $\ket{s}$ involves the simple replacement  $\smash{T_G^\varnothing \to L_G^\varnothing}$ and $\smash{\overline{T}_G \to \overline{L}_G}$ above. Similar to the observation made for \cref{eq:app/grover/weighted-hypergraph/expval-Z-e/expval-Omega-final-T-2}, a small simplification is available in this case by recalling that $\smash{\overline{L}_G(\vb{0}) = w_\varnothing}$ as in \cref{app:grover/weighted-hypergraph/mel-Omega-exp-C-A-Omega/T-L-super-special-case-gamma-0}. Thus, for hypergraphs $G$ that have zero weight on the empty edge ($w_\varnothing = 0$), or equivalently the problem instance contains no constant cost term, $\smash{\overline{Q}_{\vb{f} \vb{g}}} = 0$ whenever $\vb{f}$ and $\vb{g}$ shares the same number of trailing zeros. In particular, in \cref{eq:app/grover/weighted-hypergraph/p-layer-expval-C/result} the $\smash{\overline{Q}_{\vb{f} \vb{f}}}$ diagonal terms in the first sum all vanish, and there is a reduced number of terms in the second sum. In practice, this simplification is applicable to any hypergraph $G$ as the empty edge can always be separated and the contribution $\expval{C_\varnothing} = w_\varnothing$ separately noted; and \cref{eq:app/grover/weighted-hypergraph/p-layer-expval-C/result} which involves non-trivial computation is used for $G$ containing only the remaining non-empty edges.

\subsubsection{Evaluating \texorpdfstring{$\expval{Z_e}$}{<Ze>} for an edge \texorpdfstring{$e$}{e} on single-layer ansatz}
\label{app:grover/weighted-hypergraph/single-layer-expval-Z-e}

Here we give illustrative calculations for a single QAOA ansatz layer parametrized by angles $(\vb*{\gamma}, \beta)$. We evaluate \cref{eq:app/grover/weighted-hypergraph/expval-Z-e/expval-Omega-final-T-1,eq:app/grover/weighted-hypergraph/expval-Z-e/expval-Omega-final-T-2} with $p = 1$. We have $\vb{f}, \vb{g} \in \mathbb{F}_2^p = \mathbb{F}_2^1 = \{0, 1\}$, but we only need to enumerate $\vb{g} \leq \vb{f}$, so the relevant cases are $(\vb{f}, \vb{g}) = (0, 0), (1, 0), (1, 1)$. We establish
\begin{equation}\begin{split}
    \vb{I}(0) = (),
    \qquad
    \vb{I}(1) = (1),
    \qquad 
    \vb*{\Gamma}(0) = (\vb*{\gamma}),
    \qquad
    \vb*{\Gamma}(1) = (\vb*{\gamma}, 0).
    \label{app:grover/weighted-hypergraph/single-layer-expval-Z-e/I-Gamma}
\end{split}\end{equation}

Then
\begin{equation}\begin{split}
   Q^{(e)}_{\vb{f} \vb{g}}
   = \begin{dcases}
       T_G^e(\vb{0})
       & (\vb{f}, \vb{g}) = (0, 0) \\
       \left( e^{i \beta} - 1 \right)
       \left( e^{-i \beta} - 1 \right)
       T_G^\varnothing(\vb*{\gamma})
       T_G^\varnothing(-\vb*{\gamma})
       T_G^e(\vb{0})
       & (\vb{f}, \vb{g}) = (1, 1) \\
       \left( e^{i \beta} - 1 \right)
       T_G^\varnothing(\vb*{\gamma})
       T_G^e(-\vb*{\gamma})
       & (\vb{f}, \vb{g}) = (1, 0).
   \end{dcases}
   \label{app:grover/weighted-hypergraph/single-layer-expval-Z-e/Q-f-g}
\end{split}\end{equation}

Summing the above cases, as in \cref{eq:app/grover/weighted-hypergraph/expval-Z-e/expval-Omega-final-T-2},
\begin{equation}\begin{split}
   \expval{Z_e}_{\ket{\Omega}}
   = \left[
        1 
        + 2 \left( 1 - \cos{\beta} \right) \abs{T_G^\varnothing(\vb*{\gamma})}^2
    \right] T_G^e(\vb{0})
    + 2 \Re{ 
        \left( e^{i \beta} - 1 \right) 
        T_G^\varnothing(\vb*{\gamma}) 
        T_G^e(-\vb*{\gamma}) },
    \label{app:grover/weighted-hypergraph/single-layer-expval-Z-e/result-Omega}
\end{split}\end{equation}
where definitions for $T_G^e$ structural factors were given in \cref{app:grover/weighted-hypergraph/mel-Omega-exp-C-A-Omega/definition-loop-factors-T}, and a simplified expression for $T_G^e(\vb{0})$ is available in \cref{app:grover/weighted-hypergraph/mel-Omega-exp-C-A-Omega/T-special-case-gamma-0}. Specializing from $\ket{\Omega}$ to $\ket{s}$ involves replacing $\smash{T_G^e \to L_G^e}$ and $\smash{T_G^\varnothing \to L_G^\varnothing}$ above, giving
\begin{equation}\begin{split}
   \expval{Z_e}_{\ket{s}}
   &= \left[
        1 
        + 2 \left( 1 - \cos{\beta} \right) \abs{L_G^\varnothing(\vb*{\gamma})}^2
    \right] \delta_{e \varnothing}
    + 2 \Re{ 
        \left( e^{i \beta} - 1 \right) 
        L_G^\varnothing(\vb*{\gamma}) 
        L_G^e(-\vb*{\gamma}) } \\
    &= \begin{dcases}
        2 \Re{ 
            \left( e^{i \beta} - 1 \right) 
            L_G^\varnothing(\vb*{\gamma}) 
            L_G^e(-\vb*{\gamma}) } & \,\, \text{for} \,\, e \neq \varnothing \\
        1 & \,\, \text{for} \,\, e = \varnothing,
    \end{dcases}
    \label{app:grover/weighted-hypergraph/single-layer-expval-Z-e/result-s}
\end{split}\end{equation}
where definitions for $L_G^e$ structural factors were given in \cref{app:grover/weighted-hypergraph/mel-Omega-exp-C-A-Omega/definition-loop-factors-L}, and we have used the simplification for $L_G^e(\vb{0})$ from \cref{app:grover/weighted-hypergraph/mel-Omega-exp-C-A-Omega/L-special-case-gamma-0}.

\subsubsection{Evaluating \texorpdfstring{$\expval{C}$}{<C>} on single-layer ansatz}
\label{app:grover/weighted-hypergraph/single-layer-expval-C}

We evaluate \cref{eq:app/grover/weighted-hypergraph/p-layer-expval-C/Q-f-g-super,eq:app/grover/weighted-hypergraph/p-layer-expval-C/result} with $p = 1$ and circuit angles $(\vb*{\gamma}, \beta)$. The expressions for $\smash{\overline{Q}_{\vb{f} \vb{g}}}$ are identical to those for $\smash{Q^{(e)}_{\vb{f} \vb{g}}}$ in \cref{app:grover/weighted-hypergraph/single-layer-expval-Z-e/Q-f-g} but with $\smash{T_G^e}$ replaced by $\smash{\overline{T}_G}$. Thus
\begin{equation}\begin{split}
   \expval{C}_{\ket{\Omega}}
   = \left[
        1 
        + 2 \left( 1 - \cos{\beta} \right) \abs{T_G^\varnothing(\vb*{\gamma})}^2
    \right] \overline{T}_G(\vb{0})
    + 2 \Re{ 
        \left( e^{i \beta} - 1 \right) 
        T_G^\varnothing(\vb*{\gamma}) 
        \overline{T}_G(-\vb*{\gamma}) },
    \label{app:grover/weighted-hypergraph/single-layer-expval-C/result-Omega}
\end{split}\end{equation}
where definitions for $T_G^\varnothing$ and $\smash{\overline{T}_G}$ structural factors were given in \cref{app:grover/weighted-hypergraph/mel-Omega-exp-C-A-Omega/definition-loop-factors-T,app:grover/weighted-hypergraph/mel-Omega-exp-C-A-Omega/definition-loop-factors-T-L-super}, and a simplified expression for $\smash{\overline{T}_G(\vb{0})}$ is available in \cref{app:grover/weighted-hypergraph/mel-Omega-exp-C-A-Omega/T-L-super-special-case-gamma-0}. Specializing from $\ket{\Omega}$ to $\ket{s}$ involves replacing $\smash{T_G^\varnothing \to L_G^\varnothing}$ and $\smash{\overline{T}_G \to \overline{L}_G}$ above, giving
\begin{equation}\begin{split}
   \expval{C}_{\ket{s}}
   &= \left[
        1 
        + 2 \left( 1 - \cos{\beta} \right) \abs{L_G^\varnothing(\vb*{\gamma})}^2
    \right] w_\varnothing
    + 2 \Re{ 
        \left( e^{i \beta} - 1 \right) 
        L_G^\varnothing(\vb*{\gamma}) 
        \overline{L}_G(-\vb*{\gamma}) },
    \label{app:grover/weighted-hypergraph/single-layer-expval-C/result-s}
\end{split}\end{equation}
where definitions for $L_G^\varnothing$ and $\smash{\overline{L}_G}$  structural factors were given in \cref{app:grover/weighted-hypergraph/mel-Omega-exp-C-A-Omega/definition-loop-factors-L,app:grover/weighted-hypergraph/mel-Omega-exp-C-A-Omega/definition-loop-factors-T-L-super}, and we have invoked the simplification for $\overline{L}_G(\vb{0})$ in \cref{app:grover/weighted-hypergraph/mel-Omega-exp-C-A-Omega/T-L-super-special-case-gamma-0}. Alternatively one can evaluate explicitly the sum over all weighted edges $e$ of $\expval{Z_e}_{\ket{s}}$ in the second line of \cref{app:grover/weighted-hypergraph/single-layer-expval-Z-e/result-s}.

\subsection{Hypergraphs without phase weights}
\label{app:grover/unweighted-hypergraph}

Here we comment on the simpler setting of an undirected hypergraph $G$, as considered in \cref{app:grover/weighted-hypergraph} but with uniform phase weights $w'_e = 1$ for every edge $e \in E_G$. We still allow a generic cost weight function $w: E_G \to \mathbb{R}$. In this setting the structural factors in \cref{app:grover/weighted-hypergraph/mel-Omega-exp-C-A-Omega/definition-loop-factors-T,app:grover/weighted-hypergraph/mel-Omega-exp-C-A-Omega/definition-loop-factors-L} straightforwardly simplify,
\begin{equation}\begin{split}
    T_G^e(\vb*{\gamma})
    &= \sum_{H \in \mathcal{H}_G}
        \left( i^{m_H} \right)
        \left( \prod_{f \in E_H}
            \sin{\gamma_f} \right)
        \left( \prod_{f \in E_G \setminus E_H}
            \cos{\gamma_f} \right)
        \left( \prod_{u \in V_{H \symdiff \{e\}}^{\mathrm{odd}}}
            \cos{\omega_u} \right), \\
    L_G^e(\vb*{\gamma})
    &= \sum_{H \in \mathcal{C}_G^e}
        \left( i^{m_H} \right)
        \left( \prod_{f \in E_H}
            \sin{\gamma_f} \right)
        \left( \prod_{f \in E_G \setminus E_H}
            \cos{\gamma_f} \right).
    \label{app:grover/unweighted-hypergraph/mel-Omega-exp-C-A-Omega/definition-loop-factors-T-L}
\end{split}\end{equation}

In the case that the $\vb*{\gamma}$ angles are uniform, $\gamma_e = \gamma$ for all edges $e \in E_G$, the structural factors further simplify,
\begin{equation}\begin{split}
    T_G^e(\vb*{\gamma})
    &= \sum_{H \in \mathcal{H}_G}
        \left( i \sin{\gamma} \right)^{m_H}
        \left( \cos{\gamma} \right)^{m - m_H}
        \left( \prod_{u \in V_{H \symdiff \{e\}}^{\mathrm{odd}}}
            \cos{\omega_u} \right), \\
    L_G^e(\vb*{\gamma})
    &= \sum_{H \in \mathcal{C}_G^e}
        \left( i \sin{\gamma} \right)^{m_H}
        \left( \cos{\gamma} \right)^{m - m_H}.
    \label{app:grover/unweighted-hypergraph/mel-Omega-exp-C-A-Omega/definition-loop-factors-T-L-uniform-gamma}
\end{split}\end{equation}

With uniform $\vb*{\gamma}$ angles, the alternative expression for the super-structural factor $\overline{L}_G(\vb*{\gamma})$ in \cref{app:grover/weighted-hypergraph/mel-Omega-exp-C-A-Omega/definition-loop-factors-L-super-altform} also greatly simplifies,
\begin{equation}\begin{split}
    \overline{L}_G(\vb*{\gamma})
    &= \sum_{H \in \mathcal{C}_G} 
        \left( W_H \right)
        \left( i \sin{\gamma} \right)^{m_H - 1}
        \left( \cos{\gamma} \right)^{m - m_H + 1}
        + \left( W - W_H \right)
        \left( i \sin{\gamma} \right)^{m_H + 1}
        \left( \cos{\gamma} \right)^{m - m_H - 1},
    \label{app:grover/unweighted-hypergraph/mel-Omega-exp-C-A-Omega/definition-loop-factors-L-altform-1}
\end{split}\end{equation}
where $W = \sum_{e \in E_G} w_e$ and $W_H = \sum_{e \in E_H} w_e$ are the total cost weights on the hypergraph $G$ and subhypergraph $H$ respectively. In a case where the cost weights $w$ are also uniform, $w_e = 1$ for all edges $e \in E_G$, then $W = m$ and $W_H = m_H$ trivially, and we obtain
\begin{equation}\begin{split}
    \overline{L}_G(\vb*{\gamma})
    &= \sum_{H \in \mathcal{C}_G} 
        \left( m_H \right)
        \left( i \sin{\gamma} \right)^{m_H - 1}
        \left( \cos{\gamma} \right)^{m - m_H + 1}
        + \left( m - m_H \right)
        \left( i \sin{\gamma} \right)^{m_H + 1}
        \left( \cos{\gamma} \right)^{m - m_H - 1} \\
    &= \sum_{k = 1}^m N_G(k) \left[ 
        k \left( i \sin{\gamma} \right)^{k - 1}
        \left( \cos{\gamma} \right)^{m - k + 1}
        + \left( m - k \right)
        \left( i \sin{\gamma} \right)^{k + 1}
        \left( \cos{\gamma} \right)^{m - k - 1}
    \right],
    \label{app:grover/unweighted-hypergraph/mel-Omega-exp-C-A-Omega/definition-loop-factors-L-altform-2}
\end{split}\end{equation}
where $N_G(k)$ is the number of even-regular subhypergraphs of $G$ containing $k$ edges.

\subsection{General weighted graphs}
\label{app:grover/weighted-graph}

Here we remark on the more restrictive setting of an undirected weighted graph $G$, which is an undirected weighted hypergraph as considered in \cref{app:grover/weighted-hypergraph} but with edges comprising only pairs of vertices. That is,
\begin{equation}\begin{split}
    E_G \subseteq \{ \{u, v\}: u \in V_G, v \in V_G\}.
\end{split}\end{equation}

This setting does not yield immediate simplifications to the results derived on hypergraphs. However, more concrete interpretations of some results become possible. In particular, the set $\mathcal{C}_G$ of even-regular subgraphs, also termed Euler subgraphs, of $G$ is exactly the cycle space of $G$, which is the span of cycles in $G$ over the binary field. That is, every subgraph in $\mathcal{C}_G$ is comprised of a combination (symmetric difference) of cycles in $G$. Operationally, $\mathcal{C}_G$ is generated by the cycle basis of $G$. Correspondingly $\mathcal{C}_G^e$ as defined in \cref{app:grover/weighted-hypergraph/mel-Omega-exp-C-A-Omega/definition-C-G-e} comprises the symmetric difference between an edge $e$ and the cycle space $\mathcal{C}_G$.

\subsubsection{Forests and trees}
\label{app:grover/weighted-graph/forests}

In the special case that $G$ is a cycle-free graph, that is, $G$ is a forest, further simplifications become possible. In particular, the cycle space of $G$ contains only the empty subgraph, $\mathcal{C}_G = \{ H \}$ where $E_H = \varnothing$, and correspondingly $\mathcal{C}_G^e = \{ H \}$ where $E_H = \{ e \}$. Then \cref{app:grover/weighted-hypergraph/mel-Omega-exp-C-A-Omega/definition-loop-factors-L} reduces to
\begin{equation}\begin{split}
    L_G^e(\vb*{\gamma})
    = i \sin(\gamma_e w'_e)
        \left[ \prod_{f \in E_G \setminus \{e\}}
            \cos(\gamma_f w'_f) \right]
    = i \tan(\gamma_e w'_e)
        \left[ \prod_{f \in E_G}
            \cos(\gamma_f w'_f) \right].
    \label{app:grover/weighted-graph/loop-factor-L}
\end{split}\end{equation}

The super-structural factor in \cref{app:grover/weighted-hypergraph/mel-Omega-exp-C-A-Omega/definition-loop-factors-T-L-super} follows as
\begin{equation}\begin{split}
    \overline{L}_G(\vb*{\gamma})
    = i \left[ \sum_{e \in E_G} w_e \tan(\gamma_e w'_e) \right]
        \left[ \prod_{f \in E_G}
            \cos(\gamma_f w'_f) \right].
    \label{app:grover/weighted-graph/loop-factor-L-super}
\end{split}\end{equation}

The alternate form for $\overline{L}_G(\vb*{\gamma})$ in \cref{app:grover/weighted-hypergraph/mel-Omega-exp-C-A-Omega/definition-loop-factors-L-super-altform} can be observed to reduce to the same result.

\clearpage
\pagebreak

\end{document}